\pdfoutput=1
\documentclass[12pt]{article}%
\usepackage[numbers,sort&compress]{natbib}
\usepackage[usenames, dvipsnames]{xcolor}
\usepackage{graphicx}
\usepackage{multicol}
\usepackage{amsfonts}
\usepackage{amssymb}
\usepackage{amsmath}
\usepackage{heck}
\usepackage{afterpage}
\usepackage{setspace}
\usepackage{verbatim}
\usepackage{longtable}
\usepackage{float}
\usepackage{subcaption}
\usepackage{epsfig}
\usepackage{enumerate}
\usepackage{epstopdf}
\usepackage[enableskew, vcentermath]{youngtab}
\usepackage{adjustbox}
\usepackage{multirow}
\usepackage{tikz}
\usepackage[margin=1in]{geometry}
\usepackage{titletoc}
\usepackage{hyperref}
\usepackage{gensymb}
\usepackage{mathtools}%
\usepackage{tikz-cd}
\providecommand{\U}[1]{\protect\rule{.1in}{.1in}}
\pdfoutput=1
\newsavebox{\mysavebox}

\hypersetup{colorlinks,citecolor=black,filecolor=black,linkcolor=black,urlcolor=black}
\usetikzlibrary{decorations.markings}

\numberwithin{equation}{section}

\usetikzlibrary{chains}
\allowdisplaybreaks
\tikzset{node distance=2em, ch/.style={circle,draw,on chain,inner sep=2pt},chj/.style={ch,join},every path/.style={shorten >=4pt,shorten <=4pt},line width=1pt,baseline=-1ex}

\let\dlabel=\alabel

\newcommand{\dnode}[2][chj]{\node[#1,label={below:\dlabel{#2}}] {};
}

\newcommand{\dnodebr}[1]{\node[chj,label={below right:\dlabel{#1}}] {};
}

\newcommand{\ba}{\begin{eqnarray}}
\newcommand{\ea}{\end{eqnarray}}

\newcommand{\mf}{\mathfrak}

\newcommand{\Tr}{\, {\rm Tr}}

\newcommand{\ov}{\overset }

\newcommand{\be}{\begin{equation}}
\newcommand{\ee}{\end{equation}}

\newcommand{\kso}{\mathfrak{so}}
\newcommand{\ksp}{\mathfrak{sp}}
\newcommand{\ksu}{\mathfrak{su}}

\newcommand{\cc}{\mathbb{C}}
\newcommand{\zz}{\mathbb{Z}}

\tikzstyle{startstop} = [rectangle, rounded corners, minimum width=3cm, minimum height=1cm,text centered, draw=black, fill=blue!10]
\tikzstyle{startstop} = [rectangle, rounded corners, minimum width=3cm, minimum height=1cm,text centered, draw=black, fill=blue!10]
\tikzstyle{io} = [trapezium, trapezium left angle=70, trapezium right angle=110, minimum width=3cm, minimum height=1cm, text centered, draw=black, fill=blue!30]
\tikzstyle{process} = [rectangle, minimum width=3cm, minimum height=1cm, text centered, draw=black, fill=orange!30]
\tikzstyle{decision} = [diamond, minimum width=3cm, minimum height=1cm, text centered, draw=black, fill=green!30]
\tikzstyle{arrow} = [thick,->,>=stealth]
\tikzset{->-/.style={decoration={
  markings,
  mark=at position #1 with {\arrow[scale=2.4]{>}}},postaction={decorate}}}
\makeatletter \@addtoreset{equation}{section} \makeatother

\begin{document}

\date{January 2020}

\title{General Prescription for\\[4mm]Global $U(1)$'s in 6D SCFTs}

\institution{OXFORD}{\centerline{${}^{1}$Mathematical Institute, University of Oxford, OX2 6GG, UK}}

\institution{MILAN}{\centerline{${}^{2}$Dipartimento di Fisica, Universit\`a di Milano--Bicocca, Piazza della Scienza 3, I-20126 Milan, Italy}}

\institution{INFN}{\centerline{${}^{3}$INFN, Sezione di Milano--Bicocca, Piazza della Scienza 3, I-20126 Milan, Italy}}



\institution{PENN}{\centerline{${}^{4}$Department of Physics and Astronomy, University of Pennsylvania, Philadelphia, PA 19104, USA}}

\institution{IAS}{\centerline{${}^{5}$Institute for Advanced Study, Princeton, NJ 08540, USA}}

\authors{Fabio Apruzzi\worksat{\OXFORD}\footnote{e-mail: {\tt fabio.apruzzi@maths.ox.ac.uk}},
Marco Fazzi\worksat{\MILAN, \INFN}\footnote{e-mail: {\tt marco.fazzi@mib.infn.it}},\\[4mm]
Jonathan J. Heckman\worksat{\PENN}\footnote{e-mail: {\tt jheckman@sas.upenn.edu}},
Tom Rudelius\worksat{\IAS}\footnote{e-mail: {\tt rudelius@ias.edu}}, and
Hao Y. Zhang\worksat{\PENN}\footnote{e-mail: {\tt  zhangphy@sas.upenn.edu}}}

\abstract{We present a general prescription for determining the global $U(1)$ symmetries of
six-dimensional superconformal field theories (6D SCFTs).
We use the quiver-like gauge theory description of the tensor branch to identify candidate $U(1)$ symmetries
which can act on generalized matter. The condition that these candidate $U(1)$'s are free of
Adler-Bell-Jackiw (ABJ) anomalies provides bottom-up constraints for $U(1)$'s. This
agrees with the answer obtained from symmetry breaking patterns induced by
Higgs branch flows. We provide numerous examples illustrating the details of this proposal. In the F-theory realization of these
theories, some of these symmetries originate from deformations of non-abelian flavor symmetries localized on a component of the discriminant,
while others come from an additional generator of the Mordell-Weil group. We also provide evidence
that some of these global $U(1)$'s do not arise from gauge symmetries, as would happen in taking a decoupling limit of
a model coupled to six-dimensional supergravity.}


\maketitle

\setcounter{tocdepth}{2}

\tableofcontents


\newpage

\section{Introduction \label{sec:INTRO}}

There is a striking interplay between stringy extra-dimensional geometric
structures and low energy effective field theories. This is particularly
manifest in the context of F-theory compactifications, where intersecting
seven-branes are geometrized into elliptically fibered Calabi-Yau spaces
\cite{Vafa:1996xn, Morrison:1996na, Morrison:1996pp}. A prominent example illustrating the power
of such methods is the recent classification of six-dimensional superconformal
field theories (6D SCFTs) via F-theory \cite{Heckman:2013pva, Heckman:2015bfa}
(see also \cite{Bhardwaj:2015xxa, Bhardwaj:2019hhd}).
This provides a remarkably flexible approach to constructing 6D SCFTs which encompasses essentially all
other methods (see \cite{Witten:1995zh, Strominger:1995ac,
Ganor:1996mu, Seiberg:1996vs, Witten:1996qb, Witten:1996qz,
Bershadsky:1996nu, Brunner:1997gf, Blum:1997mm, Blum:1997fw, Aspinwall:1997ye, Intriligator:1997dh, Hanany:1997gh, Brunner:1997gk}
for a partial list of older references, references \cite{Apruzzi:2013yva, Gaiotto:2014lca, Cremonesi:2015bld, Bah:2017wxp, Apruzzi:2017nck}
for recent holographic examples, and \cite{Heckman:2018jxk} for a review).

With these results in place, it is natural to ask what detailed features of 6D
SCFTs can be extracted from the associated geometries. One piece of
information which is readily available from an F-theory model is the tensor
branch of the 6D SCFT moduli space, as this is encoded directly in terms of
K\"ahler deformations of the base of an F-theory model. Additionally, some
global symmetries correlate with the appearance of non-compact
seven-branes intersecting the localized region inhabited by a 6D SCFT. This,
in tandem with field theoretic techniques, has made it possible to tightly
constrain the structure of anomalies in 6D SCFTs
\cite{Ohmori:2014pca, Ohmori:2014kda, Heckman:2015ola, Cordova:2015fha, Shimizu:2017kzs, Cordova:2019wns}.

Even so, there are a number of outstanding open issues connected with
determining the structure of global symmetries in a 6D SCFT. In what follows we
exclusively focus on the case of continuous zero-form symmetries.
While in many cases there is a close match between the flavor symmetries expected from
geometric and field theoretic methods, there are notable examples where either
the F-theory model only captures a subset of possible flavor symmetries, and
conversely, where a ``na\"ive'' field theoretic analysis might at first suggest
a bigger flavor symmetry than what can actually be realized. Some examples of
this sort in F-theory \cite{Heckman:2015bfa} and field theory \cite{Ohmori:2015pua, Ohmori:2015pia}
are collected in the review article \cite{Heckman:2018jxk}. For the most part, these examples concentrate on
non-abelian global symmetries since these are more straightforward to identify
both in geometry and field theory.

Though only studied in a few references \cite{Bertolini:2015bwa, Heckman:2016ssk, Lee:2018ihr},
global $U(1)$ symmetries in 6D SCFTs are no less important. In
the context of 6D SCFTs, such symmetries are especially interesting because, as
opposed to non-abelian global symmetries, it is not possible to have a $U(1)$
gauge symmetry on the tensor branch of a 6D SCFT \cite{Hanany:1997gh, Cordova:2015fha}. In the F-theory
literature, some examples of $U(1)$ symmetries have also been tracked by
determining the appearance of an additional generator of the Mordell-Weil
group of the associated elliptic fibration \cite{Lee:2018ihr}.
By taking a suitable decompactification limit in which gravity decouples, this yields
a global $U(1)$ symmetry.

These results also point to several open issues, both in field theory and F-theory.
First of all, it is important to develop a systematic method to deduce the
appearance of $U(1)$ symmetries using purely field theoretic techniques.
Second, it is natural to ask whether $U(1)$ flavor symmetries in an F-theory
realization of a 6D SCFT always originate from generators of the Mordell-Weil group.

In this paper we present a general prescription for determining $U(1)$ flavor
symmetries in 6D SCFTs. Our starting point is the observation of reference
\cite{Heckman:2018pqx} that all 6D SCFTs can be viewed as either ``fission'' or
``fusion'' products obtained from a small list of progenitor theories. Fission
products are obtained by a combination of tensor branch flows accompanied by
Higgs branch deformations. Fusion products are obtained by taking some
collection of fission products and gauging a common flavor symmetry
(accompanied by introducing an additional tensor multiplet). In the language of
heterotic M-theory, these progenitor theories all arise from M5-branes probing
an ADE singularity wrapped by an $E_{8}$ nine-brane. As it arises so
frequently, we shall also view the theory of M5-branes probing an ADE
singularity as another class of progenitor theories.

We find that there are two ways in which a 6D\ SCFT\ can inherit a $U(1)$
symmetry from a progenitor theory. First of all, these symmetries can
originate from a non-abelian flavor symmetry factor in the progenitor. A
suitable Higgs branch deformation of such a theory can produce $U(1)$'s
from the Cartan subalgebra of this symmetry. Second of all, there can also be
$U(1)$ symmetries present in the progenitor theory itself. This turns out to
only occur when the progenitor has an A-type flavor symmetry, and is closely
related to the fact that an A-type finite subgroup of $SU(2)$ has a
$U(1)\subset SU(2)$ commutant group.

We also find that the resulting $U(1)$ symmetries obtained
from this process of fission and fusion typically involve a linear combination
of $U(1)$ flavor symmetries and gauged $U(1)$'s coming from the Cartan of a
non-abelian gauge group present on the tensor branch of moduli space. So,
while a purely group theoretic analysis of breaking patterns allows us to
calculate the symmetry group, it does not directly tell us much about how this
symmetry acts on the Hilbert space. Indeed, the mixing with gauge symmetries
complicates the calculation of anomaly polynomials based on anomaly matching
since the appearance of this $U(1)$ flavor symmetry is only partially
inherited from a $U(1)$ global symmetry of the progenitor theory.

To rectify this issue, we develop a more bottom-up prescription
for how to directly read off the global $U(1)$ symmetries of a given 6D SCFT
obtained from the quiver-like gauge theory arising on its tensor branch. For a set of
$N$ hypermultiplets transforming in a complex representation of a gauge group
(or a single hypermultiplet transforming in a pseudo-real representation)
we get a \textit{candidate} $U(1)$ symmetry. Most of
these $U(1)$'s turn out to suffer from Adler-Bell-Jackiw (ABJ) anomalies on the tensor branch, and
thus do not constitute genuine global symmetries. Some linear combinations,
however, do not suffer from any such anomalies and are thus valid candidate
global symmetries. This prescription also allows us to fix the $U(1)$ charge
assignments for quiver gauge theories with classical gauge groups. This turns
out to be the biggest class of examples where a $U(1)$ global symmetry
arises, and we present a general set of rules for how to read off the global
$U(1)$ symmetry in such situations, as well as in the more general case of quiver-like gauge theories
with general gauge algebras and matter content. These rules agree with the rules for $U(1)$
symmetries obtained from fission / fusion operations on progenitor theories.

Our prescription also allows us to address the geometric origin of $U(1)$
global symmetries in F-theory models decoupled from gravity. Precisely because
the $U(1)$ symmetries of a 6D\ SCFT are inherited from $U(1)$'s in a
progenitor theory, we see first of all that many candidate $U(1)$ symmetries
are only indirectly associated with the Mordell-Weil group of sections in an
F-theory model. This is best exemplified through the fact
that the only progenitor theories with a $U(1)$ symmetry are those
with an A-type global symmetry, and there are examples of 6D\ SCFTs (which we
discuss) with more than one $U(1)$ global symmetry.

This begs the question as to whether we can determine which $U(1)$'s do
originate from generators of the Mordell-Weil group. Here we present evidence
that in progenitor theories with a $U(1)$ symmetry, if they arise from M5-branes probing an ADE singularity (namely those with an
A-type non-abelian flavor symmetry), the corresponding $U(1)$ is associated in
F-theory to the appearance of an additional section. We exhibit
the form of this additional section and show that it is in standard
\textquotedblleft Morrison-Park\textquotedblright\ form \cite{Morrison:2012ei}. However, if the progenitor theories with a $U(1)$ arise from M5-branes probing an A-type singularity  which is wrapped by an $E_8$ nine-brane, we provide evidence
that this $U(1)$ is not associated to an additional section of the fibration. One issue is that
all ``natural attempts'' to find such a section appear to fail. A second issue is that in
the dual heterotic description, this $U(1)$ arises from an isometry of a non-compact K3 surface which is
destroyed by recoupling to gravity.

The rest of this paper is organized as follows. In section \ref{sec:prog} we
briefly review the structure of progenitor theories, and the fact that all 6D
SCFTs originate from a process of fission / fusion from this starting point.
In section \ref{sec:YouThaOne} we turn to a general discussion of $U(1)$
symmetries obtained from working with the tensor branch of a
6D\ SCFT. In particular, we give a general prescription for how to identify
candidate $U(1)$ symmetries and extract the corresponding 't Hooft anomalies.
Section \ref{sec:EXAMPLES} presents a number of examples illustrating our procedure.
In section \ref{sec:RG} we use this analysis to track the behavior of $U(1)$ symmetries in a
Higgs branch flow from the UV to the IR. In section \ref{sec:FTHEORY} we turn to the
geometric realization of $U(1)$ symmetries in the A-type progenitor theories.
We present our conclusions in section \ref{sec:CONC}. In Appendix \ref{app:A} we present some additional details on anomalies
with $U(1)$ symmetries. In Appendix \ref{sec:grouptheory}, we discuss how global symmetries can be obtained via group theoretic methods. In Appendix \ref{app:TBres}, we provide details of the F-theory construction of the 6D SCFT associated with heterotic $E_8$ small
instantons probing an A-type orbifold singularity.

\section{Fission, Fusion, and Progenitor Theories} \label{sec:prog}

In this section we review how all 6D\ SCFTs can be obtained from a small set
of progenitor theories \cite{Heckman:2018pqx}. The main idea is that starting from such progenitor
theories, we reach the vast majority of 6D\ SCFTs by a combination of a tensor
branch deformation followed by a Higgs branch deformation. The few 6D\ SCFTs
which cannot be obtained in this way instead result from a process of fusion
where we add an additional tensor multiplet and weakly gauge a common flavor symmetry of fission products.
For our present purposes, the main feature is that in both fission and fusion
products, there are a set of symmetries which are directly inherited from the
progenitor theory.

Recall that in the F-theory approach to constructing 6D\ SCFTs, we start with
an elliptically fibered Calabi-Yau threefold with a non-compact base. In the
base, we seek out configurations of curves which can all contract to zero size
simultaneously at finite distance in the moduli space of Calabi-Yau metrics.
This collapsing procedure results in a 6D\ SCFT. The tensor branch of moduli
space refers to instead considering K\"ahler deformations of this Calabi-Yau in
which curves of the base now have finite volume. The Higgs branch of moduli
space refers to switching on operator vevs which break the $SU(2)_{\mathcal{R}}$
R-symmetry of the SCFT. Geometrically, these are deformations in the complex
structure / intermediate Jacobian of the Calabi-Yau threefold . The results of
\cite{Heckman:2013pva, Heckman:2015bfa} (see also \cite{Tachikawa:2015wka, Bhardwaj:2018jgp})
provide a general procedure for constructing all known 6D\ SCFTs via
F-theory compactification (see \cite{Heckman:2018jxk} for a review). Here, our main interest
will be in providing a more uniform characterization of possible structures
which can appear in such theories, in particular flavor symmetries.

To track flavor symmetries in 6D\ SCFTs, we exploit the recently discovered
characterization of most 6D\ SCFTs as obtained from either fission or fusion
of a small set of progenitor theories. In M-theory language, these progenitor
theories arise from M5-branes probing a heterotic nine-brane which is in turn
wrapped by an ADE\ singularity $\mathbb{C}^{2}/\Gamma_{ADE}$, with
$\Gamma_{ADE}$ a finite order subgroup of $SU(2)$. In the F-theory
description, these theories have a partial tensor branch description given by
a configuration of curves of the form:%
\begin{equation}
\lbrack E_{8}],\underset{k}{\underbrace{\overset{\mathfrak{g}_{ADE}%
}{1},\overset{\mathfrak{g}_{ADE}}{2},...,\overset{\mathfrak{g}_{ADE}}{2}}%
},[G_{ADE}], \label{smallinst}%
\end{equation}
for the F-theory description of $k$ small instantons. Here, the notation
$\overset{\mathfrak{g}}{n}$ refers to a curve of self-intersection $-n$ in the
base of the elliptic threefold with a singular elliptic fibration over this
curve resulting in a gauge symmetry with algebra $\mathfrak{g}$. In stringy
terms, this corresponds to a seven-brane with gauge algebra $\mathfrak{g}$
wrapped over this curve. There are also pairwise intersections between the
different seven-branes which occur at a single point of normal crossing in the
base. Lastly there are non-abelian flavor symmetries associated with an
$E_{8}$ symmetry on the left, and a flavor symmetry of ADE-type $G_{ADE}$ on
the right.

Another prominent class of progenitor theories is obtained from M5-branes
probing the singular point of the geometry $\mathbb{R}_{\bot}\times
\mathbb{C}^{2}/\Gamma_{ADE}$. The resulting theories can be obtained by moving
the stack of M5-branes in the small instanton examples away from the $E_{8}$
wall. In the F-theory realization of these theories we simply decompactify the
$-1$ curve in the theory of (\ref{smallinst}). Doing so, we reach a 6D\ SCFT\ with
partial tensor branch:%
\begin{equation}
\lbrack G_{ADE}],\underset{k-1}{\underbrace{\overset{\mathfrak{g}_{ADE}%
}{2},...,\overset{\mathfrak{g}_{ADE}}{2}}},[G_{ADE}], \label{ADEprobe}%
\end{equation}

Starting from the theories of (\ref{smallinst}) and (\ref{ADEprobe}), we
reach the vast majority of 6D\ SCFTs by performing a tensor branch deformation
followed by a Higgs branch deformation. There is an algebraic characterization
of Higgs branch deformations in terms of group theoretic data associated with
the flavor symmetry factors. In the case of the theories in line (\ref{ADEprobe}),
localized deformations come from nilpotent orbits of the flavor symmetry
algebra. In the case of the deformations localized near the $-1$ curve of
(\ref{smallinst}), these deformations come from discrete group homomorphisms
$\text{Hom}(\Gamma_{ADE} , E_{8})$ \cite{DelZotto:2014hpa, Heckman:2015bfa}.
See Appendix \ref{sec:grouptheory} for additional details on the interplay between group theory and 6D SCFTs.
Let us note that there are additional Higgs branch deformations which come from more general
seven-brane recombination moves \cite{Hassler:2019eso}. For our present purposes
of viewing all 6D\ SCFTs as fission and fusion products this additional class
of flows will not play a role. See also references \cite{Cabrera:2019izd, Bourget:2019aer, Cabrera:2019dob}
for additional discussion of Higgs branch deformations of
theories with eight real supercharges.

The class of 6D~SCFTs which can be obtained in this way are referred to as
\textquotedblleft fission products.\textquotedblright\ There are some theories
from the classification of reference \cite{Heckman:2015bfa} which cannot be obtained in this
way. They can, however, all be obtained by gauging a common flavor symmetry of
such fission products, and are thus referred to as \textquotedblleft fusion
products.\textquotedblright\ Putting this together, we see that a flavor
symmetry of a progenitor theory can be mapped to both a set of fission
products as well as possible fusions thereof.

In principle, there can also be flavor symmetries which are not inherited from a progenitor theory. In the case of fission products,
these can appear due to emergent symmetries deep in the infrared of an RG flow. In the case of fusion products,
the process of consistently gauging a common flavor symmetry can also require introducing additional
matter fields (to cancel the corresponding gauge anomalies which arise). These additional flavors can in turn lead to the
appearance of additional flavor symmetries. For the most part, however, the symmetries inherited from progenitor theories
cover the vast majority of flavor symmetries which arise in 6D SCFTs.

\subsection{Generic Flavor Symmetries of Progenitor Theories}

Let us now turn to the flavor symmetries for the progenitor
theories. For the most part, there is a uniform characterization of the
expected flavor symmetries in such theories. As the rank of the flavor
symmetry and / or the number of M5-branes decreases, however, there can be
additional \textquotedblleft accidental\textquotedblright\ enhancements in the
flavor symmetry at the fixed point. One of our tasks will be to develop a
systematic prescription for dealing with such situations as well. Since it
will require additional care to treat such outlier cases, we defer a
discussion of these cases to subsequent sections.

Consider first the case of the heterotic $E_8$ small instanton probe theories. First of all,
we have a 10D\ gauge theory with $E_{8}$ gauge group and with this, a
corresponding $E_{8}$ flavor symmetry for the 6D\ theory. Similarly, from the
ADE\ singularity we get a 7D gauge theory with $G_{ADE}$ gauge group which
again leads to a flavor symmetry in the 6D\ theory. In addition to this, we
observe that in the absence of the singularity, there is a $Spin(4)$ flavor
symmetry associated with rotations transverse to the probe M5-branes but
inside the nine-brane. Writing:%
\begin{equation}
Spin(4)=SU(2)_{\mathcal{L}}\times SU(2)_{\mathcal{R}},
\end{equation}
we embed our finite order ADE\ subgroup into $SU(2)_{\mathcal{L}}$ since this has a
natural holomorphic action on $\mathbb{C}^{2}$. In this case, $SU(2)_{\mathcal{R}}$
corresponds to the R-symmetry of the 6D\ SCFT, a feature which is manifest in
the heterotic construction but not directly visible in the F-theory geometry.
Now, in the case where we have a D- or an E-type finite subgroup of
$SU(2)_{\mathcal{L}}$, the commutant subgroup is trivial, so this is the full set of
global symmetries. Additional structure appears in the A-type series. In the case
where we have an A-type subgroup $\mathbb{Z}_{N}$ with $N > 2$,
we preserve a $U(1)$ subgroup, which is an additional flavor symmetry.
In the special case $N=2$, even more is true:\ here we preserve the
$SO(4)=(SU(2)_{\mathcal{L}}\times SU(2)_{\mathcal{R}})/%
\mathbb{Z}
_{2}$ isometries, so there is an additional $\mathfrak{su}(2)$ flavor symmetry
algebra \cite{DelZotto:2014hpa, Heckman:2015bfa}. In the special case of a single
probe M5-brane, additional flavor symmetry enhancements arise. We will revisit
the analysis of flavor symmetries in these special cases in
sections \ref{sec:EXAMPLES} and \ref{sec:FTHEORY}.

\begin{figure}[t!]
\begin{center}
\includegraphics[scale = 0.5, trim = {0cm -0.5cm 0cm 0cm}]{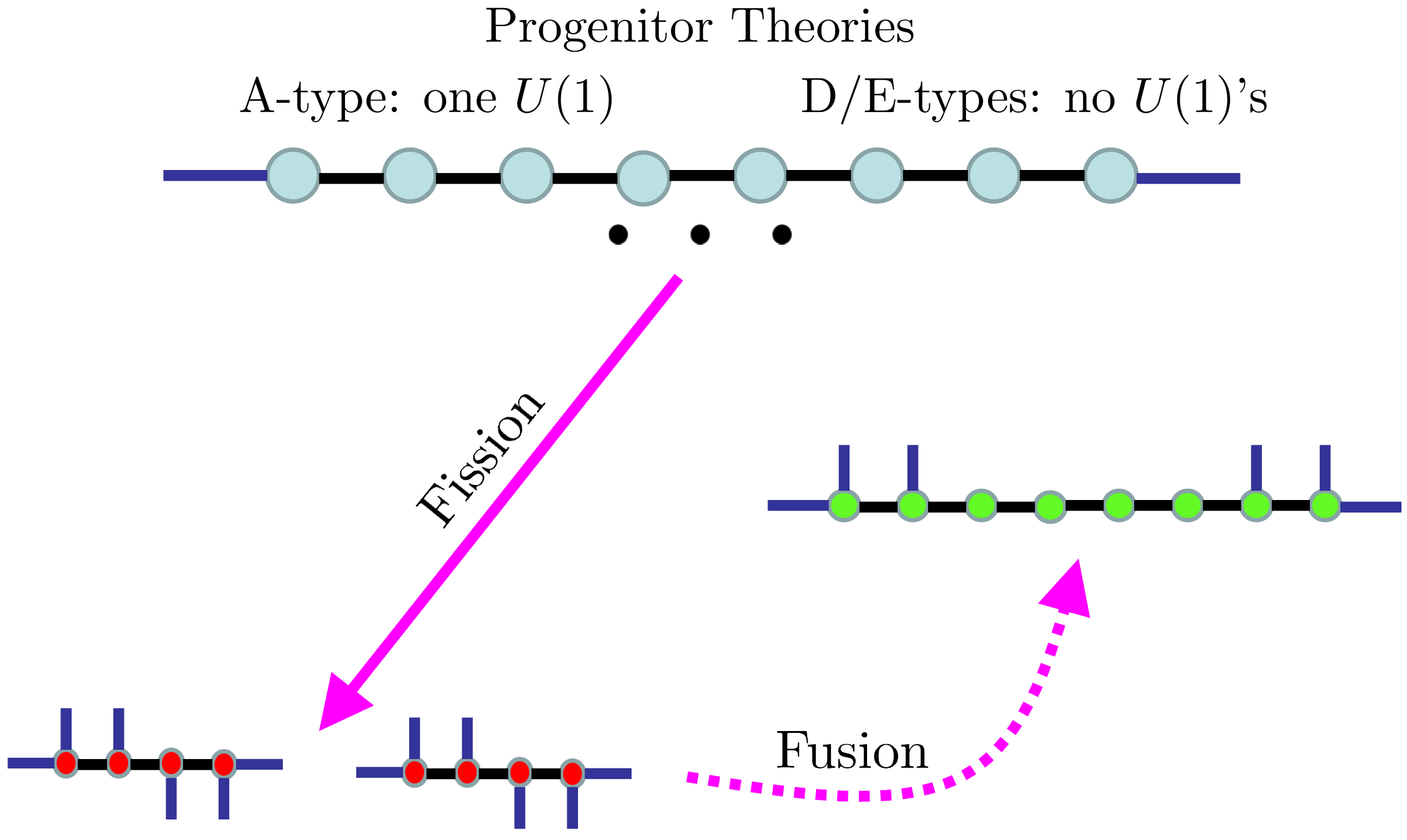}
\caption{Depiction of fission and fusion for 6D SCFTs. Progenitor theories arise from
M5-brane probes of an ADE singularity $\mathbb{C}^2 / \Gamma_{ADE}$, and can correspond
to cases with an $E_8$ nine-brane (heterotic $E_8$ small instantons)
as well as cases without such a nine-brane. In both sets of progenitor theories, there is a
$U(1)$ global symmetry factor for $\Gamma = \mathbb{Z}_N, N \geq 3$, whereas there is no abelian symmetry factor for D- and E-type singularities. For $\Gamma = \mathbb{Z}_2$, the $U(1)$ symmetry enhances to $SU(2)$. Deformations of these progenitor theories lead to ``fission'' products. Fission products can also be ``fused'' by gauging a common non-abelian global symmetry factors and adding an additional tensor multiplet.}
\label{fig:fissionfusion}
\end{center}
\end{figure}

Turning next to the theories of (\ref{ADEprobe}), we clearly observe a
$G_{L}\times G_{R}$ flavor symmetry. In the case of an A-type subgroup
$\mathbb{Z}_{N}$ with $N\geq2$, there is an additional $U(1)$ flavor group
symmetry, and it is more appropriate to write the flavor symmetry as
$S(U(N)_L \times U(N)_R)$. In the case of
an A-type subgroup $\mathbb{Z}_{2}$, we have a similar enhancement in the
flavor symmetry, this time to $SU(2)^{3}$. In
the case of a single M5-brane there are additional \textquotedblleft
accidental\textquotedblright\ enhancements in the flavor symmetries of the system.

To conclude this section, we note that here, we have focused on the
appearance of flavor symmetries which are manifest in the M-theory realization
of these systems. The non-abelian symmetries can also be extracted from the F-theory
realization of these 6D\ SCFTs, which is particularly important in
developing a uniform approach to classifying such
theories \cite{Heckman:2013pva, Heckman:2015bfa}. From the top-down perspective, however, the full flavor symmetry is not generically manifest in the complex structure moduli space of the Calabi-Yau threefolds used to engineer the theories. In some cases, the full flavor symmetry can only be realized geometrically at tuned points of the complex structure moduli space. Indeed, the typical expectation is that
these top-down constructions can sometimes \textquotedblleft
underpredict\textquotedblright\ possible flavor symmetry enhancements which
occur at the conformal fixed point. This is especially important in the
context of $U(1)$ symmetries since these factors may in fact be generators in
the Cartan of a single simple factor, for example the enhancement
$S(U(N)\times U(N))$ $\subset SU(2N)$.\footnote{See \cite{Lee:2018ihr} for more examples of this phenomenon.} As a general rule of thumb, these sorts
of accidental enhancements in the flavor symmetry tend to appear when the
number of tensor multiplets on the tensor branch is very low. For example, in
the progenitor theories, this occurs for a single small instanton of heterotic
theory next to an A-type singularity, and occurs for two M5-branes next to an
A-type singularity. The procedure we outline for extracting $U(1)$ symmetries
will provide a diagnostic for understanding when such enhancements occur.\footnote{While we can understand many such enhancements in this way, there exists at least one enhancement (namely, the $Spin(4) \rightarrow Spin(5)$ enhancement of the R-symmetry of (2,0) 6D SCFTs) that lies beyond the scope of our analysis, and we cannot be sure that a similar enhancement does not occur elsewhere.}

\section{$U(1)$ Symmetries on the Tensor Branch} \label{sec:YouThaOne}

In this section we turn to an analysis of $U(1)$ symmetries using the tensor
branch description of a 6D\ SCFT.\footnote{Note that we are interested in determining the flavor symmetries of the SCFT at the fixed point \emph{using the tensor branch} description, rather than in determining the flavor symmetries \emph{of the tensor branch theory} itself. These two flavor symmetries do not always agree, as in the case of the SCFT whose tensor branch description consists of $\mf{su}(2)$ gauge theory with four hypermultiplets, which has a $Spin(7)$ flavor symmetry at the conformal fixed point that enhances to $SO(8)$ on the tensor branch \cite{Ohmori:2015pia,Hanany:2018vph}.} From the general results of \cite{Heckman:2013pva, DelZotto:2014hpa, Heckman:2015bfa},
we already know that all 6D\ SCFTs resemble, on a partial tensor branch, a
6D\ quiver-like gauge theory, possibly with strongly coupled \textquotedblleft
conformal matter\textquotedblright\ between neighboring gauge group factors.
This fact was heavily used in references \cite{Ohmori:2014kda, Heckman:2015ola} (see also
\cite{Cordova:2015fha}) to extract the anomaly polynomial for non-abelian flavor
symmetries in 6D\ SCFTs. The main idea in this analysis is that when the
number of gauge group factors is equal to the number of tensor multiplets (on
the partial tensor branch), the Green-Schwarz-West-Sagnotti (GSWS) mechanism for
canceling gauge theoretic anomalies \cite{Green:1984sg, Green:1984bx, Sagnotti:1992qw}
leads to a unique answer for the flavor symmetry field strengths as well.
The general form of the anomaly polynomial thus obtained then takes the form:
\begin{equation}
I_{\text{full}}=I_{\text{1-loop}}+I_{\text{GS}},
\end{equation}
where $I_{\text{full}}$ denotes a formal eight-form in the flavor symmetry
field strengths, R-symmetry field strength, and background curvature, $I_{\text{1-loop}}$ denotes the one-loop contributions from
\textquotedblleft generalized matter\textquotedblright\ and $I_{\text{GS}}$
denotes the contribution from Green-Schwarz terms of the form:%
\begin{equation}
L_{\text{GS}}\supset\underset{\text{6D}}{\int}\mu_{TG}\text{ }B^{(T)}\wedge
\text{Tr}(F^{(G)}\wedge F^{(G)}).
\end{equation}
where here, the index $T$ runs over the tensor multiplets, and $G$ runs over
field strengths for both gauge and flavor symmetries. The key point is that if
we can find a presentation where the number of gauge groups and tensor
multiplets are the same, then there is a unique way (up to lattice automorphisms \cite{Apruzzi:2017iqe})
to adjust the $\mu_{TG}$ coefficients such that we cancel all gauge anomalies. To date, this sort of
analysis has been primarily carried out for non-abelian symmetries. This
includes global flavor symmetries as well as R-symmetries and diffeomorphisms.

Our main aim in this section will be to understand in general terms the
structure of $U(1)$ abelian symmetries. First of all, there are no $U(1)$
gauge symmetries on the tensor branch. This can be seen directly in the
F-theory realization of such models because these are always associated with
the existence of additional (rational) sections to the elliptic fibration \cite{Morrison:1996pp,Aspinwall:2000kf,Aspinwall:1998xj}.
This in turn requires the existence of a compact base, so in a limit where
gravity is decoupled, such $U(1)$'s are also non-dynamical. Additionally, one
can also show from purely field theoretic considerations that no $U(1)$ vector
multiplets are available on the tensor branch \cite{Hanany:1997gh, Cordova:2015fha}.

A common way to generate examples of 6D\ SCFTs with non-abelian flavor
symmetries is to first begin with a theory that has a gauge symmetry on its
tensor branch. Taking a suitable decoupling limit then produces the desired
non-abelian flavor symmetry. Such a procedure is clearly unavailable for
abelian symmetries since there are no $U(1)$ gauge symmetries available to
begin with. A related way to proceed is to consider constructing the 6D\ SCFT as
an emergent sector of a 6D\ supegravity model. This method of analysis was
used in \cite{Lee:2018ihr} to argue for the existence of $U(1)$ global
symmetries coming from the geometry of an F-theory model. While this is
definitely a way to build robust examples of global $U(1)$ symmetries, there
is the additional constraint that we couple to 6D\ supegravity, which in turn
places an upper bound on the kinds of SCFTs we can realize. For example, the
sorts of singularities which can be supported in a compact elliptic Calabi-Yau
are bounded \cite{GrassiThesis, Gross:1993fd, Kumar:2010ru}, thereby constraining the 6D\ SCFTs that can be
coupled to gravity \cite{Heckman:2019bzm}.

Our plan will be to sort out in bottom-up terms
possible couplings which could appear in our analysis of candidate $U(1)$
symmetries and their anomalies. We can, of course, consider including
additional couplings to the anti-chiral two-forms of a tensor multiplet, which
we can summarize by a general set of couplings such as:%
\begin{equation}
L_{\text{GS}}\supset\underset{6D}{\int}\mu_{T,S}\text{ }B^{(T)}_{\text{2-form}}\wedge
X^{(S)}_{\text{4-form}}(F_{\text{global}}), \label{GScoupling}%
\end{equation}
where $X^{(S)}_{\text{4-form}}(F_{\text{global}})$ is a general four-form which
depends on the field strengths of the global and gauge symmetries.

However more is possible with abelian symmetries since the first Chern class need
not vanish for the associated field strengths. In principle, then, we can also
entertain \textquotedblleft generalized Green-Schwarz
couplings\textquotedblright\ such as:%
\begin{equation}
L_{\text{gGS}}\supset\underset{6D}{\int}\kappa_{l,a}\text{ }C_{(\text{0-form)}%
}^{(l)}\wedge X_{(\text{6-form)}}^{(a)}(F_{\text{global}})+\widetilde{\kappa
}_{l,b}\text{ }\widetilde{C}_{(\text{4-form)}}^{(l)}\wedge X_{(\text{2-form)}%
}^{(b)}(F_{\text{global}}),
\end{equation}
with $X_{(\text{6-form)}}^{(a)}(F_{\text{global}})$ a general six-form built
from the global symmetry field strengths, and $X_{(\text{2-form)}}%
^{(b)}(F_{\text{global}})$ a two-form. Here, $C_{(\text{0-form)}}^{(l)}$
denotes a set of zero-forms and $\widetilde{C}_{(\text{4-form)}}^{(l)}$
denotes their magnetic duals. Note that in both the six-form and two-form
$X$'s, at least one abelian field strength must participate. In principle,
such terms might appear in the study of $U(1)$ symmetries on the tensor branch.

That being said, such terms never directly impact the structure of
anomalies in a 6D\ SCFT. To see why, observe that to get a contribution to the
anomaly polynomial, we must necessarily pair up one of the terms coming from
the $\kappa$-terms with one coming from the $\widetilde{\kappa}$-terms.
Otherwise, we cannot get a contribution to the formal eight-form. However if there
is a coupling to the four-form axion, then the associated $U(1)$ symmetry
appearing in $X_{(\text{2-form)}}^{(b)}(F_{\text{global}})$ will have already
been broken via the St\"uckelberg mechanism \cite{Park:2011wv}.
Consequently, for a genuine unbroken symmetry, such couplings play no
role in our analysis.

As a consequence, we conclude that to study the structure of anomalies with
$U(1)$ global symmetries, it is enough to consider the standard
GSWS anomaly cancelation mechanism, as well as the
resulting anomaly polynomial. With this in mind, suppose that we have a
candidate $U(1)$ global symmetry which contributes to the anomaly polynomial.
A priori, there are two sorts of terms which could be present:%
\begin{equation}
F_{\text{abelian}}F_{\text{abelian}}^{\prime}\text{Tr}(F_{\text{gauge}}%
^{2})\text{ \ \ and \ \ }F_{\text{abelian}}\text{Tr}(F_{\text{gauge}}^{3}),
\end{equation}
for some abelian symmetries and gauge symmetries. The first set of terms will
in general be canceled off by the GSWS mechanism. The second set of
terms cannot be canceled by the GSWS mechanism because the couplings
of (\ref{GScoupling}) cannot produce terms of this form.
This means that such contributions are actually generated by just the one-loop
contributions to the anomaly polynomial.

Such terms are problematic, because they are of the general Adler-Bell-Jackiw (ABJ)
type, namely they mix a candidate global $U(1)$ symmetry with a gauge
symmetry. The presence of such terms would allow us to convert global $U(1)$
charge to excitations associated with the gauge symmetry, thus violating
current conservation. This would in turn mean that the candidate $U(1)$
symmetry is not truly a symmetry. We note that this is qualitatively different
from 't Hooft anomalies involving just global symmetries.

From a bottom-up perspective, there are a number of necessary (but possibly
insufficient)\ conditions which must be met to have a $U(1)$
symmetry in a 6D SCFT. First of all, we must identify
candidate $U(1)$ symmetries from the full tensor branch description of the theory.
Next, we must provide a candidate set of charges
for the matter fields of the tensor branch theory. This includes weakly
coupled matter, but also strongly coupled generalizations such as
6D\ conformal matter (as occurs on the partial tensor branch). If a candidate
$U(1)$ symmetry and the proposed charge assignments for matter generates an
ABJ\ anomaly, then we must discard this choice of charge assignment, and if no
non-trivial assignment is available, we must discard the candidate $U(1)$
altogether. There are typically several non-trivial arithmetic
constraints which strongly limit the existence of candidate $U(1)$ symmetries.

In practice, we shall often associate a $U(1)$ with each bifundamental hypermultiplet. The overall
$U(1)$ charge of these fields is constrained by ABJ anomaly cancellation, which limits us to
rays of possible charge vectors. The overall normalization of charge assignments can then be fixed by appealing to
the string theory realization of the model, and is also inherited from charge quantization in a progenitor theory.

As an illustrative example, consider the theory of $k$ M5-branes probing a
$\mathbb{C}^{2}/\mathbb{Z}_{N}$ singularity. In the F-theory realization of
this theory, we have $k-1$ curves of self-intersection $-2$, each of which
supports an $I_{N}$ fiber. The resulting fiber types\ are given by:%
\begin{equation}
\lbrack I_{N}],\underset{k-1}{\underbrace{\overset{I_{N}}{2}%
,...,\overset{I_{N}}{2}}},[I_{N}]. \label{INITOUTIT}%
\end{equation}
In F-theory, each $I_{N}$ fiber is associated with an $\mathfrak{su}(N)$ gauge
symmetry algebra. One of the distinctions between type IIB\ and F-theory is
that the overall \textquotedblleft center of mass\textquotedblright\ $U(1)$
present in a stack of $N$ D7-branes is typically absent because this gauge
field couples to an axion. Something similar is at work in this configuration.
Indeed, from our $I_{N}$ fibers, we see $k-1$ \textquotedblleft
candidate\textquotedblright\ $U(1)$ gauge symmetries, and two candidate $U(1)$
flavor symmetries. Of these, only one linear combination turns out to be free
of ABJ\ anomalies.

To figure out the anomalies associated with the $U(1)$ symmetries,
we first need to fix a convention for the representations of our hypermultiplets. Throughout this paper,
our convention will be dictated by the topology of the associated quiver. Indexing the groups from $i=0,...,k$ from left to right, we have hypermultiplets which transform in the bifundamental representation $({\bf N_{i},\overline{N}_{i+1}})$
of neighboring gauge groups running from left to right. Note that in this convention, we have an
$SU(2)_{\mathcal{R}}$ R-symmetry doublet of scalars in the $({\bf N_{i},\overline{N}_{i+1}})$ representation,
and a fermionic superpartner which also transforms in the $({\bf N_{i},\overline{N}_{i+1}})$ representation.
We assign a candidate $U(1)$ to act on each hypermultiplet, which we take to have charge $+1$.\footnote{The
overall normalization turns out to also be fixed to be $+1$.}
Note that there is a sign convention here; we could equally well have considered
assigning multiplets to the conjugate representation. This would
not affect the calculation of ABJ anomalies because in evaluating the contributions
from fermionic loops, we note that $\mathrm{Tr}_{\bf N}F^3 = - \mathrm{Tr}_{\bf \overline{N}}F^3$,
canceling out the minus sign from the opposite charge under the candidate $U(1)$.

\begin{figure}[t!]
\begin{center}
\includegraphics[scale = 0.5, trim = {1.75cm 3cm 0cm 3cm}]{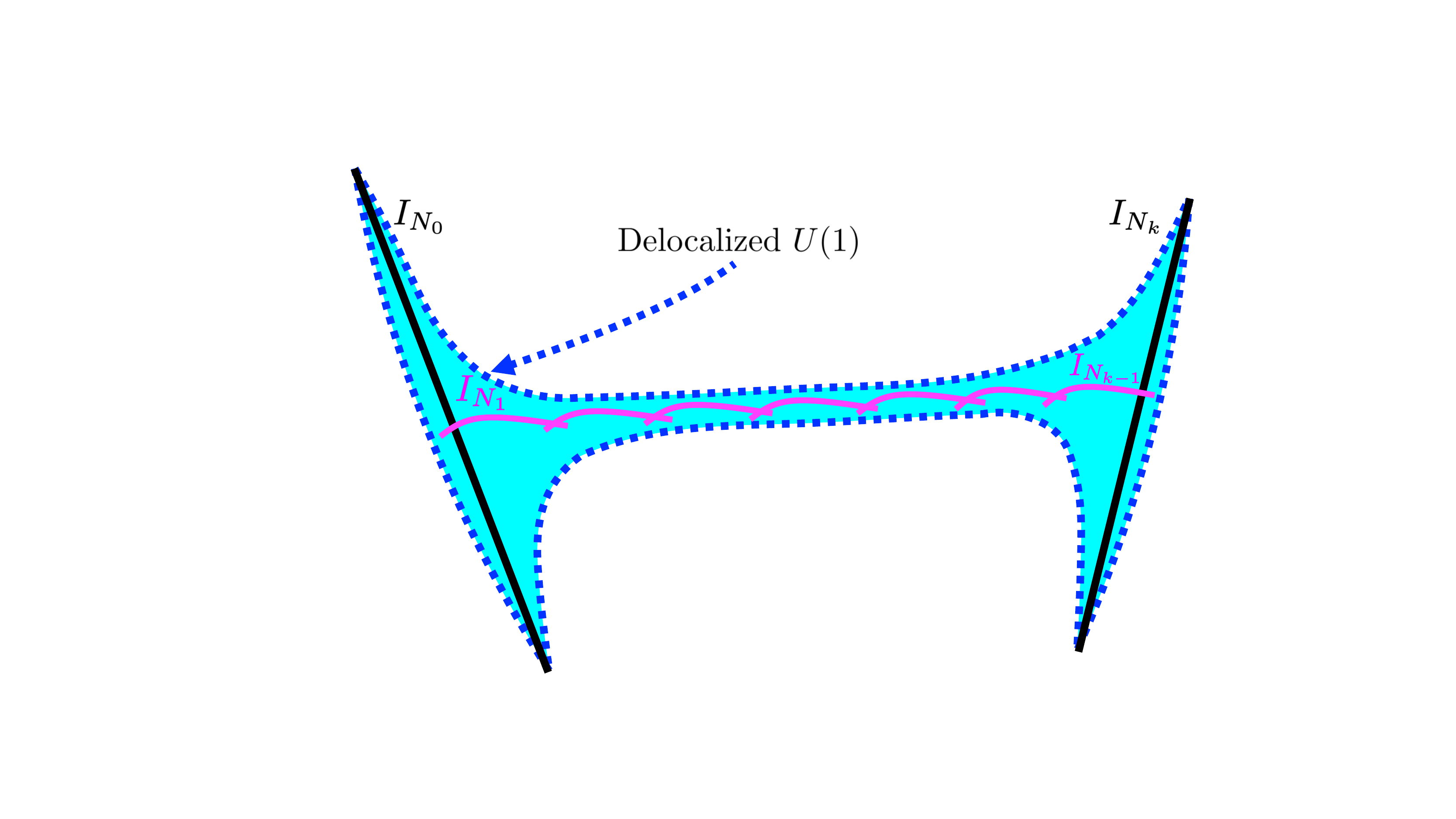}
\caption{Depiction of a delocalized $U(1)$ symmetry in the F-theory model of line (\ref{INITOUTIT}).}
\label{fig:sharedU1}
\end{center}
\end{figure}

This example illustrates a few general points. First, most \textquotedblleft
candidate\textquotedblright\ $U(1)$ symmetries will turn out to be plagued by
ABJ\ anomalies and will need to be eliminated anyway. Additionally, in the
context of F-theory, the surviving $U(1)$ is associated with the appearance of
$I_{N}$ fibers. The $U(1)$ is not really localized on any one component of the
discriminant locus but is better thought of as being shared i.e. ``delocalized''
across multiple components of the discriminant.
See figure \ref{fig:sharedU1} for a depiction.

This sort of reasoning suggests an alternative, but entirely equivalent way to
proceed in determining possible $U(1)$ symmetry factors. Starting from the
quiver-like gauge theory description of a 6D\ SCFT\ on its partial tensor
branch, we can first consider each individual gauge group factor in isolation
by taking a decoupling limit on the tensor branch. In this limit, we have a
set of global symmetries for each such gauge theory. The process of
incoporating additional gauge group factors amounts to weakly gauging a
subalgebra of the flavor symmetries, and introducing an additional tensor
multiplet to \textquotedblleft pair\textquotedblright\ with this gauge
symmetry. Returning to the example of (\ref{INITOUTIT}), each individual
$-2$ curve defines a 6D\ SCFT:%
\begin{equation}
\lbrack I_{N}],\overset{I_{N}}{2},[I_{N}].
\end{equation}
This example is well-known to have an $\mathfrak{su}(2N)$ flavor symmetry, as
opposed to the $\mathfrak{su}(N)\times\mathfrak{su}(N)\times\mathfrak{u}(1)$
global symmetry expected from the M-theory construction of the model. Observe,
however, that we can gauge the subalgebra $\mathfrak{su}(N)_L$ and thus
obtain the quiver:
\begin{equation}
\lbrack I_{N}],\overset{I_{N}}{2},\overset{I_{N}}{2},[I_{N}].
\label{AYEAYEMATEY}%
\end{equation}
The commutant of $\mathfrak{su}(N)_L$ inside of $\mathfrak{su}(2N)$ is
$\mathfrak{su}(N)\times\mathfrak{u}(1)$, which provides us with a candidate
$U(1)$ global symmetry.

We summarize these two complementary procedures as:

\begin{itemize}
\item Candidate $U(1)$'s (Method 1):\ Simply write down all candidate $U(1)$
symmetries as well as all possible charge assignments for matter fields. The
only surviving charge assignments and $U(1)$ charges are those which are free
of ABJ\ anomalies. The overall normalization of charge assignments can be fixed by appealing to Higgsing from a
another fixed point and / or by using the associated string construction for $U(1)$ charge assignments.

\item Commutant Symmetries (Method 2):\ Start with a single node of a quiver-like gauge
theory and weakly gauge the flavor symmetries of this theory. The commutant
provides a set of candidate flavor symmetries, some of which are $U(1)$
symmetries. We must again ensure that ABJ\ anomalies cancel to have a genuine
flavor symmetry.
\end{itemize}

The two procedures provide an equivalent way of generating the same
information about $U(1)$ symmetries, since any $U(1)$
obtained from the commutant procedure will necessarily be a candidate $U(1)$
of the full 6D\ SCFT, and any candidate $U(1)$ that is free of ABJ anomalies will show up in the commutant after gauging a subgroup of some flavor symmetry. There are merits to using either procedure, and in
practice it simply depends on the details of the quiver to determine which
method of extracting $U(1)$'s will be more efficient. In what follows, we shall
often emphasize that we are dealing with a flavor symmetry by writing it in capital latin
text. That being said, we will actually only discuss these symmetries at the level of algebras, rather than groups.
For this reason, we will sometimes write $U(N)$ interchangeably with $SU(N) \times U(1)$, writing $U(N) \sim SU(N) \times U(1)$
to emphasize that even though the associated groups are strictly speaking distinct, their algebras are isomorphic.

To proceed further, we now discuss in more detail the process of extracting
candidate $U(1)$'s. In section \ref{sec:EXAMPLES} we present a number of examples
illustrating our proposal. For additional details on anomaly polynomials
with global $U(1)$ symmetries, see Appendix \ref{app:A}.

\subsection{Candidate $U(1)$'s}

Our first task is to provide a precise notion of \textquotedblleft candidate
$U(1)$'s\textquotedblright\ which could appear in the tensor branch description of a
6D\ SCFT. To this end, we first review some additional elements of how all
known 6D\ SCFTs are constructed. The F-theory approach to realizing 6D\ SCFTs
proceeds in two steps. First, one specifies a choice of base with some
collection of contractible curves, and then one defines an elliptic fibration
over a given base.\ In field theory terms, the choice of base determines the
Dirac pairing for the tensor multiplets. The elliptic fibration
tells us the gauge groups and matter. In general, once the base is
specified the types of available elliptic fibrations are severely restricted.

The main building blocks in 6D\ SCFTs are curves of self-instersection $-2$
and the non-Higgsable clusters (NHCs) of reference \cite{Morrison:2012np}.
A general base is obtained either from a collection of $-2$ curves intersecting
according to to the Dynkin diagram of an ADE\ Lie algebra, or by taking NHCs
and \textquotedblleft gluing them\textquotedblright\ via curves of
self-intersection $-1$. It is also possible to sometimes append a chain of
$-2$ curves in ADE configuration to a glued configuration of NHCs.

Consider first the flavor symmetries which come from minimal fiber
enhancements. The global symmetries for the building blocks are rather
limited. By definition, the NHCs do not have a Higgs branch, and their
corresponding flavor symmetry is also trivial. The $-2$ curves with no
singular fibers produce the ADE $\mathcal{N}=(2,0)$ theories, so when viewed
as $\mathcal{N}=(1,0)$ theories we get an $\mathfrak{su}(2)_{\mathcal{L}%
}\subset\mathfrak{su}(2)_{\mathcal{L}}\times\mathfrak{su}(2)_{\mathcal{R}%
}\subset\mathfrak{so}(5)_{\text{R-symm}}$ flavor symmetry from the R-symmetry
of the $\mathcal{N}=(2,0)$ theory. When fiber enhancements are present, this
symmetry is typically destroyed but in its place we get additional non-compact
flavor symmetry factors, a point we return to shortly. Lastly, we have the
$-1$ curve theory. This realizes the E-string theory, namely the theory of a
single small instanton in heterotic M-theory. As mentioned in section
\ref{sec:prog}, this theory comes with an $\mathfrak{e}_{8}\times\mathfrak{su}%
(2)_{\mathcal{L}}$ flavor symmetry. The former comes from the $E_{8}$
nine-brane, and the latter comes from the $\mathfrak{spin}(4)\supset
\mathfrak{su}(2)_{\mathcal{L}}\times\mathfrak{su}(2)_{\mathcal{R}}$ isometries
preserved by the small instanton in $\mathbb{C}^{2}\simeq\mathbb{R}^{4}$. Let
us note that the more general theory of multiple small instantons:%
\begin{equation}
[E_8],1,2,...,2
\end{equation}
enjoys the same $\mathfrak{e}_{8}\times\mathfrak{su}(2)_\mathcal{L}$ flavor symmetry.

We obtain more intricate 6D\ SCFTs with minimal singularities in the fiber by
gauging the $\mathfrak{e}_{8}$ flavor symmetry of the small instanton theory.
This has two immediate consequences. First of all, by gauging a flavor
symmetry, we typically break the $\mathfrak{su}(2)_{\mathcal{L}}$ flavor
symmetry. In heterotic terms, this is because we have replaced the non-compact
$\mathbb{R}^{4}$ by a compact space with smaller isometry group. Second of
all, we can sometimes arrange for a $U(1)$ global symmetry to appear as the
commutant in this gauging procedure. To illustrate this point, suppose we
gauge a product subalgebra $\mathfrak{g}_{L}\times\mathfrak{g}_{R}%
\subset\mathfrak{e}_{8}$. In some cases, the commutant $H$ of $\mathfrak{g}%
_{L}\times\mathfrak{g}_{R}$ inside $\mathfrak{e}_{8}$ will contain one or more
$U(1)$ factors. This occurs in the following cases:

\begin{itemize}
\item $\mathfrak{g}_{L} = \mf{e}_{6}$, $\mathfrak{g}_{R} = \mf{su}(2)$, $H=U(1)$,

\item $\mathfrak{g}_{L} = \mf{so}(10)$, $\mathfrak{g}_{R} = \mf{su}(2)$, $H=SU(2) \times
U(1)$,

\item $\mathfrak{g}_{L} = \mf{so}(7)$, $\mathfrak{g}_{R} = \mf{so}(7)$, $H=U(1)$,

\item $\mathfrak{g}_{L} = \mf{so}(8)$, $\mathfrak{g}_{R} = \mf{su}(4)$, $H=U(1) $.

\item $\mathfrak{g}_{L} = \mf{so}(8)$, $\mathfrak{g}_{R} = \mf{su}(3)$, $H=U(1) \times
U(1)$,
\end{itemize}
where in the above, we have deferred the case of $\mathfrak{g} = \mathfrak{su}(2)$ gauging.

For theories with minimal singularity types over each curve, this is the
primary way in which flavor symmetries (including $U(1)$ symmetries)\ can
arise. Once we allow further decorations in the fiber type over each curve
obtained on the tensor branch, additional possibilities emerge. In field
theory terms, these additional decorations in the singularity type mean the
gauge group paired with a given tensor multiplet will have higher rank than
the generic NHC\ situation. This in turn leads to the presence of additional
hypermultiplets transforming in representations of the gauge groups as well as
possible flavor symmetries. In a 6D\ SCFT on its tensor branch, we can
have hypermultiplets which transform in a representation of a single gauge group, or in a bifundamental
representation. In both cases, the determining factor for the flavor symmetry
acting upon the hypermultiplets is the number of hypermultiplets. The flavor
symmetry of a set of $N$ hypermultiplets transforming in some representation
of a gauge group depends on whether that representation is real, pseudo-real,
or complex. In particular:

\begin{itemize}
\item $N$ hypermultiplets (equivalently, $2N$ half-hypermultiplets) in a
pseudo-real representation have an $SO(2N)$ flavor symmetry.

\item $N$ hypermultiplets in a complex representation have a $U(N)\sim
SU(N)\times U(1)$ flavor symmetry.

\item $N$ hypermultiplets in a real representation have an $Sp(N)$ flavor
symmetry (here $Sp(1)\sim SU(2)$).

The same rules apply when the hypermultiplets transform in a bifundamental
representation. The only issue is whether the tensor product of the two
representations $\rho_{1}\otimes\rho_{2}$ is complex, real, or
pseudo-real.\footnote{Recall that the reality conditions for tensor products of
representations $\rho_{1}\otimes\rho_{2}$ for the gauge group $G_{1}\times
G_{2}$ for real (R), pseudo-real (P), and complex (C) representations are as
follows:
\par
\begin{itemize}
\item R $\otimes$ R = R,
\par
\item P $\otimes$ R = P,
\par
\item P $\otimes$ P = R,
\par
\item C $\otimes$ Any = C.
\end{itemize}
}
\end{itemize}

Abelian flavor symmetries can arise from the above rules in one of two ways:
(1) any number of complex hypermultiplets will transform under a $U(N) \sim
SU(N) \times U(1) $ flavor symmetry, or (2) a single full hypermultiplet in a
pseudo-real representation will transform under a $SO(2) \sim U(1)$ flavor
symmetry.

In 6D SCFTs, the first of these possibilities arises for the
following representations:

\begin{itemize}
\item $N$ fundamentals of an $\mathfrak{su}(n \geq3)$ gauge algebra,

\item $N$ $\Lambda^{2}$s of an $\mathfrak{su}(n \geq5)$ gauge algebra,

\item $N$ spinors of an $\mathfrak{so}(10)$ gauge algebra,

\item $N$ fundamentals of an $\mathfrak{e}_{6}$ gauge algebra,

\item A bifundamental of an $\mathfrak{su}(n \geq3) \times\mathfrak{su}(m
\geq3)$ gauge algebra,

\item A bifundamental of an $\mathfrak{su}(n \geq3) \times\mathfrak{sp}(n
\geq1)$ gauge algebra,

\item A bifundamental of an $\mathfrak{su}(n \geq3) \times\mathfrak{so}(m)$
gauge algebra.
\end{itemize}
In practice, this last case occurs only for 6D SCFTs in the ``frozen'' phase
of F-theory \cite{Tachikawa:2015wka, Bhardwaj:2018jgp}.

The second possibility of an $SO(2) \sim U(1)$ flavor symmetry arises for the
following representations:

\begin{itemize}
\item A fundamental (two half-hypermultiplets in the fundamental) of an $\mathfrak{sp}(n \geq1)$
gauge algebra,

\item A fundamental of an $\mathfrak{e}_{7}$ gauge algebra,

\item A spinor of an $\mathfrak{so}(11)$ or $\mathfrak{so}(12)$ gauge algebra.
\end{itemize}

More representations, and hence more opportunities for abelian flavor
symmetries, arise in the case of Little String Theories (LSTs)
\cite{Bhardwaj:2015oru}.

In the above, we have omitted the case of matter in an $\mathfrak{su}(2)$
gauge theory since this case has some additional subtleties. For
$\mathfrak{su}(2)$ gauge theory paired with a tensor of charge $-2$, anomaly
cancelation considerations imply we have eight half-hypermultiplets in the
fundamental representation. Though this might suggest the matter fields
transform in the vector representation of $Spin(8)$, the F-theory
realization of this model admits only a $Spin(7)$ flavor symmetry at the superconformal fixed point
\cite{Heckman:2015bfa}, which is in fact confirmed by field theory considerations as well
\cite{Ohmori:2015pia,Hanany:2018vph}. So in this case, the matter fields transform as
half-hypermultiplets in the $(2,8)$ of $\mathfrak{su}(2)\times\mathfrak{so}%
(7)$.

This distinction between $Spin(7)$ and $Spin(8)$ is
important because it impacts what symmetries can be gauged, and
consequently, the resulting commutant flavor symmetries. Consider for example
the effects of gauging an $\mathfrak{su}(3)\subset\mathfrak{so}(7)$
subalgebra. We have the branching rules: $\mathfrak{so}(7)\supset
\mathfrak{so}(6)\supset\mathfrak{su}(3)\times\mathfrak{u}(1)$, and
consequently a commutant of $\mathfrak{u}(1)$. This is different from what we
would have obtained if we had incorrectly assumed the flavor symmetry is
$\mathfrak{so}(8)\supset\mathfrak{so}(6)\times\mathfrak{u}(1)\supset
\mathfrak{su}(3)\times\mathfrak{u}(1)\times\mathfrak{u}(1)$. Similarly, when
we gauge an $\mathfrak{su}(4)\simeq\mathfrak{so}(6)$ subalgebra of
$\mathfrak{so}(7)$, the commutant does not have any residual $\mathfrak{u}%
(1)$'s.

On a related point, an $\mathfrak{su}(2)$ gauge algebra paired with a $-2$ tensor that
meets an unpaired $-2$ tensor has a $G_{2}$ global symmetry, under which seven
half-hypermultiplets transform as the $\mathbf{{7}}$ of $G_{2}$. If six of
these half-hypermultiplets transform as a bifundamental under an
$\mathfrak{su}(3)$ gauge algebra, there is no $U(1)$ global symmetry remaining
to act on them.

Thus, in the presence of an $\mathfrak{su}(2)$ gauge algebras paired with a
$-2$ tensor, there is just one more $U(1)$ possibility to consider:

\begin{itemize}
\item A bifundamental of an $\mathfrak{su}(3)$ gauge algebra and an
$\mathfrak{su}(2)$ gauge algebra, provided the $-2$ tensor paired with the
$\mathfrak{su}(2)$ gauge algebra is not adjacent in the 6D SCFT quiver to an
unpaired $-2$ tensor.
\end{itemize}

Note that we have not discussed the possibility of a bifundamental of
$\mathfrak{su}(2)\simeq \mathfrak{sp}(1)$ and $\mathfrak{su}(2)\simeq \mathfrak{sp}(1)$. Quivers with these bifundamentals
often have additional, delocalized $SU(2)$ global symmetries. We will study
these quivers and work out their global symmetries later in subsection
\ref{sec:SU2only}.

So far, our discussion has focused on obtaining a set of candidate global
$U(1)$'s. We now turn to the associated ABJ\ anomalies coming from such
symmetries. The ABJ anomalies for a hypermultiplet of global $U(1)$ charge $q$
in a representation $\rho$ of a non-abelian gauge algebra $\mathfrak{g}$ are
given by (see Appendix \ref{app:A})
\begin{equation}
I_{\text{ABJ}} \supset \frac{1}{6}qF_{U(1)}\,\mathrm{Tr}_{\rho}F_{\mathfrak{g}}%
^{3},\label{eq:ABJ}%
\end{equation}
where $F_{U(1)}$ and $F_{\mathfrak{g}}$ are the field strengths for the $U(1)$
global symmetry and the gauge algebra $\mathfrak{g}$, respectively, and
$\,\mathrm{Tr}_{\rho}$ is the trace in the representation $\rho$. To avoid cluttering
later expressions, in what follows we shall often leave the subscript for the choice of
representation for the trace implicit, but will instead indicate it as appropriate.

To get an ABJ anomaly, we must, by necessity, have a gauge algebra which supports
representations with a non-vanishing cubic Casimir. For simple Lie algebras,
this only occurs in the case of $\mathfrak{g}=\mathfrak{su}(N)$ for $N\geq3$.
Additionally, we know the $U(1)$ charges for our hypermultiplets. This is
because each fundamental of $SU(N)$ for $N\geq3$ carries charge $1$ under its
associated \textquotedblleft candidate $U(1)$.\textquotedblright\ More
generally, however, we may determine the charges of all hypermultiplets under
the ABJ anomaly-free $U(1)$'s using the Lie algebra branching rules explained
above. Namely, we may decompose the maximal flavor symmetry associated with a
given node of the quiver into a gauged part and a global part, the latter of
which may involve $U(1)$ factors. The branching rules for this decomposition
allow us determine the $U(1)$ charges for each component, up to overall normalization.

So, in all of these cases we have a set of well-posed constraints, as obtained
from equation (\ref{eq:ABJ}). Consequently, we learn that of our candidate
$U(1)$'s, only the appearance of an $\mathfrak{su}(N)$ gauge algebra can
impose a non-trivial constraint. These are necessary conditions, and also
appear to be sufficient. The total number of $U(1)$'s in a 6D SCFT are thus
given by:
\begin{equation}
\text{\# $U(1)$'s = \# candidate $U(1)$'s $-$ \# }\mathfrak{su}\text{$(N)$'s
with }N\geq3\text{,}%
\end{equation}
as computed on the full tensor branch of the 6D\ SCFT.\footnote{This number is correctly reproduced in the $AdS_7$ gravity duals \cite{Apruzzi:2013yva} of ``holographic'' SCFTs with only $\mathfrak{su}(N_i)$ algebras on their tensor branch (with variable $N_i$), and a large number of such gauge algebras. (Note that, because of the presence of D8-branes in the Type IIA construction, these SCFTs are not engineered by M5-brane probes, but do admit a dual F-theory engineering \cite{DelZotto:2014hpa}.) The number of ABJ anomaly-free global $U(1)$'s in field theory matches the number of massless abelian gauge bosons in the supergravity reduction on $AdS_7$
\cite{bergman-fazzi-rodriguezgomez-tomasiello}.} Again, the
\textquotedblleft candidate\textquotedblright\ $U(1)$'s are given by first
specifying the number of $-1$ curve theories as well as the flavor symmetries
which can act upon a set of weakly coupled hypermultiplets, in accord with the
discussion given above. The  \# $\mathfrak{su}(N)$'s are simply all gauge
group factors which can introduce a non-trivial constraint, as per equation
(\ref{eq:ABJ}). Moreover, the above procedure based on the branching of
representations fixes the overall $U(1)$ charge assignments for all
hypermultiplets appearing on the tensor branch.

\begin{figure}[t!]
\begin{center}
\includegraphics[scale = 0.4, trim={0 2.5cm 0 1.5cm}]{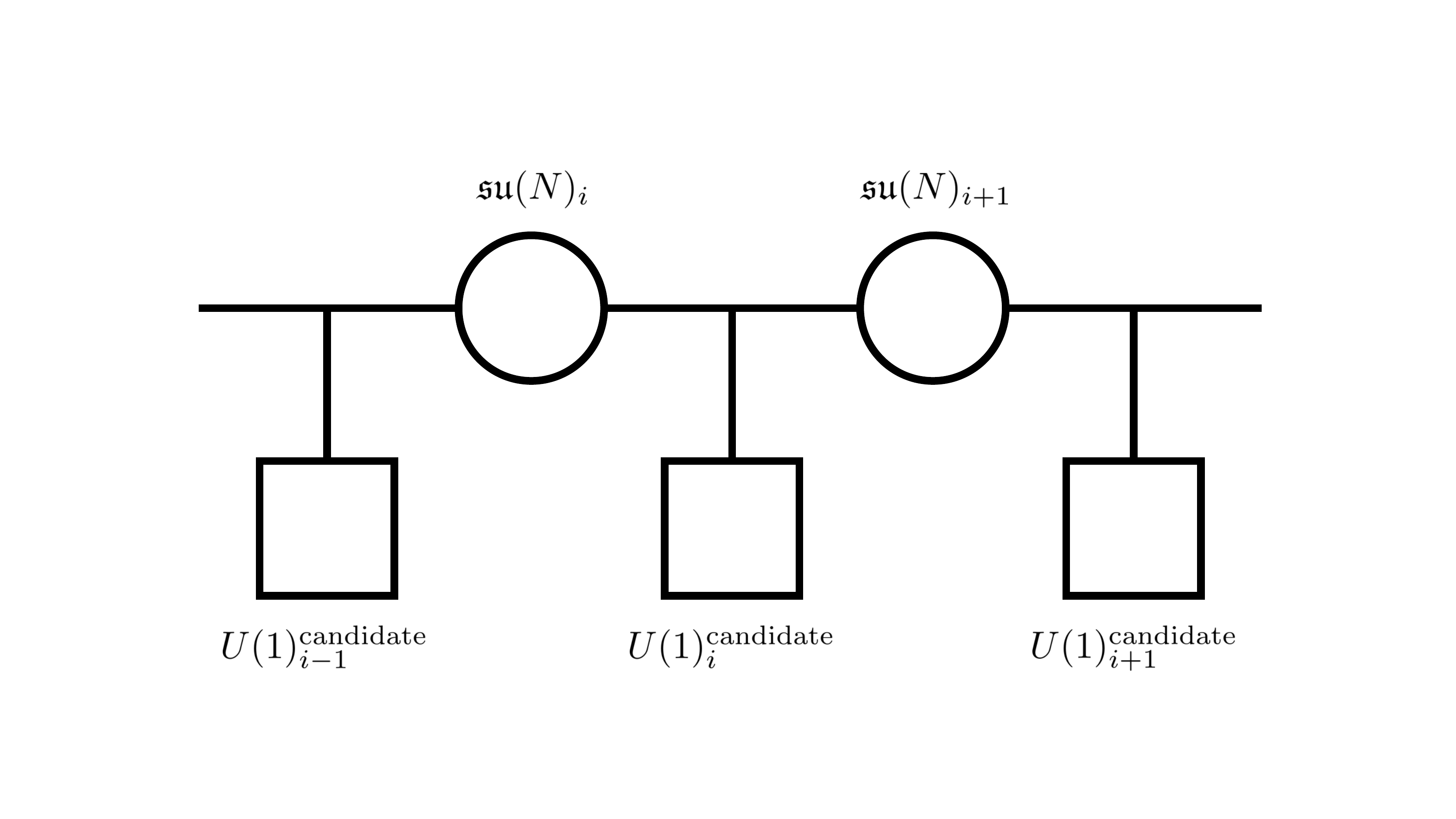}
\caption{Depiction of the local quiver gauge theory associated with the tensor branch of the 6D SCFT described by
line (\ref{quiverbeforemepitifulmortals}). We have also indicated the appearance of the candidate $U(1)$ global symmetries
which act on bifundamental hypermultiplets.}
\label{fig:mwahaha}
\end{center}
\end{figure}

As an instructive example for how these constraints appear, consider the
theory of $k+1$ M5-branes probing a $\mathbb{C}^{2}/\mathbb{Z}_{N}$ orbifold
singularity with $N\geq3$. On the tensor branch, we have a quiver gauge theory
given by:%
\begin{equation}\label{quiverbeforemepitifulmortals}
\underset{\lbrack N_{f}=N]}{\overset{\mathfrak{su}(N)_{1}}{2}}%
\,\,{\overset{\mathfrak{su}(N)}{2}}\,\,...\,\,\overset{\mathfrak{su}%
(N)}{2}\,\,\underset{[N_{f}=N]}{\overset{\mathfrak{su}(N)_{k}}{2}},
\end{equation}
By the rules above, we expect that each set of $N$
fundamentals will transform in a $U(N)\sim SU(N)\times U(1)$ flavor
symmetry, while each bifundamental will transform under a $U(1)$ flavor
symmetry. Altogether, this gives $1+1+k-1=k+1$ candidate $U(1)$ flavor symmetries. See figure
\ref{fig:mwahaha} for a depiction of the quiver, including the candidate $U(1)$ charge assignments.

However, some candidate symmetries will suffer from ABJ anomalies:
\begin{equation}
I_{\text{ABJ}}\supset\frac{1}{6}\sum_{i = 1}^{k} \sum_{J=0}^{k}q_{i,J}F_{J}\,\mathrm{Tr}%
_{\text{fund}}F_{i}^{3}.
\end{equation}
Here, the summation is over the gauge group factors indexed by $i=1,...,k$ and the
candidate $U(1)$ symmetries indexed by $J = 0,...,k$. Each $q_{i,J}$ denotes the
contribution to the anomaly from summing over all hypermultiplets charged under the $i$th
gauge group $\mathfrak{su}(N)_{i}$
and the $J$th candidate symmetry $U(1)_{J}$. In the present example where
we have hypermultiplets in bifundamental representations
of neighboring gauge groups of the quiver, $q_{i,J} = -N$ if $J = i-1$, $q_{i,J} = +N$
if $J = i$ and $q_{i,J} = 0$ otherwise.
Observe that the sum over just the $J$ index immediately tells that the linear combination of $U(1)$'s
$\sum_{J=0}^{k}q_{i,J}F_{J}$ suffers from an ABJ anomaly for $i=1,...,k$.
Thus, in total, $k$ $U(1)$'s will be anomalous, and only one of the $k+1$
candidate $U(1)$'s is free of ABJ\ anomalies.

So far, we have left all of the $U(1)$ charges generic, denoted as $q_{i}$.
For fundamentals of $SU(N\geq3)$, these charges $q_{i}$ are readily determined:
any fundamental of $SU(N)$ carries charge 1 under its associated
\textquotedblleft candidate $U(1)$.\textquotedblright\ More generally,
however, we may determine the charges of all hypermultiplets under the ABJ
anomaly-free $U(1)$'s using Lie algebra branching rules. Namely, we may
decompose the maximal flavor symmetry associated with a given node of the
quiver into a gauged part and a global part, the latter of which may involve
$U(1)$ factors. The branching rules for this decomposition allow us determine
the $U(1)$ charges for each component, up to overall normalization.

This overall normalization cannot be fixed by ABJ anomaly cancellation, but it can be determined from a top-down perspective: in the string theory construction of A-type progenitor theories, bifundamentals are associated with strings between neighboring stacks of D6-branes, which carry charge $1$ under the $U(1)$ global symmetry of the theory (as we will see in more detail in the following section). In a more general theory, one may either appeal to a similar string construction to normalize charges, or alternatively, one can track the $U(1)$ charges from the Higgsing of a progenitor theory.

Thus, we have two methods for determining the $U(1)$ charges of
hypermultiplets in a quiver, which were also stated below
(\ref{AYEAYEMATEY}). We reproduce them here for the ease of the reader:

\begin{itemize}
\item Candidate $U(1)$'s (Method 1):\ Simply write down all candidate $U(1)$
symmetries as well as all possible charge assignments for matter fields. The
only surviving charge assignments and $U(1)$ charges are those which are free
of ABJ\ anomalies. The overall normalization of charge assignments can be fixed by appealing to Higgsing from a
another fixed point and / or by using the associated string construction for $U(1)$ charge assignments.
See the analysis of the theory in (\ref{eq:TBAinst}) below for an example.

\item Commutant Symmetries (Method 2):\ Start with a single node of a quiver-like gauge
theory and weakly gauge the flavor symmetries of this theory. The commutant
provides a set of candidate flavor symmetries, some of which are $U(1)$
symmetries. We must again ensure that ABJ\ anomalies cancel to have a genuine
flavor symmetry. See the analysis of the theory in (\ref{eqn:ex3}) below for an example.
\end{itemize}

In some cases, the first of these methods is easier to implement, while in
other cases the second method is preferable. In what follows, we will see
instances of each, and we will demonstrate the equivalence of these two
methods in an illustrative example in the theory of line (\ref{eqn:compareTwoMethods}).

\section{Examples} \label{sec:EXAMPLES}

In the previous section we provided a general procedure for determining the
$U(1)$ global symmetries of a 6D\ SCFT. This amounts to listing all $U(1)$'s
which can act on our \textquotedblleft matter fields\textquotedblright%
\ (including E-string theories), including the associated $U(1)$ charges, as
dictated by the branching of the flavor symmetry after gauging a subalgebra.
After this, we can determine which $U(1)$'s are compatible with the
constraints of ABJ\ anomaly cancelation. Our plan in this section will be to
explain how these rules work in practice by presenting some illustrative
examples. In addition to determining the global $U(1)$ symmetries, we also
work out the associated anomaly polynomials for global symmetries in these
theories. These examples are not meant to be exhaustive, but rather to exhibit
the different possible phenomena which can occur. As a point of terminology, we shall often refer to
the candidate $U(1)$ associated with a bifundamental between two gauge groups as a ``baryonic $U(1)$.''

In what follows, we shall make use of some earlier results on the structure of global symmetries
obtained in references \cite{Heckman:2015bfa} and \cite{Heckman:2016ssk}. In the context of
heterotic $E_8$ small instanton probes of an ADE singularity, there is a tight correspondence between
Higgs branch deformations and discrete group homomorphisms $\text{Hom}(\Gamma_{ADE}, E_8)$. Given a homomorphism
$\rho \in \text{Hom}(\Gamma_{ADE}, E_8)$, the commutant of the image $[\text{Im}(\rho),E_8]$ determines a flavor symmetry.

In the context of M5-brane probes of ADE singularities, there is a close correspondence between certain Higgs branch deformations of
the field theory and nilpotent orbits of the flavor symmetry algebra. We collect some of the necessary information about
the resulting global symmetries in Appendix \ref{sec:grouptheory}, to which we refer the interested reader for additional details.

The rest of this section is organized as follows. We begin by analyzing the
progenitor theories with an A-type flavor symmetry. Indeed, we have already argued using
M-theory and heterotic M-theory realizations of these theories that they must possess a $U(1)$
global symmetry. Here, we directly establish this using our proposal for reading off $U(1)$ symmetries
via the tensor branch description. The defining feature of these examples is the appearance of
$\mathfrak{su}(N)$ gauge algebras with complex representations. We then turn to more elaborate examples
with $\mathfrak{su}(N)$ gauge algebras, following this with the special case of theories with primarily
$\mathfrak{su}(2)$ gauge algebras. With this in place, we next analyze theories which do not have hypermultiplets
in a complex representation of $\mathfrak{su}(N)$, as will occur in theories with 6D conformal matter and in the
D- and E-type progenitor theories. Finally, we also consider theories involving frozen phases of F-theory featuring adjacent $SO$-$SU$ gauge groups.

\subsection{A-type Progenitor Theories}

We now turn to the A-type progenitor theories. Recall that the tensor branch description for this class of theories is of one of two types:
\begin{align}
& [SU(N)]~\underset{k}{\underbrace{\overset{\mathfrak{su}({N})}{2} \,
\overset{\mathfrak{su}({N})}{2} ~ ...~\overset{\mathfrak{su}({N})}{2}}%
}~[SU(N)],\label{AtypeM5probe}\\
& [E_{8}]~\underset{k}{\underbrace{\overset{\mathfrak{su}({N})}{1} \,
\overset{\mathfrak{su}({N})}{2}~...~\overset{\mathfrak{su}({N})}{2}}%
}~[SU(N)].\label{AtypeSmallInst}%
\end{align}
In M-theory terms, the theories of (\ref{AtypeM5probe}) describe the partial tensor branch of M5-branes probing a $\mathbb{C}^2 / \mathbb{Z}_N$ singularity and the theories of (\ref{AtypeSmallInst}) describe the partial tensor branch of M5-branes probing a heterotic $E_8$ nine-brane wrapping the singularity $\mathbb{C}^2 / \mathbb{Z}_N$. In this subsection we shall assume $N \geq 3$ since there are some additional
subtleties which arise in the special case $N = 2$.

A pleasant feature of the M-theory realization of these SCFTs is that the $U(1)$ symmetry
is directly visible as an isometry of the geometry probed by the M5-branes. Here, we would like to see how this comes about
by directly analyzing the partial tensor branch of the theory.

\begin{table}
    \centering
    \begin{tabular}{|c|c|c|c|c|c|}\hline
       Representation  & Fund & Adj & $\mathcal{S}^{2}$ & $\Lambda^2$ & $\Lambda^3$ \\ \hline
     $c_\rho$    & 1 & 0 & $N+4$ & $N-4$ & $\frac{1}{2}(N^2 - 9 N + 18)$ \\ \hline
    \end{tabular}
    \caption{Group theory factors for cubic Casimirs of $SU(N)$ with $N \geq 3$. Here, ``Fund'' refers to the fundamental representation, ``Adj'' to the adjoint representation, $\mathcal{S}^2$ to the two-index symmetric representation, and $\Lambda^{2}$ and $\Lambda^{3}$ to the two-index anti-symmetric and three-index anti-symmetric representations. The parameter $c_\rho$ is defined by $\Tr_\rho F^3 := c_\rho \Tr_{\textrm{fund}} F^3$. The value of $c_{\overline{\rho}}$ in a complex conjugate representation $\overline{\rho}$ is related
    as $c_{\rho} = - c_{\overline{\rho}}$.}
    \label{tab:GTCC}
\end{table}

Since the analysis is somewhat simpler in the case of M5-branes probing an A-type singularity,
we start with the theories of line (\ref{AtypeM5probe}), and then turn to the theories of
line (\ref{AtypeSmallInst}). As a warmup, consider first the
theory of three M5-branes probing a $\mathbb{C}^2/\mathbb{Z}_N$ singularity, with $N \geq 3$:
\begin{equation}
    [SU(N)_L] \,\, \overset{\mathfrak{su}(N)_1}{2}\,\, \overset{\mathfrak{su}(N)_2}{2} \,\, [SU(N)_R],~~~N \geq 3.
    \label{eqn:ex2}
\end{equation}
This theory has three candidate $U(1)$ global symmetries: one associated with the $N_L$ hypermultiplets charged under $\mathfrak{su}(N)_1$, one associated with the $N_R$ hypermultiplets charged under $\mathfrak{su}(N)_2$, and one baryonic $U(1)$ associated with the bifundamental hypermultiplet $({\bf N}, \overline{{\bf N}})$ of $\mathfrak{su}(N)_1 \times \mathfrak{su}(N)_2$. We denote these three $U(1)$'s as $U(1)_L$, $U(1)_R$, and $U(1)_B$, respectively. We then assign the bifundamental hypermultiplets $({\bf N}, \overline{{\bf N}})$ of $SU(N)_L-\mathfrak{su}(N)_1$, $\mathfrak{su}(N)_1-\mathfrak{su}(N)_2$, and $\mathfrak{su}(N)_1-SU(N)_R$ charges $q_L$, $q_R$, and $q_B$ under their respective $U(1)$ symmetries. As we have already mentioned in our discussion of the theory in line
(\ref{quiverbeforemepitifulmortals}), we can, without loss of generality, set $q_L$ = $q_R$ = $q_B$ = 1.
Thus, the total ABJ anomaly involving gauge symmetries can be written as
\begin{equation}
    I_{\mathrm{ABJ}}^\text{tot} = \frac{N}{6} \Big(  - F_{U(1)_L} \Tr (F_{\ksu(N)_1}^3) +  F_{U(1)_B} \Tr (F_{\ksu(N)_1}^3) -   F_{U(1)_B} \Tr (F_{\ksu(N)_2}^3) + F_{U(1)_R} \Tr (F_{\ksu(N)_2}^3) \Big),
\end{equation}
where $\Tr F^3:=\Tr_{\text{fund}} F^3$, in accordance with Table \ref{tab:GTCC}. We see that two linear combinations of $U(1)$'s have ABJ anomalies, one for each gauge algebra $\ksu(N)_1$ and $\ksu(N)_2$:
\begin{equation}
    \begin{aligned}
    &\ksu(N)_1~:~~~   -F_{U(1)_L} + F_{U(1)_B} \\
    &\ksu(N)_2~:~~~ -F_{U(1)_B} + F_{U(1)_R} .
    \end{aligned}
\end{equation}
Thus, there is one surviving $U(1)$, whose generator can be written as:
\begin{equation}
    T_{\mathrm{survive}} \propto t_{L} + t_{B} + t_{R}.
\end{equation}
The matter content of the theory, which transforms under $SU(N)_L \times \mf{su}(N)_1 \times \mf{su}(N)_2 \times SU(N)_R \times U(1)$, is given by
\begin{equation}
    ({\bf N}, \overline{\bf N}, \textbf{1}, \textbf{1})_{1} \oplus (\textbf{1}, {\bf N}, \overline{\bf N}, \textbf{1})_{1} \oplus (\textbf{1}, \textbf{1}, {\bf N}, \overline{\bf N})_{1}.
\end{equation}

More generally, for a theory of $k+1$ M5-branes probing a $\mathbb{C}^2/\mathbb{Z}_N$ singularity, the tensor branch of this theory
is given by:
\begin{equation}
    [SU(N)_L] \,\, \overset{\mathfrak{su}(N)_1}{2}\,\,...\,\, \overset{\mathfrak{su}(N)_k}{2} \,\, [SU(N)_R],~~~N \geq 3,
\end{equation}
and the global symmetry is $SU(N)_L \times SU(N)_R \times U(1)$. Here, each bifundamental $({\bf N}, \overline{\bf N})$ has charge 1.
The global symmetry is as expected from the M-theory construction.

Let us turn to the other class of A-type progenitor theories, which involve $k$ M5-branes probing an $E_8$ wall and a $\mathbb{C}^2/\mathbb{Z}_N$ singularity. Again, we shall assume $N \geq 3$. These theories take the form
\begin{equation}
    [E_8] \,\, 1  \,\, \overset{\ksu(1)}{2} \,\, \overset{\ksu(2)}{2} \,\, \cdots \,\, \underset{[N_f=1]}{\overset{\ksu(N)_1}{2}} \cdots \overset{\ksu(N)_k}{2} \,\, [SU(N)].
     \label{eq:TBAinst}
\end{equation}
In terms of the 6D SCFT / group theory correspondence reviewed in Appendix \ref{sec:grouptheory}, these are associated with
the trivial homomorphism from $\mathbb{Z}_N$ into $E_8$.

Let us consider candidate $U(1)$ symmetries as associated with the bifundamentals between the different $\mathfrak{su}$ factors. To aid in this analysis, we split up our indexing of the gauge algebras into those which are on the ``ramp'' of gauge algebra factors with increasing rank and those which are on the ``plateau'' of gauge algebra factors which all have the same rank.

Starting at the very left of the ramp, we consider the bifundamentals attached to the $\mathfrak{su}(2)$ gauge algebra. Observe that since we have fixed the location of two of the eight half hypermultiplets (in the collision with the $\mathfrak{su}(1)$ factor), the only global symmetry available is $\mathfrak{su}(3)$, all of which is gauged. We conclude that none of these hypermultiplets can actually be charged under a candidate $U(1)$.

Turning next to the gauge algebras $\mathfrak{su}(i)$ of the ramp with $i > 2$, we see that the bifundamental $(\mathbf{i}, \overline{\mathbf{i+1}})$ of $\ksu(i)-\ksu(i+1)$ has an associated baryonic $U(1)$ global symmetry. Similarly, the bifundamental $(\mathbf{N-1} , \overline{\mathbf{N}})$ between $\mathfrak{su}(N-1)$ and $\mathfrak{su}(N)_{1}$, as well as the bifundamentals $(\mathbf{N}_{j}, \overline{\mathbf{N}}_{j+1})$ of $\ksu(N)_j$ and $\ksu(N)_{j+1}$ all have candidate $U(1)$'s. Hence, there are $(N-3)$ candidate $U(1)$'s from the ramp, and $k - 1$ candidate $U(1)$'s from the plateau, for a total of $k+ N - 4$ $U(1)$'s associated with bifundamentals between gauge algebras. In addition, there is a $U(1)_M$ associated with the ``middle'' anti-fundamental $\overline{\mathbf{N}}$ of $\ksu(N)_1$, and another $U(1)_R$ associated with the $U(N)_R \sim SU(N)_R \times U(1)_R$ flavor symmetry acting on the $N$ fundamentals of $\ksu(N)_k$. This gives a total of $N+k-2$ candidate $U(1)$'s.

By a similar analysis as equation (\ref{eqn:su2noU1}), we know that the $\mathfrak{su}(2) - \mathfrak{su}(3)$ bifundamental is not charged under a baryonic $U(1)$ symmetry. Therefore, as we will show below, there is only one $U(1)$ surviving in the above theory. More surprisingly, the ``ramp" is not charged under the $U(1)$ symmetry at all: only the plateau is charged under this $U(1)$. We will demonstrate this using the method of candidate $U(1)$'s (method 1).

Let us now turn to the total ABJ anomaly of this theory (see Appendix \ref{app:A}). We have already mentioned our indexing convention for the gauge algebras, as implicit in being on the ramp or the plateau. To label the $U(1)$'s we shall reference them as $U(1)_{i,i+1}$ to indicate that it acts on the bifundamental between $\mathfrak{su}(i)$ and $\mathfrak{su}(i+1)$ on the ramp, and similarly $U(1)_{N_i , N_{i+1}}$ that it acts on the bifundamental between $\mathfrak{su}(N_i)$ and $\mathfrak{su}(N_{i+1})$. We indicate the two additional $U(1)$'s by $U(1)_M$ and $U(1)_R$.
The total ABJ anomaly of this theory is then conveniently organized according to contributions from gauge groups on the ramp and those on the plateau (see also Appendix \ref{app:A}):
\begin{equation}
I_{ABJ} = I_{\mathrm{ramp}} + I_{\mathrm{plateau}}.
\end{equation}
Here, there are contributions from gauge groups on the ramp, as indicated by $I_{ramp}$, as well as contributions from
gauge groups on the plateau, as indicated by $I_{plateau}$. For the contributions on the ramp, the candidate $U(1)$'s which appear
all act on bifundamentals between gauge algebras. For the contributions on the plateau, the candidate $U(1)$'s include those which act
on bifundamentals between gauge algebras, as well as $U(1)_M$ and $U(1)_R$. Reading from left to right across the quiver, the contributions to each term are:
\begin{align}
  I_{\mathrm{ramp}} & =  \frac{1}{6} \left(4 F_{U(1)_{3,4}} \Tr ( F_{\mathfrak{su}(3)}^3) \right) + \\
   & + \frac{1}{6} \sum_{j = 4}^{N-1} \left[-(j-1) F_{U(1)_{j-1,j}} \Tr (F_{\mathfrak{su}(j)}^3) + (j+1) F_{U(1)_{j,j+1}} \Tr (F_{\mathfrak{su}(j)}^3)\right],\\
  I_{\mathrm{plateau}} & =  \frac{1}{6} \left( - F_{U(1)_{M}} \mathrm{Tr}(F^{3}_{\mathfrak{su}(N)_{1}}) - (N - 1) F_{U(1)_{N-1,N_{1}}} \mathrm{Tr}(F^3_{\mathfrak{su}(N)_{1}}) + N F_{U(1)_{N_{1},N_{2}}} \mathrm{Tr} (F^{3}_{\mathfrak{su}(N)_{1}}) \right) +\\
  & +  \frac{1}{6} \sum_{i = 2}^{k - 1} \left[-N F_{U(1)_{N_{i-1},N_{i}}} \Tr(F_{\mathfrak{su}(N)_i}^3) + N F_{U(1)_{N_i,N_{i+1}}} \Tr(F_{\mathfrak{su}(N)_i}^3) \right] + \\
  & + \frac{1}{6} \left( -N F_{U(1)_{N_{k-1},N_{k}}} \Tr(F_{\mathfrak{su}(N)_k}^3) + N F_{U(1)_{R}} \Tr(F_{\mathfrak{su}(N)_k}^3) \right).
\end{align}
Of course, the contribution from the plateau is is absent when $k=0$.
As before, each gauge field strength in this expression removes one linear combination of $U(1)$'s from the theory. Thus, the $U(1)$ counting rules gives exactly one ABJ anomaly-free $U(1)$ symmetry.

We now determine the charge of each hypermultiplet under this $U(1)$. To begin, as noted above, the bifundamental of $\mf{su}(2)$ and $\mf{su}(3)$ cannot carry any $U(1)$ charge, since $\mf{su}(3) \oplus \mf{u}(1)$ is not a subalgebra of $\mf{g}_2$. To satisfy ABJ anomaly constraints, this means that the bifundamental of $\mf{su}(3)$ and $\mf{su}(4)$ must carry charge $q_{3,4} =0$. This constraint propagates to the rest of the ramp via the ABJ anomaly constraint equations:
\begin{equation}
    \begin{aligned}
    F_{\mathfrak{su}(3)}: &\quad 4 q_{3,4} = 0 \\
    F_{\mathfrak{su}(j)}: &\quad -(j-1) q_{j-1,j} + (j+1)q_{j,j+1} = 0 \quad (4 \leq j \leq N-1) \\
    F_{\mathfrak{su}(N)_1}: &\quad -(N-1) q_{N-1,N} - q_M + N q_{N_1,N_2} = 0 \\
    F_{\mathfrak{su}(N)_j}: &\quad -N q_{N_{j-1},N_{j}} + Nq_{N_{j},N_{j+1}} = 0 \quad (2 \leq j \leq k - 1) \\
    F_{\mathfrak{su}(N){k}} : &\quad -N q_{N_{k-1},N_{k}} + N q_{R} = 0 \, .
    \end{aligned}
\end{equation}
Solving these equations gives
\begin{equation}
q_{i,i+1} = 0\ \  (3 \leq i \leq N-1), \quad q_M = N, \quad q_{R} = +1 , \quad q_{N_j,N_{j+1}} =1\ \  (1 \leq j \leq k-1 ).
\end{equation}
We see that, indeed, the entire ramp is uncharged under the $U(1)$ global symmetry, and only the matter charged under the plateau of $\mf{su}(N)$ gauge algebras carries $U(1)$ charge.

The above analysis depends on the assumption that $k > 1$. For $k=1$, the theory has a ramp, but no plateau. Let us take the example of $N = 5$ and compare the $k=2$ theory with the $k=1$ theory:
\begin{equation}
\begin{aligned}
& [E_8] \,\, 1  \,\, \overset{\ksu(1)}{2} \,\, \overset{\ksu(2)}{2} \,\, \overset{\ksu(3)}{2} \,\, \overset{\ksu(4)}{2} \,\, \underset{[N_f=1]}{\overset{\ksu(5)_1}{2}}\,\, \overset{\ksu(5)_2}{2} \,\, [SU(5)] \\
& [E_8] \,\, 1  \,\, \overset{\ksu(1)}{2} \,\, \overset{\ksu(2)}{2} \,\, \overset{\ksu(3)}{2} \,\, \overset{\ksu(4)}{2} \,\, {\overset{\ksu(5)}{2}}\,\, [SU(6)].
\end{aligned}
\end{equation}
The theory with $k=2$ falls in the category that we have already analyzed above, for which have a single $U(1)$. The second theory with $k=1$, however, experiences an accidental flavor symmetry enhancement, $SU(5) \times U(1) \rightarrow SU(6)$, leaving us without an abelian flavor symmetry. Such symmetry enhancement occurs frequently when dealing with ``short SCFT quivers,'' which are characterized by the absence of a plateau of gauge algebras \cite{Heckman:2016ssk,Heckman:2018pqx,Hassler:2019eso}.

\subsection{Quivers with $\mf{su}(N)$ Matter with $N \geq 3$}

In this subsection we present a broader class of examples in which the gauge groups on the tensor branch are again $\mathfrak{su}(N)$ gauge
algebras, with hypermultiplets in complex representations, which we refer to as ``$\mathfrak{su}(N)$ matter.''
We defer the discussion of the $\mathfrak{su}(2)$ case to a later subsection.

To begin, we consider the theory of a single $-1$ tensor multiplet paired with an $\mf{su}(N)$ gauge algebra:
\begin{equation}
     \underset{[N_{\Lambda^2} = 1]}{\overset{\mathfrak{su}(N)}{1}}\,\,[SU(N+8)],~~~N \geq 5.
    \label{eqn:ex1}
\end{equation}
The $N=5$ version of this theory was considered in detail from both the field theory and F-theory perspectives in \cite{Lee:2018ihr}. The theory has $N+8$ fundamentals and one anti-symmetric hypermultiplet charged under $\mathfrak{su}(N)$. The former transform under an $SU(N+8)$ flavor symmetry. Since both the fundamental and the anti-symmetric representations are complex for $N \geq 5$, each of them gives a candidate $U(1)$ global symmetry. We denote these $U(1)$'s as $U(1)_F$ and $U(1)_{\Lambda^2}$, respectively.

Let us use the candidate $U(1)$ method (method 1) to determine the $U(1)$ charges. The fundamentals of $\mf{su}(N)$ carry charge $q_{F}$ under $U(1)_{F}$, whereas the anti-symmetric carries charge $q_{\Lambda^2}$ under $U(1)_{\Lambda^2}$. We take both charges to be $+1$, and determine, up to normalization, the linear combinations free of ABJ anomalies. The overall normalization can also be fixed by the lattice of charges obtained from a progenitor theory. The theory has an ABJ anomaly of the form:

\begin{equation}
    I_{\mathrm{ABJ}} = \frac{1}{6} \Big((N+8) F_{U(1)_F} \Tr_{\textrm{fund}} (F_{\ksu(N)}^3)  + 1 F_{U(1)_{\Lambda^2}} \Tr_{\Lambda^2} (F_{\ksu(N)}^3) \Big).
\end{equation}
Using Table \ref{tab:GTCC}, we note that we may write $\Tr_{\Lambda^2}F_{\ksu(N)}^3 = (N-4) \Tr_{\textrm{fund}}F_{\ksu(N)}^3$. Thus, we see that the linear combination
\begin{align}
    (N+8) F_{U(1)_F}  + (N-4) F_{U(1)_{\Lambda^2}}
\end{align}
suffers from an ABJ anomaly, and it is not a good global symmetry of the theory. One $U(1)$ remains a valid symmetry of the theory, and it is generated by the linear combination
\begin{equation}
    T_{\mathrm{survive}} \propto (N-4) t_{F}- (N+8) t_{\Lambda^2}.
\end{equation}
The full global symmetry of the theory is therefore $SU(N+8) \times U(1)$, and the matter content under $\mf{su}(N) \times SU(N+8) \times U(1)$ is given by:
\begin{equation}
    (\textbf{N}, \overline{\textbf{N+8}})_{(N-4)} \oplus  ( \textbf{(N(N-1)/2}, \textbf{1})_{-(N+8)}.
\end{equation}

Note that this analysis applies only for $N \geq 5$. For $N=4$, the anti-symmetric is real rather than complex, and the flavor symmetry enhances to $SU(12) \times Sp(1)$. For $N=3$, the anti-symmetric is simply an anti-fundamental, so we get another hypermultiplet in the fundamental and the flavor symmetry enhances to $SU(12) = SU(N+9)$. For $N=2$, the anti-symmetric representation is trivial, whereas the fundamental representation is pseudo-real, and the flavor symmetry is given by $SO(20)$.

\subsubsection{Comparison of $U(1)$ Identification Methods}

Up to this point, we have presented some examples which illustrate the merits of the two methods for determining two $U(1)$'s. We now present an
illustrative example which shows how the two arrive at the same answer. With this in mind, consider the theory
\begin{equation}
\underset{[N_f=2]}{\overset{\mf{su}(3)}2} \,\, \underset{[N_f=1]}{\overset{\mf{su}(4)_1}2}  \underset{[N_f=4]}{\overset{\mf{su}(4)_2}2} \,\,.
\label{eqn:compareTwoMethods}
\end{equation}
Our plan will be to first analyze candidate $U(1)$ symmetries by listing all possible global symmetries which could act on matter fields (method 1). We will then compare this to the $U(1)$ symmetries obtained by treating each quiver node in isolation, and computing the branching rules associated with weakly gauging some of the flavor symmetries of each node.

We begin with method 1, listing all possible global symmetries which could act on matter fields. By inspection, there is a $U(2)_L \sim SU(2)_L \times U(1)_L$ flavor symmetry rotating the two fundamentals of $\mf{su}(3)$, a $U(4)_R \sim SU(4)_R \times U(1)_R$ rotating the four fundamentals of $\mf{su}(4)_2$, a $U(1)_M$ acting on the fundamental of $\mf{su}(4)_1$, and two baryonic $U(1)$'s, which we denote $U(1)_{B,L}$ and $U(1)_{B,R}$ acting on the bifundamentals.

The theory has ABJ anomalies of the form:
\begin{equation}
\begin{aligned}
I_{ABJ} = \frac{1}{6} \Big( -2 F_{U(1)_L} \Tr (F_{\ksu(3)}^3) + 4 F_{U(1)_{B, L}} \Tr (F_{\ksu(3)}^3) -  3 F_{U(1)_{B, L}} \Tr (F_{\ksu(4)_1}^3)
-  F_{U(1)_M} \Tr (F_{\ksu(4)_1}^3) \\
\quad  + 4 F_{U(1)_{B, R}} \Tr (F_{\ksu(4)_1}^3) - 4 F_{U(1)_{B, R}} \Tr (F_{\ksu(4)_2}^3) +  4 F_{U(1)_{R}} \Tr (F_{\ksu(4)_2}^3) \Big).
\end{aligned}
\end{equation}
Let us denote a basis of the candidate $U(1)$ space as $(F_{U(1)_L}, F_{U(1)_{B, L}}, F_{U(1)_M}, F_{U(1)_{B, R}}, F_{U(1)_R})$. The three gauge groups each have an ABJ anomaly, which are associated with the following linear combinations of $U(1)$'s:
\begin{equation}
\begin{aligned}
{\mathfrak{su}(3)}:&\ \  (-2, 4, 0, 0, 0)\\
{\mathfrak{su}(4)_1}:&\ \ (0, -3, -1, 4, 0)\\
{\mathfrak{su}(4)_2}:&\ \ (0, 0, 0, -4, 4)
\end{aligned}
\end{equation}
The subspace of ABJ anomaly-free $U(1)$'s is the two-dimensional null space of these three vectors,\footnote{That is to say, we consider the space of vectors which are orthogonal to these three vectors inside our five-dimensional space.} and is spanned by:
\begin{equation}
\begin{aligned}
T_a & \propto (2, 1, 1, 1, 1) \\
T_b & \propto (2, 1, -3, 0, 0)
\end{aligned}
\label{eq:nullspace}
\end{equation}

Next, we analyze the abelian symmetries of the same theory of line \eqref{eqn:compareTwoMethods} using branching rules of symmetries (method 2). Each $-2$ tensor carrying an $\mf{su}(4)$ gauge algebra has an associated flavor symmetry $SU(8)$, under which 8 fundamentals of $\mf{su}(4)$ transform. For $\mf{su}(4)_1$, this $SU(8)$ is broken according to:
\begin{equation}
\begin{aligned}
SU(8) & \supset SU(4) \times U(1)  \times SU(4) \rightarrow (SU(3) \times U(1)) \times U(1) \times SU(4)  \\
\mathbf{8} & \rightarrow  (\mathbf{4},\mathbf{1})_{-1} + (\mathbf{1},\mathbf{4})_{1} \rightarrow (\mathbf{1}, \mathbf{3})_{(0,-1)} + (\mathbf{3}, \mathbf{1})_{(1, 1)} + (\mathbf{1}, \mathbf{1})_{(-3, 1)}.
\label{eq:firstSU}
\end{aligned}
\end{equation}
 Here, the $SU(3)$ factor is identified with the $\mf{su}(3)$ gauge symmetry, while the $SU(4)$ factor is identified with the $\mf{su}(4)_2$ gauge symmetry.

Consider next the flavor symmetries associated with the $\mf{su}(4)_2$ gauge algebra. In isolation from the other
parts of the quiver, there is again a $SU(8)$ flavor symmetry. Weakly gauging appropriate subalgebras leads to the branching rules:
\begin{equation}
\begin{aligned}\label{eqn:treeoflife}
SU(8) & \supset SU(4) \times U(1) \times SU(4)  \\
\mathbf{8} & \rightarrow  ( \mathbf{4}, \mathbf{1})_{1}  + ( \mathbf{1}, \mathbf{4})_{-1}
\end{aligned}
\end{equation}
Note that the first $SU(4)$ here is identified with the $\mf{su}(4)_1$ gauge symmetry, so a state of charge $+1$ under the $U(1)$ of line \eqref{eqn:treeoflife} will correspond to a state of charge $(0,1)$ under the $U(1) \times U(1)$ symmetries appearing in line (\ref{eq:firstSU}).

Finally, there are the global symmetries associated with the $\mf{su}(3)$ gauge algebra on a $-2$ curve.
In isolation, this theory has an $SU(6)$ global symmetry. We weakly gauge an $\mathfrak{su}(4)_{1}$ subalgebra to reach the
theory of line (\ref{eqn:compareTwoMethods}). The branching rules are:
\begin{equation}
\begin{aligned}
SU(6) & \supset SU(2) \times U(1) \times SU(4)  \\
\mathbf{6} & \rightarrow ( \mathbf{1}, \mathbf{4})_{-1}  + (\mathbf{2}, \mathbf{1})_{2}.
\end{aligned}
\label{eq:branch3}
\end{equation}
The $SU(4)$ factor here is identified with the $\mf{su}(4)_1$ gauge symmetry, so the $U(1)$ charge $-1$ here is identified with the $U(1)$ charge vector $(0,-1)$ in line (\ref{eq:firstSU}). Thus, the full matter content under the $(SU(2), \mathfrak{su}(3),\mathfrak{su}(4)_L,\mathfrak{su}(4)_R, SU(4))_{(U(1)_a, U(1)_b)}$ symmetry is given by
\begin{equation}\label{eq:matter}
\begin{aligned}
(\mathbf{2}, \overline{\mathbf{3}}, \mathbf{1}, \mathbf{1}, \mathbf{1})_{(2, 2)} \oplus
(\mathbf{1}, \mathbf{3}, \overline{\mathbf{4}}, \mathbf{1}, \mathbf{1})_{(1, 1)} \oplus
(\mathbf{1}, \mathbf{1}, \overline{\mathbf{4}}, \mathbf{1}, \mathbf{1})_{(1, -3)}  \\
\oplus
(\mathbf{1}, \mathbf{1}, \mathbf{4}, \overline{\mathbf{4}}, \mathbf{1})_{(1, 0)}
\oplus
(\mathbf{1}, \mathbf{1}, \mathbf{1}, \mathbf{4}, \overline{\mathbf{4}})_{(1, 0)}
\end{aligned}
\end{equation}
Note that we have added an additional minus sign relative to the charges in the branchings (\ref{eq:firstSU})-(\ref{eq:branch3}) whenever taking the complex conjugate.

Let us compare this result to that obtained using method 1 above. There, each of these five multiplets carries charge $1$ under a different candidate $U(1)$ and charge 0 under the other four candidate $U(1)$'s. However, projecting the charge vectors onto the basis vectors in (\ref{eq:nullspace}), which span the ABJ anomaly-free subspace of candidate $U(1)$'s, we find precisely the charges in (\ref{eq:matter}), demonstrating agreement between the two methods.

Having worked out the full matter content of the theory, we may use the prescription of Appendix \ref{app:A} to compute the full anomaly polynomial of this theory, including the $U(1)$ field strengths $F_a, F_b$ corresponding to $U(1)_a, U(1)_b$, respectively:
\begin{equation}
\begin{aligned}
I_8 =&  \frac{395}{12} c_2(R)^2 - \frac{35}{48} c_2(R) p_1(T) + \frac{181}{5760} p_1(T)^2 - \frac{103}{1440} p_2(T) + c_2(R)\Big( -\frac{21}{16} (\Tr F_{SU(2)}^2) \\
& - \frac{23}{16} (\Tr F_{SU(4)}^2) - \frac{169}{2} F_a^2 - 63 F_a F_b - \frac{153}{2} F_b^2\Big) + \frac{5}{128} (\Tr F_{SU(2)}^2)^2 + \frac{3}{128} (\Tr F_{SU(4)}^2)^2\\
& + \frac{1}{6} (\Tr F_{SU(4)}^4) + \frac{1}{64} (\Tr F_{SU(2)}^2)(\Tr F_{SU(4)}^2) - \frac{2}{3} F_a \Tr F_{SU(4)^3} + \Tr F_{SU(2)}^2 \Big(  \frac{27}{8} F_a^2 + \frac{21}{4} F_a F_b\\
& + \frac{27}{8} F_b^2 \Big) + \Tr F_{SU(4)}^2 \Big(  \frac{17}{8} F_a^2 + \frac{3}{4} F_a F_b + \frac{9}{8} F_b^2 \Big) \\
& + p_1(T)\Big( \frac{1}{32} (\Tr F_{SU(2)}^2) + \frac{1}{24} (\Tr F_{SU(4)}^2) + \frac{3}{2} F_a^2 + F_a F_b + \frac{3}{2} F_b^2 \Big) \\
& + \frac{119}{2} F_a^4 + 106 F_a^3 F_b + 189 F_a^2 F_b^2 + 90 F_a F_b^3 + \frac{135}{2} F_b^4.
\end{aligned}
\end{equation}

\subsection{Examples with $\mathfrak{su}(2)$ Gauge Symmetry}

As we have already mentioned, the appearance of $\mathfrak{su}(2)$ gauge algebras complicates the analysis because
the flavor symmetry of the associated SCFT is slightly smaller than what might appear possible from a ``naive'' analysis
of the tensor branch description. To illustrate, consider the theory
\begin{equation}
     \underset{[N_f=1]}{\overset{\mathfrak{su}(2)}{2}}\,\, \overset{\mathfrak{su}(3)}{2} \,\, [SU(4)].
    \label{eqn:ex3}
\end{equation}


The hypermultiplet of $\mf{su}(2)$ is pseudo-real, so it transforms under $SO(2) \sim U(1)_L$. The four fundamentals of $\mf{su}(3)$ transform under $U(4)_R \sim SU(4) \times U(1)_R$. One might have also expected a baryonic $U(1)_B$ under which the bifundamental of $\ksu(2)$ and $\ksu(3)$ transforms. However, we recall that eight half-hypermultiplets in the fundamental representation of $\ksu(2)$ transform as a spinor of $Spin(7)$, rather than the na\"ively expected vector of $SO(8)$. Since $Spin(7)$ decomposes as $Spin(7) \supset SU(3) \times U(1)$, it gives only a single $U(1)$, rather than the pair of $U(1)$'s which would have been expected from the decomposition $SO(8) \supset SU(3) \times U(1)^2$. As a result, the $SO(2) \sim U(1)_L$ and the $U(1)_B$ are condensed into a single $U(1)$, which in conjunction with $U(1)_R$ gives two candidate $U(1)$ flavor symmetries. However, the $\ksu(3)$ gauge symmetry introduces an ABJ anomaly, which removes one linear combination of $U(1)$'s and ultimately leaves a global symmetry of $SU(4) \times U(1)$.

The $U(1)$ charges of the various matter multiplets can be determined by the branching rule method. The $\textbf{8}$ of $Spin(7)$ decomposes under $SU(3) \times U(1)$ according to
\begin{equation}
    \textbf{8} \rightarrow \textbf{3}_1 \oplus \bar{\textbf{3}}_{-1} \oplus \textbf{1}_{-3} \oplus \textbf{1}_{3}.
\end{equation}
The $\textbf{6}$ of $SU(6)$ decomposes under $SU(4) \times SU(2) \times U(1)$ according to
\begin{equation}
    \textbf{6} \rightarrow (\textbf{4},\textbf{1})_{-2} \oplus (\textbf{1},\textbf{2})_{4}.
\end{equation}
Putting these together, we find that the full matter content of the theory under $ \mf{su}(2) \times \mf{su}(3) \times SU(4) \times U(1)$ is:
\begin{equation}
 (\textbf{2}, \overline{\textbf{3}}, \textbf{1})_{2} \oplus (\textbf{2}, \textbf{1}, \textbf{1})_{-6}  \oplus (\textbf{1}, \textbf{3},  \overline{\textbf{4}} )_{1}.
\end{equation}
Note that with these charge assignments, the $U(1)$ is indeed free of any ABJ anomaly associated with the $\mf{su}(3)$ gauge algebra.

Given the full matter content of the theory, we may again use the prescription of Appendix \ref{app:A} to compute the full anomaly polynomial of this theory, including the $U(1)$ field strength, which we denote by $F$:
\begin{equation}
\begin{aligned}
I_8 =&  \frac{143}{24} c_2(R)^2 - \frac{3}{16} c_2(R) p_1(T) + \frac{109}{5760} p_1(T)^2 - \frac{67}{1440} p_2(T) +c_2(R)\Big(  -72 F^2 - \frac{2}{3} \Tr F_{SU(4)}^2 \Big) \\
& + \frac{1}{8} \Tr F_{SU(4)}^4 + \frac{1}{48} (\Tr F_{SU(4)}^2)^2  -  \frac{1}{2} F \Tr F_{SU(4)}^3 + \frac{27}{8} F^2 \Tr F_{SU(4)}^2 + \frac{729}{2}F^4 \\
& + p_1(T) \Big( \frac{9}{4} F^2 + \frac{1}{32} \Tr F_{SU(4)}^2 \Big).
\end{aligned}
\end{equation}

More generally, by a similar analysis, the theory
\begin{equation}
     \underset{[N_f=1]}{\overset{\mathfrak{su}(2)}{2}}\,\, \underset{[N_f=1]}{\overset{\mathfrak{su}(3)}{2}} \,\, ... \,\, \,\, \underset{[N_f=3]}{\overset{\mathfrak{su}(3)}{2}}
     \label{eqn:su2noU1}
\end{equation}
will have a global symmetry of $SU(3) \times U(1)^2$, in agreement with our general prescription for $U(1)$ counting.
The $U(1)$ charges of the hypermultiplets in the theory may similarly be determined using the branching rules of $SO(7)$ and $SU(6)$.

Let us next consider the theory of line (\ref{eqn:ex3}), but with an unpaired $-2$ tensor added to the left of the quiver:
\begin{equation}
     2 \,\, \overset{\mathfrak{su}(2)}{2}\,\, \overset{\mathfrak{su}(3)}{2} \,\, [SU(4)].
    \label{eqn:ex4}
\end{equation}
where on the unpaired tensor there is formally an $\mathfrak{su}(1)$ gauge algebra, $\overset{\mathfrak{su}(1)}{2}$, meaning that there is an hypermultiplet in the fundamental of $\mathfrak{su}(2)$ of the neighbor $-2$ curve. In what follows, every time an unpaired tensor is coupled this will be implied.
Once again, there is a $U(1)_R$ from coming from the four fundamentals of $\mf{su}(3)$, which transform under $U(4) \sim SU(4) \times U(1)_R$. However, this is the only $U(1)$ that shows up: the unpaired $-2$ tensor effectively reduces the flavor symmetry of the eight half-hypermultiplets in the fundamental representation of $\mf{su}(2)$ to $G_2$, which decomposes as $G_2 \rightarrow SU(3)$: there is no $U(1)$ factor under which the half-hypermultiplets in the fundamental of $\mf{su}(2)$ (or the bifundamental of $\mf{su}(2) \times \ksu(3)$) transform. Thus, there is only a single candidate $U(1)$, namely $U(1)_R$. However, this $U(1)$ is removed by the $\mf{su}(3)$ ABJ anomaly, leaving only an $SU(4)$ global symmetry remaining.

More generally, by a similar analysis, the theory
\begin{equation}
    2\,\, \overset{\mathfrak{su}(2)}{2}\,\, \underset{[N_f=1]}{\overset{\mathfrak{su}(3)}{2}} \,\, ... \,\, \,\, \underset{[N_f=3]}{\overset{\mathfrak{su}(3)}{2}}
\end{equation}
will have a global symmetry of $SU(3) \times U(1)$.

This theory also arises from Higgsing of the A-type progenitor theory obtained from M5-branes probing a $\mathbb{C}^2 / \mathbb{Z}_{3}$ singularity. In that case, we label the resulting theory by nilpotent orbits of $\mathfrak{su}(3)_L$ and $\mathfrak{su}(3)_R$. These are associated with associated partitions $\mu_L=[3]$, $\mu_R=[1^3]$, which have commutants $H_L = \emptyset$, $H_R=SU(3)$.
We thus expect a global symmetry of $SU(3) \times U(1)$, which indeed matches our field theory analysis.

As another example, consider the theory
\begin{equation}
    {\overset{\mathfrak{su}(2)}{2}}\,\, \underset{[N_f=6]}{\overset{\mathfrak{su}(4)}{2}}.
\end{equation}
There is a $U(1)_R$ coming from the six fundamentals of $\mf{su}(4)$, which transform under $U(6) \sim SU(6) \times U(1)_R$. However, this is the only $U(1)$ that shows up: the flavor symmetry $SO(7)$ of the leftmost $-2$ tensor is gauged by $\mf{su}(4) \simeq \mf{so}(6)$, but $U(1) \times SU(4)$ is not a subgroup of $SO(7)$. As a result, there is no $U(1)$ factor under which bifundamental of $\mf{su}(2) \times \ksu(4)$ transforms. Thus, there is only a single candidate $U(1)$, namely $U(1)_R$, and this $U(1)$ is removed by the $\mf{su}(4)$ ABJ anomaly,
leaving only an $SU(6)$ global symmetry remaining.

More generally, by a similar analysis, the theory
\begin{equation}
     {\overset{\mathfrak{su}(2)}{2}}\,\, \underset{[N_f=2]}{\overset{\mathfrak{su}(4)}{2}}\,\, {\overset{\mathfrak{su}(4)}{2}} \,\, ... \,\, \,\, \underset{[N_f=4]}{\overset{\mathfrak{su}(4)}{2}}.
\end{equation}
will have a global symmetry of $SU(4) \times SU(2) \times U(1)$. This theory corresponds to a Higgs branch flow obtained from the
A-type progenitor theory of M5-branes probing the singularity $\mathbb{C}^2 / \mathbb{Z}_4$. Higgsing of the flavor symmetries on the left and right is characterized by nilpotent orbits of $\mathfrak{su}(4)_L$ and $\mathfrak{su}(4)_R$ associated with partitions $\mu_L=[2^2]$, $\mu_R=[1^4]$, which have commutants $H_L = SU(2)$, $H_R=SU(4)$. From our discussion in Appendix B, line (\ref{eqn:nilpotentglobal}), we expect a global symmetry of $SU(4) \times SU(2) \times U(1)$, which indeed matches our field theory analysis.

Next, consider the theory
\begin{equation}
    [SO(14)] \,\, {\overset{\mathfrak{sp}(1)}{1}}\,\, {\overset{\mathfrak{su}(3)}{2}} \,\, [SU(4)].
\end{equation}
This theory shows up as (part of) a quiver in section 5.3 of \cite{Mekareeya:2017jgc}. There is a $U(1)_R$ coming from the four fundamentals of $\mf{su}(3)$, which transform under $U(4) \sim SU(4) \times U(1)_R$. There is also a $U(1)_B$ associated with the bifundamental of $\mf{sp}(1)$ and $\mf{su}(3)$: the flavor symmetry $SO(20)$ of $\ksp(1)$ associated with the $-1$ tensor decomposes into $SO(14) \times SO(6)$, and this $SO(6)$ further decomposes into $SU(3) \times U(1)_B$. The $SU(3)$ factor is gauged, but the $U(1)_B$ remains. Similarly, the $U(6)$ flavor symmetry of the six fundamentals of $\mf{su}(3)$ decomposes into $U(4) \times U(2)$, which further decompose as $U(4) \sim SU(4) \times U(1)_R $ and $U(2) \sim SU(2) \times U(1)_B$. This $SU(2) \simeq Sp(1)$ is then gauged. As a result, we have two candidate $U(1)$'s: $U(1)_B$ and $U(1)_R$, but the ABJ anomaly associated with $\mf{su}(3)$ eliminates one linear combination of them. Therefore, in the final analysis, the flavor symmetry is $SO(14) \times SU(4) \times U(1)$.

\subsection{Examples with only $\mf{su}(2)$ Gauge Symmetries}\label{sec:SU2only}

Having discussed in great detail situations where we have hypermultiplets in complex representations of $\mathfrak{su}(N)$
we now turn to some cases where the gauge algebra consists solely of $\mathfrak{su}(2)$ on the tensor branch. These
cases are interesting because our methods predict that there are no $U(1)$ symmetries in such situations. Additionally, we
can use our approach to also extract the non-abelian symmetries from these cases.

As a first example, we begin with the theory
\begin{equation}
     \underset{[N_f=2]}{\overset{\ksu(2)_1}{2}} \,\, \underset{[N_f=2]}{\overset{\ksu(2)_2}{2}}.
     \label{eq:su2ex1}
\end{equation}
There is a bifundamental $({\bf 2}_1,{\bf2}_2)$ of the two $\ksu(2)$ algebras, and there are two additional fundamentals for each of these algebras. The fact that the eight half-hypermultiplets in the fundamental representation of a given $\mf{su}(2)$ gauge algebra transform as a spinor of $SO(7)$ is crucial for determining the global symmetry of this theory: namely, $SO(7)$ decomposes as $SO(7) \rightarrow SU(2)^3$, and the spinor obeys the branching rule
\begin{equation}
    {\bf 8} \rightarrow (\bf{2}, \bf{2}, \bf{1}) \oplus (\bf{2},\bf{1}, \bf{2}).
    \label{eq:spinbranch}
\end{equation}
In the theory at hand, one of the $SU(2)$'s is gauged by the other $-2$ tensor. As a result, the full matter content of the theory is given by
\begin{equation}
   \frac{1}{2} ({\bf 2}_1, {\bf 2}_2, {\bf 1}_L,{\bf 2}_B,{\bf 1}_R) \oplus  \frac{1}{2} ({\bf 2}_1, {\bf 1}_2, {\bf 2}_L,{\bf 2}_B,{\bf 1}_R) \oplus  \frac{1}{2} ({\bf 1}_1, {\bf 2}_2, {\bf 1}_L,{\bf 2}_B,{\bf 2}_R).
\end{equation}
Here, the subscripts $1$ and $2$ represent the gauge algebras $\ksu(2)_1$ and $\ksu(2)_2$, respectively, while the subscripts $L$, $B$, $R$ represent $SU(2)$ global symmetries associated with the two fundamentals of $\ksu(2)_1$, the bifundamental, and the two fundamentals of $\ksu(2)_2$, respectively. We see that the theory has an $SU(2)^3$ global symmetry.

According to the matter content and its charge under the global symmetry factors, we are able to compute the full anomaly polynomial including the global symmetry field strengths, denoted as $F_{SU(2)_L}, F_{SU(2)_R}$ and $F_{SU(2)_B}$:
\begin{equation}
\begin{aligned}
I_8 =&  \frac{23}{6} c_2(R)^2 - \frac{1}{12} c_2(R) p_1(T) + \frac{11}{720} p_1(T)^2 - \frac{2}{45} p_2(T)+ c_2(R) \Big( -2 \Tr F_{SU(2)_B}^2 - \frac{1}{2} \Tr F_{SU(2)_L}^2\\
& - \frac{1}{2} \Tr F_{SU(2)_R}^2  \Big)  + \frac{9}{32} (\Tr F_{SU(2)_B}^2)^2+ \frac{1}{32}(\Tr F_{SU(2)_L}^2)^2+ \frac{1}{32}(\Tr F_{SU(2)_R}^2)^2\\
& + p_1(T) \Big( \frac{1}{48} \Tr F_{SU(2)_L}^2 + \frac{1}{48} \Tr F_{SU(2)_R}^2 + \frac{1}{16}\Tr F_{SU(2)_B}^2 \Big) \\
& + \frac{3}{16} \Tr F_{SU(2)_B}^2 \Tr F_{SU(2)_L}^2 + \frac{3}{16} \Tr F_{SU(2)_B}^2 \Tr F_{SU(2)_R}^2 + \frac{1}{48} \Tr F_{SU(2)_L}^2 \Tr F_{SU(2)_R}^2
\end{aligned}
\end{equation}


Next, let us consider the theory of (\ref{eq:su2ex1}), but with an unpaired $-2$ tensor added to the left-hand side. Now we have
\begin{equation}
    2 \,\, \underset{[N_f=1]}{\overset{\ksu(2)_1}{2}} \,\, \underset{[N_f=2]}{\overset{\ksu(2)_2}{2}}.
\end{equation}

In this case, the global symmetry associated with the middle $-2$ tensor is $G_2$, which decomposes as $G_2 \rightarrow SU(2)^2$. The ${\bf 7}$ of $G_2$ obeys the branching rule
\begin{equation}
    {\bf 7} \rightarrow (\bf{2}, \bf{2}) \oplus (\bf{1},\bf{3}).
\end{equation}
The first $SU(2)$ factor here is gauged by the $\mf{su}(2)_2$ gauge algebra.
As a result, the full matter content of the theory is given by
\begin{equation}
   \frac{1}{2} ({\bf 2}_1, {\bf 2}_2, {\bf 2}_B,{\bf 1}_R) \oplus  \frac{1}{2} ({\bf 2}_1, {\bf 1}_2, {\bf 3}_B,{\bf 1}_R) \oplus \frac{1}{2} ({\bf 2}_1, {\bf 1}_2, {\bf 1}_B,{\bf 1}_R) \oplus \frac{1}{2} ({\bf 1}_1, {\bf 2}_2, {\bf 2}_B,{\bf 2}_R).
\end{equation}
We see that relative to the previous example, the unpaired $-2$ tensor has effectively combined the $SU(2)_L$ symmetry and the $SU(2)_B$ symmetry, resulting in a total global symmetry of $SU(2)^2$. This is quite similar to what happened in the example in (\ref{eqn:ex4}) above: the addition of the unpaired $-2$ tensor in that case combined two $U(1)$ global symmetries into one.

What happens if we add an unpaired tensor to the other side of the quiver as well? We then have
\begin{equation}
    2 \,\, \underset{[N_f=1]}{\overset{\ksu(2)_1}{2}} \,\, \underset{[N_f=1]}{\overset{\ksu(2)_2}{2}} \,\, 2.
\end{equation}
Now, using the $G_2$ branching rule for both gauge algebras, we find a total matter content of
\begin{equation}
   \frac{1}{2} ({\bf 2}_1, {\bf 2}_2, {\bf 2}_B) \oplus  \frac{1}{2} ({\bf 2}_1, {\bf 1}_2, {\bf 3}_B)\oplus  \frac{1}{2} ({\bf 2}_1, {\bf 1}_2, {\bf 1}_B) \oplus  \frac{1}{2} ({\bf 1}_1, {\bf 2}_2, {\bf 3}_B) \oplus  \frac{1}{2} ({\bf 1}_1, {\bf 2}_2, {\bf 1}_B).
\end{equation}
Now, the unpaired $-2$ tensor on the right-hand side has combined the $SU(2)_R$ symmetry and the $SU(2)_B$ symmetry. The theory has only an $SU(2)$ global symmetry.

More generally, a theory of the form
\begin{equation}
     \underset{[N_f=2]}{\overset{\ksu(2)_1}{2}} \,\, {\overset{\ksu(2)_2}{2}} \,\, ... \,\, \underset{[N_f=2]}{\overset{\ksu(2)_k}{2}}
\end{equation}
has an $SU(2)^3$ global symmetry provided $k \geq 2$ (for $k=1$ it has an $SO(7)$ global symmetry). The reason why we find only $SU(2)^3$ rather than $SU(2)^{k+3}$ is due to the branching rule (\ref{eq:spinbranch}). Concentrating on eight half-hypermultiplets charged under the the $i$th gauge symmetry factor $\mf{su}(2)_i$, we find a decomposition
\begin{equation}
\frac{1}{2} (\mathbf{2}_i, \mathbf{8}) \rightarrow \frac{1}{2} (\mathbf{2}_i, \mathbf{2}_{B}, \mathbf{2}_{i-1}, \mathbf{1}) \oplus (\mathbf{2}_i, \mathbf{2}_{B}, \mathbf{1}, \mathbf{2}_{i+1}).
\end{equation}
Crucially, the bifundamental $( \mathbf{2}_{i-1}, \mathbf{2}_i)$ and the bifundamental $(\mathbf{2}_i, \mathbf{2}_{i+1})$ transform under the \emph{same} $SU(2)_B$ baryonic symmetry. This propagates down the entire quiver, so there is only one baryonic $SU(2)_B$ in addition to the $SU(2)_L$ and $SU(2)_R$ symmetries, rather than the $k+1$ baryonic symmetries we would have found if the flavor symmetry of the eight half-hypermultiplets of each $\mf{su}(2)_i$ were $SO(8)$ rather than $Spin(7)$. Note that for $k=2$, this theory actually shows up in the 6D SCFT-group theory correspondence discussed in Appendix \ref{sec:grouptheory} as the $E_7$ nilpotent orbit of Bala-Carter label $A_2+2A_1$ (see Appendix A.2 of \cite{Heckman:2016ssk}), and its global symmetry is indeed $SU(2)^3$.

By a similar analysis, a theory of the form
\begin{equation}
   2 \,\,  \underset{[N_f=1]}{\overset{\ksu(2)_1}{2}} \,\, {\overset{\ksu(2)_2}{2}} \,\, ... \,\, \underset{[N_f=1]}{\overset{\ksu(2)_k}{2}}
   \label{eq:examp}
\end{equation}
has an $SU(2)^2$ global symmetry provided $k \geq 2$ (for $k=1$ it has a $G_2$ global symmetry). For $k=2, 3$, this theory shows up in the 6D SCFT-group theory correspondence with the $E_8$ nilpotent orbits of Bala-Carter labels $D_5(a_1)+A_1$ and $A_4+A_2$, respectively (see Appendix A.3 of \cite{Heckman:2016ssk}). For $k=4$, the theory shows up in the correspondence with $\text{Hom}(SL(2,5), E_8)$ (see Appendix B.5 of \cite{Frey:2018vpw}). In all three of these cases, its global symmetry is indeed $SU(2)$.

Additionally,
\begin{equation}
   2 \,\,  \underset{[N_f=1]}{\overset{\ksu(2)_1}{2}} \,\, {\overset{\ksu(2)_2}{2}} \,\, ... \,\, \underset{[N_f=2]}{\overset{\ksu(2)_k}{2}} \,\, 2
\end{equation}
has an $SU(2)$ global symmetry provided $k \geq 2$ (for $k=1$ it has an $SU(3)$ global symmetry). For $k=2$, this theory shows up in the 6D SCFT-group theory correspondence with the $E_8$ nilpotent orbit of Bala-Carter label $A_4+A_2+A_1$ (see Appendix A.3 of \cite{Heckman:2016ssk}), and its global symmetry is indeed $SU(2)$.

We can also consider D-type quivers, such as the theory:
\begin{equation}
   2 \,\,  \overset{2}{\overset{\ksu(2)_1}{2}} \,\, {\overset{\ksu(2)_2}{2}} \,\, ... \,\, \underset{[N_f=2]}{\overset{\ksu(2)_k}{2}}
\end{equation}
Now, since the tensor carrying $\mf{su}(2)_1$ meets two unpaired $-2$ tensors, its flavor symmetry is reduced to $SU(3)$, of which $SU(2)$ is gauged by $\mf{su}(2)_2$. However, since $\mf{su}(2) \oplus \mf{su}(2)$ is not a subalgebra of $\mf{su}(3)$, there is no additional $SU(2)_B$ global symmetry acting on the hypermultiplets charged under $\mf{su}(2)_1$. Thus, the full global symmetry of the theory is simply $SU(2)_R$, coming from the two fundamental hypermultiplets of $\mf{su}(2)_k$, in contrast with the $SU(2)^2$ we saw in (\ref{eq:examp}).\footnote{One might have thought that the symmetry would be $SU(2)_R \times U(1)_B$ rather than simply $SU(2)_R$, since $SU(2) \times U(1) \subset SU(3)$. However, the branching rule $\mathbf{3} \rightarrow \mathbf{2}_1 \oplus \mathbf{1}_{-2}$ for $SU(3) \rightarrow SU(2) \times U(1)$ associated with the flavor symmetry of $\mf{su}(2)_1$ is incompatible with the rule $\mathbf{8} \rightarrow (\mathbf{2}, \mathbf{1})_1 \oplus (\mathbf{2}, \mathbf{1})_{-1} \oplus  (\mathbf{1}, \mathbf{2})_1 \oplus (\mathbf{1}, \mathbf{2})_{-1}$ of $Spin(7) \rightarrow SU(2) \times SU(2) \times U(1)$ for the flavor symmetry of $\mf{su}(2)_2$, indicating that a $U(1)$ global symmetry cannot exist.} For $k=2$, the full matter content is given by
\begin{equation}
   ({\bf 2}_1, {\bf 2}_2, {\bf 1}_R) \oplus  2 ({\bf 2}_1, {\bf 1}_2, {\bf 1}_R) \oplus   ({\bf 1}_1, {\bf 2}_2, {\bf 2}_R).
\end{equation}
For $k=3$, it is given by
\begin{equation}
   ({\bf 2}_1, {\bf 2}_2, {\bf 1}_3, {\bf 1}_R) \oplus  2  ({\bf 2}_1, {\bf 1}_2, {\bf 1}_3, {\bf 1}_R) \oplus   ({\bf 1}_1, {\bf 2}_2, {\bf 2}_3, {\bf 1}_R)  \oplus ({\bf 1}_1, {\bf 1}_2, {\bf 2}_3, {\bf 2}_R).
\end{equation}

By similar reasoning,
\begin{equation}
   2 \,\,  \overset{2}{\overset{\ksu(2)_1}{2}} \,\, {\overset{\ksu(2)_2}{2}} \,\, ... \,\, \underset{[N_f=1]}{\overset{\ksu(2)_k}{2}}\,\, 2
\end{equation}
has no global symmetry at all. For $k=2$, the full matter content is given by
\begin{equation}
   ({\bf 2}_1, {\bf 2}_2) \oplus  2 ({\bf 2}_1, {\bf 1}_2) \oplus  2 ({\bf 1}_1, {\bf 2}_2).
\end{equation}

\subsection{An Example with $\mathfrak{sp}(N)$ Matter}

Finally, let us consider the theory
\begin{equation}
      \underset{[N_f=8]}{\overset{\ksp(1)_1}{1}} \,\, \underset{[N_f=2]}{\overset{\ksu(2)_2}{2}}.
\end{equation}
The theory has an $SO(16) \times SU(2) \times SU(2)$ global symmetry, and the matter content is
\begin{equation}
   \frac{1}{2} ({\bf 2}_1, {\bf 1}_2,{\bf 16}, {\bf 1} , {\bf 2}) \oplus  \frac{1}{2} ({\bf 2}_1, {\bf 2}_2, {\bf 1}, {\bf 2}, {\bf 1}) \oplus  \frac{1}{2} ({\bf 1}_1, {\bf 2}_2, {\bf 1}, {\bf 2}, {\bf 2}).
\end{equation}
More generally, a theory of the form
\begin{equation}
      \underset{[N_f=8]}{\overset{\ksp(1)}{1}} \,\, {\overset{\ksu(2)_1}{2}} \,\, {\overset{\ksu(2)_2}{2}} \,\, ... \,\, \underset{[N_f=2]}{\overset{\ksu(2)_k}{2}}
\end{equation}
has an $SO(16) \times SU(2)^2$ global symmetry. This theory shows up in the 6D SCFT-group theory correspondence via the homomorphism $\mathbb{Z}_2 \rightarrow E_8$ with Dynkin label $2'$, which has a commutant of $SO(16)$.

Finally, we can consider a similar theory of the form
\begin{equation}
     {\overset{\mathfrak{sp}(N)}{1}}\,\, \underset{[N_f=8]}{\overset{\mathfrak{su}(2N+8)_1}{2}}\,\, \overset{\mathfrak{su}(2N+8)_2}{2} \,\,...\,\, {\overset{\mathfrak{su}(2N+8)_k}{2}} \,\, [SU(2N+8)].
\end{equation}
Theories of this form are discussed at length in section 5.4 of \cite{Mekareeya:2017jgc}. There is a $U(1)_R$ on the right coming from the $2N+8$ fundamentals of $\mf{su}(2N+8)_k$, which transform under $U(2N+8) \sim SU(2N+8) \times U(1)_R$. There are $k-1$ baryonic $U(1)$'s associated with the bifundamentals of $\mf{su}(2N+8)_i$ and $\mf{su}(2N+8)_{i+1}$, and there is another $U(1)$ associated with the $8$ fundamentals of $\mathfrak{su}(2N+8)_1$, which transform under $U(8) \sim SU(8) \times U(1)$. Finally, there is a baryonic $U(1)$ associated with the bifundamental of $\mf{sp}(N)$ and $\mf{su}(N+8)_1$: the flavor symmetry $SO(4N+16)$ of the $-1$ tensor decomposes into $U(2N+8)$, of which $\mf{su}(2N+8)$ is gauged, leaving behind a $U(1)$ flavor symmetry. Thus, in total, there are $k+2$ candidate $U(1)$'s, $k$ linear combinations of which are eliminated by ABJ anomalies of $\ksu(2N+8)_i$, $i=1,...,k$. This leaves two $U(1)$'s, for a final global symmetry of $SU(8) \times SU(2N+8) \times U(1)^2$.

A separate analysis is needed for the case $N=0$. Now, the $-1$ tensor is unpaired, and there is no bifundamental of $\mf{sp}(N)$ and $\mf{su}(2N+8)_1$, hence no baryonic $U(1)$ associated with this bifundamental. However, as emphasized in \cite{Mekareeya:2017jgc}, there are two choices for the embedding of the gauge symmetry $\mf{su}(8)_1$ into the $E_8$ symmetry of the $-1$ tensor, which have commutants $SU(8) \times SU(2)$ and $SU(8) \times U(1)$, respectively (in field theory terms, these are distinguished by a choice of discrete $\theta$ angle). As a result, the flavor symmetry of these theories is given by $SU(8)^2 \times SU(2) \times U(1)$ and $SU(8)^2 \times U(1)^2$, respectively.

\subsection{Examples with no $\mathfrak{su}(N)$ Matter with $N \geq 3$}

Let us now consider examples of $U(1)$ global symmetries that do not involve hypermultiplets of $\mf{su}(N)$ gauge symmetries with $N \geq 3$. These examples are simpler in that they do not involve any ABJ anomalies. We first consider theories with classical $SO$ and $Sp$ gauge algebras, and then turn to examples with more general exceptional gauge algebras.

As a first example without $\mathfrak{su}(N)$ matter, consider an alternating $\kso$-$\ksp$ quiver of the form:
\begin{equation}
... \,\, \ov{\mf{so}(n_L)}4 \,\, \underset{[N_f=2n +8-(n_L+n_R)/2]}{\ov{\mf{sp}(n)}1} \,\, \ov{\mf{so}(n_R)}4 \,\, ....
\end{equation}
The $\mf{sp}(n)$ gauge algebra here requires $2n+8-(n_L+n_R)/2$ fundamental hypermultiplets to cancel gauge anomalies. The fundamental representation of $\mf{sp}(n)$ is pseudo-real, so we should really think of $4n+16-n_L-n_R$ half-hypermultiplets transforming under an $\mf{so}( 4n+16-n_L-n_R)$ global symmetry. This does not generically produce a $U(1)$ global symmetry, but for $4n+16-n_L-n_R=2$, we get an $\mf{so}(2) \simeq \mf{u}(1)$ global symmetry algebra. Thus, a single fundamental hypermultiplet of $\mf{sp}(n)$ is associated with a $U(1)$ global symmetry at the superconformal fixed point.

Next, consider a quiver of the form
\begin{equation}
... \,\,  \underset{[N_f= 6-n]}{\ov{\mf{e}_6}n} \,\, ...
\label{eq:E65}
\end{equation}
Here, $n$ represents a curve of self-intersection $-n$, with $1 \leq n \leq 6$. This theory has $6-n$ fundamental hypermultiplets charged under the $\mf{e}_6$ gauge algebra, which transform under a $U(n) \sim SU(n) \times U(1)$ global symmetry.

Similarly, for
\begin{equation}
... \,\,  \underset{[N_f= (8-n)/2]}{\ov{\mf{e}_7}n} \,\, ...
\label{eq:E76}
\end{equation}
the theory has $8-n$ half-hypermultiplets transforming in the fundamental representation of the $\mf{e}_7$ gauge algebra, which also transform under a $SO(8-n)$ global symmetry. For $n=6$, we get an $SO(2) \sim U(1)$ global symmetry.

Another class of global $U(1)$'s arise when the $E_8$ global symmetry of a curve of self-intersection $-1$ is partially gauged, and a $U(1)$ factor is left over. Given a 6D SCFT quiver of the form
\begin{equation}
...  \,\, \ov{\mf{g}_L}m \,\, 1 \,\, \ov{\mf{g}_R}n \,\, ...
\end{equation}
we must have $\mf{g}_L \times \mf{g}_R \subset \mf{e}_8$, which is interpreted as weakly gauging part of the $E_8$ global symmetry.
The commutant subgroup of $E_8$ left ungauged is a global symmetry of the theory at the conformal fixed point.

One example of a $U(1)$ global symmetry of this type occurs when $\mf{g}_L = \mf{e}_6$, $\mf{g}_R = \mf{su}(2)$:
\begin{equation}
... \,\,  \ov{\mf{e}_6}6 \,\, \underset{[U(1)]}1 \,\, \ov{\mf{su}(2)}2 \,\, ...
\label{eq:E6comm}
\end{equation}
This follows from $\mf{e}_6 \times \mf{su}(2) \times \mf{u}(1) \subset \mf{e}_6 \times \mf{su}(3) \subset \mf{e}_8$. One could also have a quiver of the form
\begin{equation}
\ov{\mf{su}(4)}2 \,\, \underset{[U(1)]}1 \,\, \ov{\mf{so}(8)}4 \,\, ...
\end{equation}
Here, $\mf{so}(8) \times \mf{su}(4) \times \mf{u}(1) \simeq \mf{so}(8) \times \mf{so}(6) \times \mf{so}(2) \subset \mf{e}_8$, so we indeed get a $U(1)$ global symmetry associated with this gauging.

Finally, we note that one can construct theories with multiple $U(1)$'s by combining the above examples into a single theory. For instance, one could consider a theory of the form
\begin{equation}
     \underset{[U(1)]}{\ov{\mf{e}_6}5} \,\, \underset{[U(1)]}1 \,\, \ov{\mf{su}(2)}2 \,\, \ov{\mf{so}(7)}3 \,\, \ov{\mf{su}(2)}2 \,\, 1 \,\, \underset{[U(1)]}{\ov{\mf{e}_7}6}
\end{equation}
This theory has three $U(1)$'s: one associated with the $\mf{e}_6$ gauge algebra on the $-5$ curve (as in line (\ref{eq:E65})), one associated with the $\mf{e}_7$ gauge algebra on the $-5$ curve (as in line (\ref{eq:E76})), and one associated with the $U(1)$ left ungauged on the leftmost $-1$ curve (as in line (\ref{eq:E6comm})).

\subsection{Frozen SCFT Examples}

Let us now turn to some examples which arise from the frozen phase of F-theory \cite{McOrist:2010jw, Tachikawa:2015wka, Bhardwaj:2015oru, Bhardwaj:2018jgp}. Recall that these SCFTs still arise from an elliptically fibered Calabi-Yau threefold, but in which the physical interpretation of singular elliptic fibers is different from that which is assigned in the geometric phase of F-theory.

In frozen SCFTs, we may see bifundamentals of $\kso$ and $\ksu$ gauge algebras,
which introduce candidate $U(1)$ global symmetries. We illustrate below with a pair of examples.

To begin, consider the theory with $\kso(20) \oplus \ksu(10)$ gauge algebra and Dirac pairing
\begin{equation}
    \Omega = \left(\begin{array}{cc}
    -4 & 2 \\
    2 & -2
    \end{array}\right).
\end{equation}
The matter content of this theory is
\begin{equation}
   2 ({\bf 20}, {\bf 1}) \oplus ({\bf 20}, \overline{\bf 10}).
\end{equation}
The two fundamentals of $\kso(20)$ transform under an $Sp(2)$ global symmetry. There is also a baryonic $U(1)$ global symmetry associated with the bifundamental of $\mf{so}(20)$ and $\ksu(10)$. This follows from the fact that the full $Sp(12)$ symmetry associated with 12 fundamentals of $\kso(20)$ decomposes as $Sp(10) \times Sp(2)$, the former which further branches to $U(10) \sim SU(10) \times U(1)$. This $SU(10)$ is gauged, leaving behind a $U(1)$ baryonic symmetry. However, there is also an ABJ anomaly associated with the $\ksu(10)$ gauge symmetry:
\begin{equation}
I_{\mathrm{ABJ}} \supset  - \frac{10}{3} F_{U(1)} \Tr(F_{\mathfrak{su}(10)}^3)
\end{equation}
As a result, the $U(1)$ is removed from the spectrum, leaving just an $Sp(2)$ global symmetry.

Next, we consider the theory with $\kso(20) \oplus \ksu(12)$ gauge algebra and Dirac pairing
\begin{equation}
    \Omega = \left(\begin{array}{cc}
    -4 & 2 \\
    2 & -2
    \end{array}\right).
\end{equation}
The matter content of this theory is
\begin{equation}
   4 ({\bf 1}, {\bf 12}) \oplus ({\bf 20}, \overline{\bf 12}).
\end{equation}
Here, the four fundamentals of $\ksu(12)$ transform under a $U(4)_R \sim SU(4)_R \times U(1)_R$ global symmetry. There is also a baryonic $U(1)$ global symmetry associated with the bifundamental of $\mf{so}(20)$ and $\ksu(12)$, as above. However, there is also an ABJ anomaly associated with the $\ksu(12)$ gauge symmetry:
\begin{equation}
I_{\mathrm{ABJ}}    \frac{1}{6} \Big( -20 F_{U(1)_B} \Tr(F_{\mathfrak{su}(12)}^3) + 4 F_{U(1)_R} \Tr(F_{\mathfrak{su}(12)}^3) \Big).
\end{equation}
As a result, the linear combination $-20 F_{U(1)_B} + 4 F_{U(1)_R}$ is removed from the spectrum, leaving an $SU(4) \times U(1)$ global symmetry.

\section{$U(1)$'s and RG Flows \label{sec:RG}}

In previous sections, we presented a general prescription for how to read off the $U(1)$ global symmetries of a 6D SCFT
from data associated with its tensor branch description. We have, in particular, explained how to read off these symmetries for the A-type progenitor theories and implicitly argued that no other progenitor theories possess such symmetries. Additionally, we have directly extracted
the $U(1)$ global symmetries and their associated anomalies in a number of examples.

As we have already mentioned, one reason to focus on the symmetries of the progenitor theories is that all known 6D SCFTs
arise from a combination of fission and fusion moves from this single uniform starting point. This in turn motivates a more general question concerning the fate of various $U(1)$ symmetries under deformations from one UV fixed point to another fixed point in the IR.

As shown in references \cite{Louis:2015mka,Cordova:2016xhm, Cordova:2016emh}, supersymmetric deformations of 6D SCFTs
arise from vevs of operators, either coming from tensor branch deformations or Higgs branch deformations.
The structure of such RG flows has been analyzed, for example, in references \cite{Heckman:2015ola,
Cordova:2015fha, Heckman:2015axa, Heckman:2016ssk, Heckman:2018pqx}.

Tensor branch deformations are somewhat simpler in that they preserve the global symmetries of the UV theory. However, such a flow often introduces new, emergent global symmetries to the IR SCFT, which result from ungauging a symmetry of the (tensor branch description of the) UV theory. Additional global symmetry enhancements may also occur, as discussed above for progenitor theories with a small number of M5-branes.

For Higgs branch flows, we expect that giving vevs to hypermultiplets or their strongly coupled analogs (namely
6D conformal matter) will generically break both global and gauge symmetries of the quiver-like gauge theory obtained from the partial tensor branch of a 6D SCFT. A non-trivial consequence of this observation is that a global symmetry of an IR fixed point (obtained after a Higgs branch deformation) could in fact originate from a linear combination of gauge and global symmetries in the tensor branch description of the UV parent theory.

To illustrate, consider a 6D SCFT with partial tensor branch containing a product gauge group $G_1 \times G_2$ with terms in the Lagrangian description on the tensor branch:
\begin{equation}
\mathcal{L}_{\mathrm{partial \, tensor}} \supset \phi_{(1)} \mathrm{Tr} F_{(1)} \wedge \ast F_{(1)} + \phi_{(2)} \mathrm{Tr} F_{(2)} \wedge \ast F_{(2)},
\end{equation}
with $\phi_i$ the vevs for the tensor multiplet scalars. Suppose we now consider a breaking patter for our matter (be it hypermultiplets or conformal matter) which only retains a diagonal symmetry $G_{\mathrm{diag}} \subset G_{1} \times G_{2}$. The resulting kinetic term for the gauge fields is:
\begin{equation}
\mathcal{L}_{\mathrm{partial \, tensor}} \supset (\phi_{(1)} + \phi_{(2)}) \mathrm{Tr} F_\text{diag} \wedge \ast F_\text{diag}.
\end{equation}
Of course, if we consider the limit where one of the $\phi_{(i)}$'s becomes infinite, we effectively convert one of the gauge symmetries into a flavor symmetry. This in turn means that the diagonal symmetry will also become a global symmetry. From the perspective of a 6D SCFT, the origin of the diagonal symmetry may therefore seem ``mysterious'' seeing as it originates from degrees of freedom which are most apparent on the tensor branch. That being said, the above considerations make it quite clear where these symmetries originate from. In particular, they also show that a na\"ive application of 't Hooft anomaly matching between the UV and the IR is simply inappropriate because the original global symmetries of a 6D SCFT may in fact be broken by a given Higgs branch deformation.

To illustrate this phenomenon, we examine the Higgs branch flow between the following theories, working on the tensor branch of each theory:
\begin{equation}
\underset{[N_f=14,\, N_{\Lambda^2} = 1]}{\ov{\mf{su}(6)}1} ~~~ \rightarrow ~~~\underset{[N_f=13,\, N_{\Lambda^2} = 1]}{\ov{\mf{su}(5)}1} .
\end{equation}
The UV theory, consisting of $\mf{su}(6)$ on a $-1$ curve, has the following matter content under $\mf{su}(6) \times SU(14) \times U(1)$:
\begin{equation}
({\bf 6}, \overline{\bf 14})_{1} + ({\bf 15}, {\bf 1})_{-7}.
\label{eq:UVmat}
\end{equation}
The $-7$ charge is necessary for ABJ anomaly cancelation, using the fact (from Table \ref{tab:GTCC}) that $c_\rho = 2$ for the $\Lambda^2$ representation of $\mf{su}(6)$. The anomaly polynomial of the UV theory is
\begin{align}
I_{\text{UV}} &= \frac{199 c_2(R)^2}{12}-\frac{3}{2} c_2(R) \Tr F_{SU(14)}^2-630 c_2(R) F^2-\frac{53 c_2(R) p_1(T)}{24}+27 F^2 \Tr F_{SU(14)}^2 \nonumber \\
&-F \Tr F_{SU(14)}^3+\frac{1}{8} p_1(T) \Tr F_{SU(14)}^2+\frac{1}{4} \Tr F_{SU(14)}^4+\frac{1}{32} (\Tr F_{SU(14)}^2)^2+\frac{56133 F^4}{8} \nonumber \\
&+\frac{693 F^2 p_1(T)}{16}+\frac{217 p_1(T)^2}{1920}-\frac{31 p_2(T)}{480}.
\label{eq:UVanompoly}
\end{align}

The $\mf{su}(6) \times SU(14) \times U(1)$ symmetry is broken to $\mf{su}(5) \times SU(12) \times U(1)^2$ by giving vevs to the scalars in two different hypermultiplets in the fundamental of $\mathfrak{su}(6)$. Indeed, this follows because the triplet of D-term constraints for the $\mathfrak{su}(6)$ gauge symmetry cannot be satisfied by giving a vev to a single hypermultiplet (see e.g. \cite{Danielsson:1996uv, Morrison:2012np}). Denoting the associated hypermultiplets in the fundamental of $\mathfrak{su}(6) \times SU(14)$ as $Q_{i} \oplus \widetilde{Q}^{\dag}_{i}$ for $i=1,...,14$, we give a vev to $Q_{1}$ and $\widetilde{Q}^{\dag}_{2}$. This breaks the non-abelian flavor symmetry
to $SU(12)$.

In addition, the anti-symmetric tensor {\bf 15} decomposes under $\mf{su}(6) \rightarrow \mf{su}(5) \times \mf{u}(1)$ as follows:
\begin{equation}
{\bf 15} \rightarrow {\bf 10}_2 + {\bf 5}_{-4}.
\label{eq:antisymdecomp}
\end{equation}
So, in total, the matter in (\ref{eq:UVmat}) decomposes as
\begin{align}
({\bf 6}, \overline{\bf 14})_{1}&\rightarrow   ({\bf 5}, \overline{\bf 12})_{(1, 1, 1)}  +  ({\bf 1}, \overline{\bf 12})_{(-5, 1, 1)} + 2 ({\bf 5}, {\bf 1})_{(1, -6, 1)}  + 2 ({\bf 1}, {\bf 1})_{(-5, -6, 1)} \nonumber \\
 ({\bf 15}, {\bf 1})_{-7} &\rightarrow    ({\bf 10}, {\bf 1})_{(2, 0, -7)}  + ({\bf 5}, {\bf 1})_{(-4, 0, -7)}, \label{eq:totaldecomp}
\end{align}
where the three $U(1)$ charges correspond respectively to the charges under the $\mf{su}(6)$ Cartan $H_{\mf{su}(6)} = \text{diag}(-5, 1, 1, 1, 1, 1)$, the $SU(14)$ Cartan $H_{SU(14)} = \text{diag}(-6, -6, 1, 1, ..., 1, 1)$, and the UV $U(1)$ generator $H_{\text{UV}}$, respectively. One of the three $U(1)$'s suffers from an ABJ anomaly, which leaves a $\mf{su}(5) \times SU(12) \times U(1)^2$ symmetry manifest. In the IR, however, the fundamental from the anti-symmetric tensor combines with the 12 fundamentals from the breaking of the $\mf{su}(6) \times SU(14)$ bifundamental, and the flavor symmetry enhances from $SU(12) \times U(1)^2$ to $SU(13) \times U(1)$.

From ABJ anomaly cancelation in the IR, we know that the matter of the IR theory transforms under $\mf{su}(5) \times SU(13) \times U(1)$ as
\begin{equation}
({\bf 5}, \overline{\bf 13})_1 + ({\bf 10}, {\bf 1})_{-13}.
\end{equation}
The anomaly polynomial of this theory is
\begin{align}
I_{\text{IR}} & = \frac{277 c_2(R)^2}{24}-1300 c_2(R) F^2-\frac{5}{4} c_2(R) \Tr F_{SU(13)}^2-\frac{83 c_2(R) p_1(T)}{48}+\frac{365625 F^4}{8} \nonumber \\
&+\frac{1625 F^2 p_1(T)}{16}+\frac{525}{8} F^2 \Tr F_{SU(13)}^2+\frac{11}{96} p_1(T) \Tr F_{SU(13)}^2+\frac{5}{24} \Tr F_{SU(13)}^4\nonumber \\
&-\frac{5}{6} F \Tr F_{SU(13)}^3+\frac{1}{32} (\Tr F_{SU(13)}^2)^2+\frac{7 p_1(T)^2}{72}-\frac{p_2(T)}{18}.
\label{eq:IRanompoly}
\end{align}
The mismatch in gravitational anomalies is given by
\begin{equation}
\Delta I := I_{\text{UV}} - I_{\text{IR}} = \frac{91 p_1(T)^2}{5760}-\frac{13 p_2(T)}{1440} +...,
\end{equation}
which implies that there are 13 additional, free hypermultiplets arising in the IR of the RG flow. Part of the goal of our analysis is to determine the $U(1)$ charges of these free hypermultiplets. Before we can do this, however, we must first address our main question of interest: writing the IR $U(1)$ as a linear combination of $U(1)$'s in the UV.

To determine the coefficients of the appropriate linear combination $a H_{\mf{su}(6)} + b H_{SU(14)} + c H_{\text{UV}}$, we impose the following constraints:
\begin{enumerate}[1)]
\item The correct linear combination must not be broken by Higgsing. From the specified vevs for $Q_1$ and $\tilde Q_2$ above, each of which carry charge $1$ under $H_{\text{UV}}$, this gives
\begin{equation}
-5 a  -6 b + c = 0.
\label{eq:const1}
\end{equation}
\item The correct linear combination must assign charge 1 to the bifundamental of $\mf{su}(5) \times SU(13)$. Concentrating on a fundamental of $\mf{su}(5)$ that originates in the fundamental of $\mf{su}(6) \times SU(13)$, this gives
\begin{equation}
 a +  b + c = 1.
\end{equation}
\item  Additionally, the correct linear combination must assign charge 1 to the fundamental of $\mf{su}(5)$ that originates in the anti-symmetric tensor of $\mf{su}(6)$. This fundamental is not charged under $SU(14)$, but from line (\ref{eq:totaldecomp}) it carries charge $-4$ under $H_{\mf{su}(6)}$, and charge $-7$ under $H_{\text{UV}}$. Thus,
 \begin{equation}
 -4 a - 7 c = 1.
\end{equation}
\item Finally, the correct linear combination must assign charge $-13$ to the anti-symmetric tensor of $\mf{su}(5)$. This comes from the anti-symmetric tensor of $\mf{su}(6)$, and from (\ref{eq:totaldecomp}) it carries charge $2$ under $H_{\mf{su}(6)}$, and charge $-7$ under $H_{\text{UV}}$. Thus
 \begin{equation}
 2 a - 7 c = -13.
 \label{eq:const4}
\end{equation}
\end{enumerate}
Equations (\ref{eq:const1})-(\ref{eq:const4}) represent four equations for three unknowns, but is not over-constrained.
Indeed, there is a non-trivial solution:
\begin{equation}
a = - \frac{7}{3}\,,~~~ b = \frac{15}{7} \,,~~~ c = \frac{25}{21}.
\label{eq:lincomb}
\end{equation}
The appearance of rational numbers (rather than integers) is inconsequential. What is important is that all physical states have properly quantized charges. Indeed, the denominators have been chosen so that the $U(1)$ charge of the bifundamental $(\mathbf{5}, \overline{\mathbf{12}})$ is normalized to $1$.

The fact that $a \neq 0$ here is of great importance: it shows that the IR $U(1)$ is a linear combination of not only global symmetries, but also \emph{gauge} symmetries in the UV. This is a perfectly sensible statement on the tensor branch, where the $\mf{su}(6)$ and $\mf{su}(5)$ gauge theories are weakly coupled, but it is not clear how one is supposed to interpret this $U(1)$ at the SCFT fixed point. Furthermore, since the UV $U(1)$ symmetry does not match the IR $U(1)$ symmetry, there is no $U(1)$ symmetry preserved along the flow, so $U(1)$ anomalies cannot be matched between the two theories in any straightforward manner.

Let us now return to the 13 free hypermultiplets in the IR of the RG flow. Goldstone bosons associated with the two $(\mathbf{5}, \mathbf{1})_{(1,-6,1)}$ hypermultiplets as well as one of the $(\bf{1}, \bf{1})_{-5, -6, 1}$ hypermultiplets are eaten due to the Higgsing of $\mf{su}(6) \rightarrow \mf{su}(5)$. From (\ref{eq:totaldecomp}), we see that the 13 remaining free hypermultiplets correspond to
\begin{equation}
 ({\bf 1}, \overline{\bf 12})_{(-5, 1, 1)}  + ({\bf 1}, {\bf 1})_{(-5, -6, 1)} \rightarrow  ({\bf 1}, \overline{\bf 12})_{15}  + ({\bf 1}, {\bf 1})_{0} ,
\end{equation}
where on the right-hand side, we have plugged in the UV charges to the linear combination of (\ref{eq:lincomb}). We see that the IR matter consists of a fundamental of $SU(12)$ of charge $15$ and an uncharged singlet. Comparing (\ref{eq:UVanompoly}) and (\ref{eq:IRanompoly}), one sees that the fundamental of $SU(12)$ is actually necessary to match the $\Tr F_{SU(12)}^4$ anomalies between the UV and the IR, which must agree because the $SU(12)$ symmetry is preserved along the flow. Note, however, that these 13 free hypermultiplets do not carry the same $U(1)$ charge, so unlike the 13 hypermultiplets of $\mf{su}(5)$, they do not assemble into an enlarged $SU(13)$ flavor symmetry. This is not a problem for anomaly matching because, as we have seen, the full $SU(13)$ flavor symmetry is not preserved along the flow, but appears only at the UV and IR fixed points.

\section{Geometry of Global $U(1)$'s in F-theory}
\label{sec:FTHEORY}

In this section we turn to the geometric origin of $U(1)$ global symmetries in F-theory realizations of 6D SCFTs. For previous work on global $U(1)$'s in F-theory in the context of 6D SCFTs see \cite{Lee:2018ihr}.\footnote{In the examples of \cite{Lee:2018ihr}, the F-theory base is compact, and gravity is decoupled in a second step by taking a suitable decoupling limit. On the contrary, in our examples below we work with non-compact, local bases from the start.}
Rather than analyzing the corresponding Weierstrass model for each of the examples presented in the previous sections,
we shall instead analyze the origin of abelian symmetries in the progenitor theories.
Indeed, since the progenitor theories are a common starting point
for all 6D SCFTs, it is at some level enough to understand the geometric origin
of $U(1)$ symmetries in these theories and then track their behavior under fission and fusion moves.

Recall that the two general classes of progenitor theories are given by a quiver-like partial tensor branch:
\begin{align}
& [G_{ADE}]~\underset{k}{\underbrace{\overset{\mathfrak{g}_{ADE}}{2}%
~\overset{\mathfrak{g}_{ADE}}{2}~...~\overset{\mathfrak{g}_{ADE}}{2}}%
}~[G_{ADE}],\label{m5o}\\
& [E_{8}]~\underset{k}{\underbrace{\overset{\mathfrak{g}_{ADE}}{1}%
~\overset{\mathfrak{g}_{ADE}}{2}~...~\overset{\mathfrak{g}_{ADE}}{2}}%
}~[G_{ADE}].\label{heto}%
\end{align}
The theories of (\ref{heto}) arise from M5-branes probing a heterotic nine-brane wrapping an ADE singularity \cite{Aspinwall:1997ye, DelZotto:2014hpa}. The theories of (\ref{m5o}) arise from M5-branes probing an ADE singularity. Note that the theories of (\ref{m5o}) can be viewed as the result of a tensor branch deformation of the theories of (\ref{heto}).

In the F-theory realization of these 6D SCFTs, we specify a non-compact base $B$ with contractible configurations of
curves, and then supplement this by a suitable elliptic fibration so that the total space is a non-compact
Calabi-Yau threefold. In the theories of (\ref{m5o}), we work with a singular
base $\mathbb{C}^2 / \mathbb{Z}_{k}$ and a pair of seven-branes with
$G_{ADE}$ flavor seven-branes which intersect transversely over the orbifold fixed point. In the theories of (\ref{heto}),
we instead work with a base $\mathbb{C}^2$ and a transverse collision of a non-compact $E_8$ seven-brane with a $G_{ADE}$ seven-brane.\footnote{Heterotic small instantons on $\mathbb{C}^2$ have a tensor branch $1,2,...,2$. It is well-known that the F-theory realization of the corresponding 6D SCFT does not generically exhibit an $E_8$ flavor symmetry or the additional $SU(2)_L$ flavor symmetry which are both manifest in the heterotic description. These symmetries are typically viewed as ``emergent'' at the fixed point. The $E_8$ flavor symmetry can be made manifest by a suitable tuning of complex structure moduli.} This is accompanied by a $k$-fold intersection of an $I_1$ locus of the discriminant. For the explicit presentations of the associated Weierstrass models in these cases, see for example \cite{Aspinwall:1997ye, DelZotto:2014hpa, Heckman:2014qba}, as well as the review article of reference \cite{Heckman:2018jxk}. In each of these cases with a non-abelian flavor symmetry, we can see a potential origin for a $U(1)$ global symmetry in a fission or fusion product, simply by performing a suitable smoothing deformation of the singular Calabi-Yau geometry. In this sense, these cases provide a uniform perspective on many candidate $U(1)$ symmetries for 6D SCFTs.

There can also be additional $U(1)$ symmetries which originate from the Mordell-Weil group of the elliptic fibration. Indeed, in models with a \textit{compact} base, namely those in which the 6D theory is coupled to gravity, there is a well-defined procedure for extracting \textit{candidate} $U(1)$ gauge symmetries. One way to obtain global $U(1)$ symmetries is to now take a limit in which the base is non-compact. This decouples gravity and also makes the $U(1)$ into a global symmetry \cite{Lee:2018ihr}. As we explain later, however, not all global $U(1)$ symmetries need to originate from such a procedure, though when available, it is clearly a useful way to build possible examples.

To construct such examples, we begin with a generic point of the base and ask whether the corresponding elliptic curve has an additional generator in its Mordell-Weil group (the group law for a given elliptic curve). This defines an additional marked point on the elliptic curve, and thus an additional candidate section for the elliptic fibration. As we move to different points of the base, the fate of this marked point may either persist, or disappear. If it persists, we have a candidate for an extra section of the elliptic fibration. The appearance of at least two candidate sections, say $\sigma$ and $\sigma^{\prime}$ means that we have two distinct divisors $\sigma_{\ast}(B)$ and $\sigma^{\prime}_{\ast}(B)$ in the Calabi-Yau threefold. In the M-theory reduction, this additional divisor provides a candidate $U(1)$ boson, and consequently, there is a close interplay between the Mordell-Weil group of the family of elliptic curves and $U(1)$ gauge symmetries \cite{Morrison:1996pp,Aspinwall:2000kf,Aspinwall:1998xj}.\footnote{For an incomplete list of references on $U(1)$ gauge symmetries in F-theory see e.g. \cite{Morrison:2012ei,Grimm:2010ez,Krause:2011xj,Braun:2011zm,Grimm:2011tb,Braun:2014nva,Borchmann:2013jwa,Cvetic:2013nia,
Cvetic:2013uta,Borchmann:2013hta,Cvetic:2013qsa,Klevers:2016jsz,Klevers:2017aku,Klevers:2014bqa,Cvetic:2015ioa,Kuntzler:2014ila,
Lawrie:2015hia,Dolan:2011iu,Marsano:2011nn,Raghuram:2017qut,Collinucci:2018aho,Collinucci:2019fnh}. See also the reviews \cite{Cvetic:2018bni,Weigand:2018rez} for a more recent account on this vast subject.}

Returning to the case of \textit{global} $U(1)$'s, it is natural to ask whether some non-compact remnant of the Mordell-Weil group
will persist in these cases as well. In global F-theory models, abelian gauge symmetries act on charged matter fields, which are localized at codimension-two singularities of the Calabi-Yau threefold. However, even in compact models some of these abelian factors may acquire a mass and become non-dynamical via a ``geometric St\"uckelberg mechanism.'' Those that stay massless (dynamical) are associated to singularities which admit a small K\"ahler resolution \cite{Grimm:2011tb,Braun:2014nva}. This is clearly the case if the Weierstrass model can be put into conifold form (i.e. $ab=cd$, with $a,b,c,d$ polynomials in the base and fiber coordinates) \cite{Braun:2014nva}, or if it admits a ``matrix factorization'' of a certain form \cite{Collinucci:2019fnh} (see also \cite{Collinucci:2018aho} for the relevant terminology). We will see that this construction carries over in certain non-compact F-theory models as well.

We begin by first adopting a convenient parametrization of elliptically fibered Calabi-Yau threefolds which is particularly amenable to the analysis of possible $U(1)$ symmetries. After this, we analyze the (infinite class of) examples of (\ref{m5o}) which in M-theory terms are given by M5-brane probes of an A-type singularity. We then turn to the theories of (\ref{heto}) which in heterotic terms are given by small $E_8$ instanton probes of an A-type singularity. Finally, we will focus on cases where the non-abelian global symmetry contains an $SU(N)$ factor for $N \geq 3$ and a sufficient number of collapsing curves in the base geometry. We will comment on these special cases, as appropriate.

\subsection{Geometric Preliminaries} 
\label{sec:setup}

In this subsection, we state some of the geometric preliminaries we will make use of to analyze the abelian symmetries of the A-type
progenitor theories. For our purposes, it will be helpful to describe the base of the F-theory model in the limit where all the curves participating in the 6D SCFT have collapsed to zero size. In this case, we obtain the tensor branch by performing blowups of the base.

For the heterotic small instanton probes of an A-type singularity, the base is given by $\mathbb{C}^{2}$.
For the M5-brane probes of an A-type singularity, the base is given by $\mathbb{C}^{2} / \mathbb{Z}_{k}$ where the group action on the
local base coordinates is $(u,v) \mapsto (\omega u , \omega^{-1} v)$ with $\omega = \exp(2 \pi i / k)$.

The local presentation of the elliptic fibration is customarily written in general form as:
\begin{equation}\label{eq:yxfg}
y^2 = x^{3} + fx z^{4} + g z^{6}
\end{equation}
where $[x:y:z]$ are weighted homogeneous coordinates of $\mathbb{P}^{2}_{[2,3,1]}$. For the prescribed group actions
(trivial or by $\mathbb{Z}_k$), $f$ and $g$ are given by polynomials in the local coordinates $(u,v)$ as:
\begin{equation}
f = \sum_{i,j} f_{i,j} u^{i} v^{j}, \qquad g = \sum_{i,j} g_{i,j} u^{i} v^{j}.
\end{equation}
See references \cite{Morrison:2016nrt,DelZotto:2014fia,Heckman:2018jxk} for a discussion of Weierstrass models over a base $B$ with more general orbifold singularities. In the case of the $\mathbb{Z}_{k}$ group action, we need to ensure that each monomial we
retain is invariant under the group action. This restricts us to the invariant terms $(uv)$, $u^{k}$, $v^{k}$ and products thereof.
As standard, the components of the discriminant locus $\Delta = 4 f^3 + 27 g^2$ correspond to degeneration loci of the fibration, i.e. matter curves in the base.

Now, for the purposes of analyzing possible global symmetries of the model,
it will be convenient to parameterize our elliptic fibration as:
\begin{equation}\label{eq:W}
y^2 = s^3 +b_2 s^2 z^2 +2b_4 s z^4 + b_6 z^6 ,
\end{equation}
where the usual $x$ fiber coordinate and the $f,g$ coefficients in \eqref{eq:yxfg} are recovered via
\begin{equation}\label{eq:fg}
f=-\frac{1}{3}(b_2^2-6b_4), \quad g=\frac{1}{27}(2 b_2^3-18 b_2 b_4+27b_6), \quad x =s+\frac{1}{3}b_2 z^2.
\end{equation}

We are interested in models which have an additional generator of the Mordell-Weil group. In what follows, it will suffice to consider
the Morrison-Park (MP) construction \cite{Morrison:2012ei} of additional sections. The MP construction requires specific coefficients $b_2,b_4,b_6$ in order to engineer an extra section of the elliptic fibration.

In a \textit{compact} model this section would yield an extra generator of the global Picard group of divisors of the threefold which would then yield an extra $U(1)$ gauge boson in the effective theory by the usual logic of reducing the supergravity three-form along the Poincar\'e dual of the new divisor \cite{Morrison:1996pp}.

In a \textit{non-compact} model, additional care is needed because Poincar\'e duality will not produce a dynamical vector boson in this case. Note, however, that in any such model which \textit{could} be recoupled to gravity, obtaining a local fibration in MP form would be a necessary condition for this to extend to a global model.\footnote{See sections 1.1 and 1.2 of reference \cite{Collinucci:2019fnh} for a discussion on the local vs. global Picard group of divisors.}

Our strategy should thus be clear: We will first seek out necessary conditions to have an additional section in MP form. When this condition is met, we can conceive of the existence of a compact F-theory model defined on a compact base which has an additional gauged $U(1)$.\footnote{Similarly, reference \cite{Lee:2018ihr} provides necessary conditions (on local bases) which come from anomaly constraints as well as constraints ensuring the existence of an extra section.} On the other hand, since our progenitor theories fit into infinite families (as specified by the number of collapsing curves) the existence of even one globally consistent model is enough to establish the pattern for all generic members of the family. This will be the sense in which we show that the progenitor theories of \eqref{m5o} do indeed have a global $U(1)$ symmetry, namely by establishing the existence of an additional section in MP form.

To reach a fibration in MP form, the $b_2,b_4,b_6$ coefficients are restricted as follows:
\begin{equation}\label{eq:bMP}
b_2^\text{MP}=c_2, \quad b_4^\text{MP}=\frac{1}{2}(c_1 c_3-b^2c_0), \quad b_6^\text{MP}=c_0 c_3^2-b^2c_0 c_2+\frac{1}{4}b^2 c_1^2,
\end{equation}
with $b,c_0,\ldots,c_3$ base polynomials, i.e. sections of line bundles (over the base) of prescribed degree. Over each point of the base, the extra section \emph{generically} cuts the fiber at a point $Q \in \mathbb{P}^2_{[2,3,1]}$ with fiber coordinates
\begin{equation}\label{eq:QMP}
Q: [s:y:z] = \left[ c_3^2 -b^2 c_2 : -c_3^3+b^2 c_2 c_3 -\tfrac{1}{2} b^4c_1 : b\right].
\end{equation}
The zero section, which always exists, sits at $Z:[s:y:z]=[1:1:0]$. The above restrictions enforce singularities in codimension-two (in the threefold) where the matter charged under the $U(1)$ symmetry is localized; in the generic MP model there are two such loci, which intersect each other. (See \cite{Morrison:2012ei,Morrison:2014era,Mayrhofer:2014laa} for details.)

Our plan will be to consider further
tunings in the Weierstrass models of the progenitor theories, and thus obtain the
requisite additional section in MP form.

\subsection{Warmup: Bifundamentals of $SU(M) \times SU(N)$} \label{ssec:warm}

Though our primary interest is in the case of interacting SCFTs, we have also seen in section \ref{sec:YouThaOne}
that, to a large extent, the global symmetries of a bifundamental hypermultiplet provide a helpful
guide to the structure of global symmetries in interacting 6D SCFTs. As a warmup to our general question, we first analyze the
behavior of an additional MP section in this special case. We note that this does not produce an interacting
SCFT since we have no collapsing curves in this case. Even so, the geometry still contains most of the relevant features we will
need in the case of interacting SCFTs.

The local geometry for a hypermultiplet in the bifundamental representation of $SU(M) \times SU(N)$ involves the transverse collision of
two components of the discriminant with respective singularity types $I_M$ and $I_N$ on an F-theory base $B = \mathbb{C}^2$ with local
coordinates $(u,v)$. We let $u=0$ denote the $I_M$ locus and $v = 0$ denote the $I_N$ locus. The corresponding Weierstrass model can be obtained by taking
\begin{equation}\label{eq:ImIn}
b_2=1+u^M v^N, \quad b_4=u^M v^N, \quad b_6=u^M v^N,
\end{equation}
and has discriminant
\begin{equation}\label{eq:ImOut}
\Delta = u^M v^N(4-u^M v^N).
\end{equation}
This indeed yields the correct orders of vanishing of $(f,g,\Delta)$ over $u=0$ and $v=0$. There is an extra $I_1$ locus which does not carry any gauge algebra and does not intersect any of the branes with a non-abelian flavor group (hence it is just a spectator).

The above presentation of \eqref{eq:ImIn} fits into a (non-generic) Morrison-Park model upon putting
\begin{equation}\label{eq:MPbsInIm}
b=v^{N/2}\ , \quad c_0 = 0\ , \quad c_1 = 2 u^{M/2}\ , \quad c_2= 1+u^M v^N\ , \quad c_3 = u^{M/2} v^N.
\end{equation}
Note that other choices are possible, e.g.  $c_0 \propto u^M$, $c_1 \propto u^M v^{N/2}$, $c_3 = 0$ and the rest unchanged. The above Ansatz is clearly valid for $M$ and $N$ even. In fact, away from the locus $u = 0$ ($v=0$), we can always expand around some fixed value $u = u_{\mathrm{fixed}}$ (likewise for $v$), and obtain a power series expansion for the parameter $u$ (likewise for $v$).
Let us assume that $M,N$ are both even, and comment on the other possibilities later.

The extra rational point generically sits at $Q:[s:y:z]=[-v^N:0:v^{N/2}]$ which is the same as $[-1:0:-1]$ acting with the $\cc^*$ action of $\mathbb{P}^2_{[2,3,1]}$;\footnote{That is, we work in a patch where $v \neq 0$, (and where $z \neq 0$), which does not contain the zero section but contains the new rational point Q.} this shows that $Q$ is always distinct from $Z$. As a further check, note that with the above choices the MP fibration can be put into ``conifold form'' (simply because $b_6^\text{MP}$ turns into a perfect square for $c_0=0$),\footnote{This situation is commonly referred to in the F-theory literature as $U(1)$ restriction \cite{Grimm:2010ez}.} corresponding to the tranverse collision of $A_{M-1}$ and $A_{N - 1}$ singularities, as expected from e.g. \cite{DelZotto:2014fia}. Given the criterion of \cite{Braun:2014nva}, in a compact setting we would interpret this fact as saying that the F-theory model has an extra (massless) $U(1)$ gauge symmetry.

Indeed, in the $z\neq 0$ patch of \eqref{eq:W} (which does not contain the zero section $Z$ but contains $Q$), with the coefficients on the righthand side of \eqref{eq:bMP} restricted as in \eqref{eq:MPbsInIm}, we have
\begin{align}\label{eq:conifold-InIm}
y^2-\frac{1}{4}b^2 c_1^2 &= s(s^2+c_2 s +c_1c_3) \nonumber \\
(y-u^{M/2} v^{N/2})(y+u^{M/2} v^{N/2}) &=s(s^2 +(1+u^M v^N)s + 2 u^M v^N),
\end{align}
which can be put into standard form $\tilde{x}\tilde{y}=\tilde{u}^M \tilde{v}^N$ by shifting back $s\mapsto x$ and applying an analytic change of variables. 
The conifold form makes it easy to identify the extra generator of the Mordell-Weil group (i.e. a new rational section) \cite{Braun:2014nva,Collinucci:2019fnh}. Note also that in this new presentation of the singularity, all appearance of ``fractional powers'' such as $M / 2$ has disappeared, as anticipated.

Taking stock of the above example, we see that in the geometry there is a manifest $SU(M) \times SU(N) \times U(1)$ flavor symmetry. This is in accord with the general unfolding of $I_{M+N}$ to a pair of colliding $I_M$ and $I_N$ components of the discriminant. In group theory terms, we also have the maximal subalgebra $\mathfrak{su}(M+N) \supset \mathfrak{su}(M) \times \mathfrak{su}(N) \times \mathfrak{u}(1)$.

This also illustrates another general point that an F-theory model only tends to make manifest a subset of the full flavor symmetries of the field theory. For example, the free hypermultiplets in question clearly transform in the fundamental representation of $U(M+N)$. The ``off-diagonal'' terms of this symmetry are absent. Additionally, the overall center of mass $U(1)$ is not present in our analysis. In interacting 6D SCFTs, we expect this center of mass $U(1)$ to decouple, but that the other $U(1)$ will persist.

Turning to the case where either $M$ or $N$ (or both) is odd, we expect on physical grounds that the above analysis will still hold. Namely, one should be able to find identifications of the form \eqref{eq:MPbsInIm} which do not involve any roots, thus avoiding unwanted branch cuts in the local model over $\cc^2$. In fact, if one thinks of the A-type progenitors of \eqref{m5o} as an infinite family as $M,N$ grow large, the physics should not distinguish between the $M,N$ and $M+1,N+1$ representatives. Starting from the case where $M,N$ are both even, it is reasonable to expect a global $U(1)$ symmetry to be present in the odd case as well, with the would-be branch cuts of $u^{M/2}$, $v^{N/2}$ simply appearing as  an artifact of the chosen presentation. This is in accord with the fact that in the local conifold presentation of the singularity, we saw that these branch cuts eventually disappeared.\footnote{Alternatively, one may stick to the latter presentation and work in the double cover of the base branched over $u=0$ and $v=0$, by setting $u=U^2$ and $v=V^2$ (with a $\zz_2$ involution acting as $U\mapsto -U$ and $V \mapsto -V$). Pulling back from $\cc[u,v]$ to $\cc[U,V]$, one ends up with a representative of the extra rational point $Q$ over each copy of the base.}

\subsection{M5-branes at A-type Singularities}

Let us now turn to an F-theoretic analysis of the global $U(1)$ symmetry for $k$ M5-branes probing an A-type singularity. We divide our discussion up into the generic case with three or more M5-branes, and the special case with two M5-branes. The case of a single M5-brane has
already been covered in subsection \ref{ssec:warm}.

In the case of $k \geq 2$ M5-branes, the F-theory base is $\mathbb{C}^{2} / \mathbb{Z}_{k}$, and there are $k-1$ collapsing curves of self-intersection $-2$. From our discussion of Weierstrass models on a singular base and the transformation properties of $f$ and $g$ under the
$\mathbb{Z}_{k}$ group action, we observe that each of the coefficients $b_{i}$ appearing in the shifted Weierstrass model must also be invariant under the group action (see e.g. section 4 of reference \cite{DelZotto:2017pti} for the details). This in turn restricts the power series expansion for each term:
\begin{equation}
b_{i}(u,v) = \sum_{a,b} (b_{i})_{a,b} u^{a} v^{b}, \quad i=2,4,6;
\end{equation}
namely all terms are obtained from the $\mathbb{Z}_{k}$ invariant monomials $uv$, $u^{k}$ and $v^{k}$. From this, it follows at once that the ansatz \eqref{eq:ImIn} corresponds to the tuned choice
\begin{subequations}
\begin{align}
&{(b_2)}_{0,0}=1\ , \quad {(b_2)}_{a,b} = \delta_{a,M}\delta_{b,N}\ , \\
&{(b_4)}_{0,0}=0\ , \quad {(b_4)}_{a,b} = \delta_{a,M}\delta_{b,N}\ , \\
&{(b_6)}_{0,0}=0\ , \quad {(b_6)}_{a,b} = \delta_{a,M}\delta_{b,N}\ ,
\end{align}
\end{subequations}
for fixed $M,N>0$, which is clearly compatible with the constraints of the $\mathbb{Z}_{k}$ group action. In particular,
this means that all statements regarding the extra section and $Q$ carry over unchanged.

Our analysis applies equally well to the case of $N = 2$, as well as the case of $k = 2$. In both limits, however, we have already seen several indications from our field theory analysis that additional enhancements in the flavor symmetries are to be expected. We interpret this to mean that the F-theory geometry only makes manifest a subset of all the symmetries.

For the case of $k = 2$, we can actually anticipate what we must do to produce this higher enhancement. In this case, we have a single $-2$ curve which supports an $I_N$ fiber. There are two marked points on this curve, indicating collisions with two distinct $I_N$ components of the discriminant locus. If we consider the tuned limit where we move these two points on top of each other, we observe that the fiber type from the ``flavor branes'' enhances to $I_{2N}$, anticipating an $SU(2N)$ flavor symmetry, which is the answer expected from our analysis of section \ref{sec:YouThaOne}. In the limit where we collapse the $-2$ curve to zero size, we also see that we no longer have a transverse intersection of two $I_N$ components of the discriminant locus. This tuning is a special feature of having a single $-2$ curve.

The case of $N = 2$ is even more subtle in the corresponding geometry. In this case, the tensor branch deformation consists of only $\mathfrak{su}(2)$ gauge algebras. There is in this case an enhancement in the global symmetry beyond $SU(2) \times SU(2) \times U(1)$, which is in accord with the fact that the geometry only tends to see a subset of possible flavor symmetries. That being said, we see that for generic choices of $N$ and $k$, the F-theory analysis correctly predicts the global symmetries expected both from M-theory and from field theory.

\subsection{Small Instanton Probes of A-type Singularities}
\label{sec:E8sum}

We now turn to the other class of A-type progenitor theories given by heterotic $E_8$ small instanton probes of A-type singularities.
Again, we divide our discussion up into the generic case where the A-type non-abelian flavor symmetry is $SU(N)$ for $N \geq 3$ and there are a sufficient number of curves on the tensor branch, and less generic situations in which further enhancements in the flavor symmetry are possible.

The Weierstrass model for this case is defined over a base $B = \mathbb{C}^2$ with $f,g$ and $\Delta$ given by \cite{Aspinwall:1997ye}:
\begin{align}\label{eq:fgDeltaE8SUn}
f&=-3u^4(1-v^N) \ , \\
g&=2u^5(u+v^m)\ , \\
\Delta &= 108 u^{10} v^N (3u^2-3u^2 v^N+u^2 v^{2N}+2uv^{m-N}+v^{2m-N})\ .
\end{align}
It corresponds to the collision of an $I_N$ fiber located at $v=0$ against a $I\! I^*$ one at $u=0$ for generic positive $m,N$. We assume $m \geq  N \geq 1$. In heterotic language, the number of mobile small $E_8$ instantons is given by $k \equiv m-N$.

To produce an F-theory model with fibers on curves in Kodaira-Tate form, we must perform successive blowups of the base. This procedure is presented in reference \cite{Aspinwall:1997ye} and is also reviewed in reference \cite{Heckman:2018jxk}. The partial tensor branch is obtained by performing $k$ blowups in the base. Additional blowups ($N$ of them) are necessary because of the intersection of an $I_N$ fiber type with a $I\!I^*$ fiber type, which yields the full tensor branch. There are three distinguished cases for the number of small instantons: $k=0$, $k = 1$, and $k \geq 2$. Additionally, $N \geq 3$ provides a generic A-type singularity whereas $N = 2$ leads to additional enhancements in the flavor symmetry. From our field theory analysis we expect a $U(1)$ global symmetry to be there for generic $N,k$, but to be absent for generic $N$ and $k=0,1$, as well as for $N=2$. See Appendix \ref{app:TBres} for some additional details.

An important point in this class of examples is that for sufficiently low values of $k$, we can obtain these same 6D SCFTs
by starting with a global model in the $E_8 \times E_8$ heterotic string and taking a suitable decoupling limit \cite{Witten:1995gx,Ganor:1996mu,Seiberg:1996vs}. Observe, however, that in the heterotic model on a compact, singular K3 surface,
there is no $U(1)$ symmetry; it does not descend from the unbroken $E_8$'s, and there are no isometries of the K3 surface. This illustrates
that although the local A-type singularity may possess such an isometry, coupling to gravity will remove it. From this perspective, we ought not to expect our global $U(1)$ symmetry to arise from taking a decoupling limit of a gauged $U(1)$ symmetry in a 6D supergravity model.
In this sense, the analysis of our local model for \eqref{heto} in MP form need not work. We will see that this expectation is borne out.

The local model can be engineered by taking
\begin{equation}
b_2 = 3 u^2, \quad b_4= \frac{3}{2} u^4 v^N, \quad b_6 =3 u^6 v^N+2u^5 v^m,
\end{equation}
which reconstruct the $f,g$ polynomials of \eqref{eq:fgDeltaE8SUn}. If we insist on fitting this choice of $b_2,b_4,b_6$ into a MP model, we always encounter branch cuts. For example, one may take
\begin{equation}\label{eq:bcE8}
c_0=0, \quad c_1=u^2, \quad c_2=3u^2,\quad c_3=3u^2 v^N,
\end{equation}
which however requires the non-single-valued identification $b^2=4u(3uv^N+2v^m)$. We have tried multiple polynomial ans\"atze for $b,c_0,\ldots,c_3$ but all seem to involve an unwanted branch cut in the local base.
\begin{figure}[t!]
\centering
\includegraphics[scale = 0.45, trim = {0.5cm 5.5cm 0 4.5cm}]{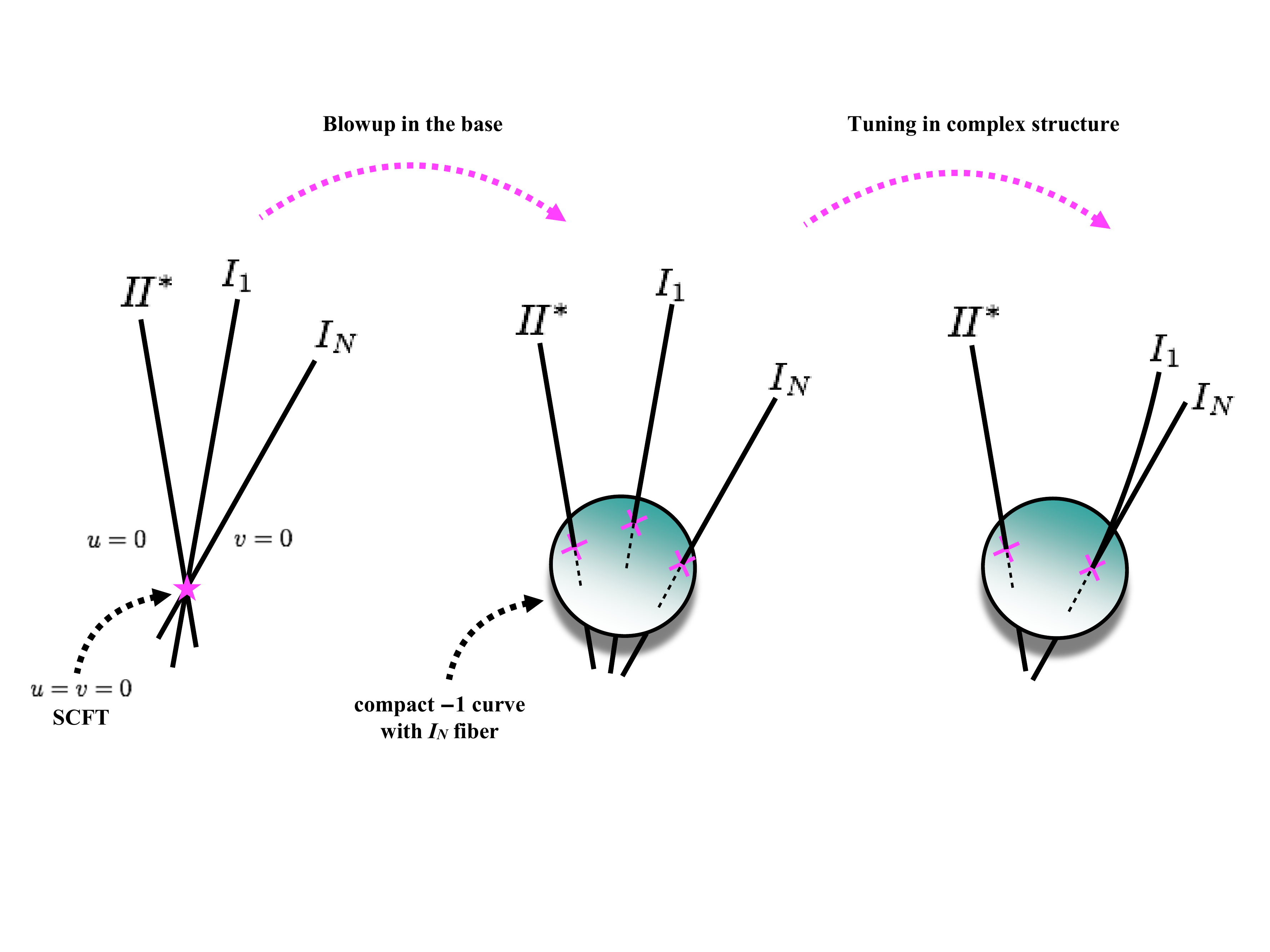}
\caption{The $k=1$ case of the $(E_8,SU(N))$ collision.}
\label{fig:kone}
\end{figure}

Let us now turn to the special case where $k = 1$. From our field theory analysis, we expect that in this case there is in fact an enhancement in the flavor symmetry factor $SU(N) \times U(1)$ to $SU(N+1)$. On the partial tensor branch of this theory,
\begin{align}
[E_{8}]~\overset{\mathfrak{su}(N)}{1}~[SU(N)],
\end{align}
 we have a single $-1$ curve with $I_{N}$ fiber type. This curve is met by three distinct collisions with the rest of the discriminant. First, there is the $E_8$ locus. Second, we have a non-compact component with $I_N$ fiber type. Third, there is another non-compact component with $I_{1}$ fiber type. By a suitable tuning of the Weierstrass model in this case, we observe that the $I_{1}$ and $I_{N}$ points can be merged to produce a single non-compact $I_{N+1}$ component of the discriminant. Note that in doing so, we no longer have a transverse intersection of these two components of the discriminant; it is a special feature of the $k = 1$ case. See figure \ref{fig:kone} for a depiction and Appendix \ref{app:TBres} for a calculation, where this becomes more manifest in the partial tensor branch resolution.

Consider next the case of $k = 0$, i.e. the case of $(E_8, SU(N))$ conformal matter. We have a full tensor branch description given by
\begin{equation}
[E_{8}]~1~\overset{\mathfrak{su}(1)}{2}~\overset{\mathfrak{su}(2)}{2}~\ldots~ \overset{\mathfrak{su}(N-1)}{2}~[SU(N)].
\end{equation}
In this case, the additional $I_1$ locus $2u+u^2(3-3v^N+v^{2N})+v^m=0$ intersects the $I_N$ locus $v=0$ at two points, and the $I\! I^*$ locus $u=0$ at the origin, generating a decoupled ``$SU(1)\times SU(N)$'' bifundamental hypermultiplet (see figure \ref{fig:kzero}). In Appendix \ref{app:TBres} we indeed show that in a partial resolution of the base, the extra $I_1$ locus intersects only the $I_N$ locus. In other words, there is a global $U(1)$ which acts trivially on the Hilbert space of the SCFT (since no matter is charged under it), hence the flavor symmetry at the fixed point is just $SU(N)$, as expected from the field theory analysis.
\begin{figure}[t!]
\centering
\includegraphics[scale = 0.5, trim = {1.5cm 8.5cm 0 8.5cm}]{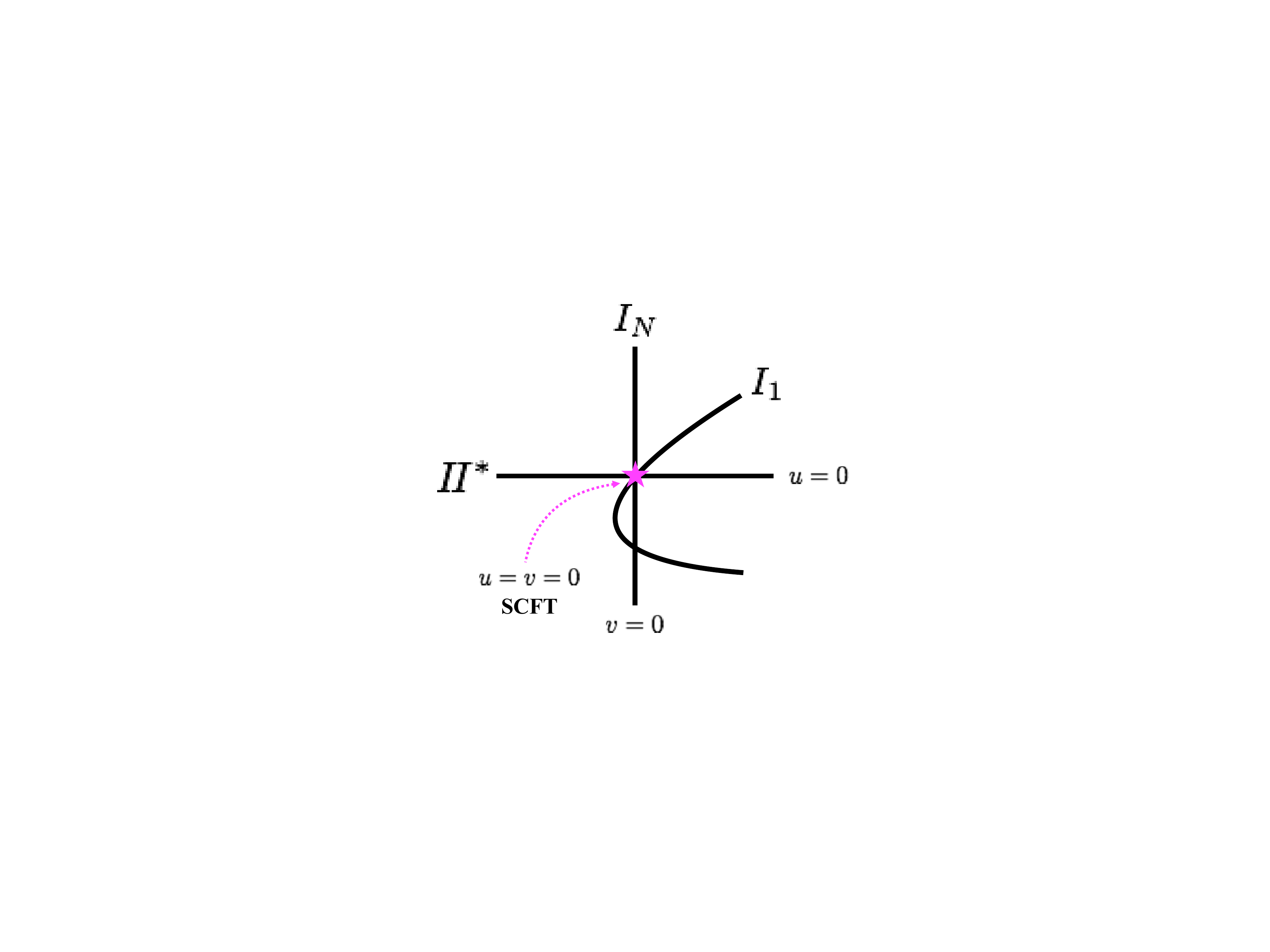}
\caption{The $k=0$ case of the $(E_8,SU(N))$ collision, also known as $(E_8,SU(N))$ conformal matter.}
\label{fig:kzero}
\end{figure}

Lastly, consider the models obtained from taking $N = 2$. In this case, all of the tensor branch gauge algebras are $\mathfrak{su}(2)$, and a formal ``$\mathfrak{su}(1)$'' gauge algebra obtained from an $I_1$ fiber over a compact $-2$ curve. In these cases, the expectation from field theory is that there is a generic enhancement in the flavor symmetry, which is not visible in the F-theory presentation of the model.

For completeness, let us briefly comment on some of the D- and E-type progenitor theories, where we do not generically expect any $U(1)$ factor in the flavor symmetry. As an example, the local models for the theories of \eqref{m5o} with one M5-brane probing an E-type singularity have local Weierstrass models \cite{DelZotto:2014hpa, Heckman:2014qba}:
\begin{align}
&(E_6,E_6): y^2=x^3+u^4v^4, \\
&(E_7,E_7): y^2=x^3+u^3v^3 x, \\
&(E_8,E_8): y^2=x^3+u^5v^5.
\end{align}
In the first and third case we have $f=0$ identically, which implies $b_4=\frac{1}{6}b_2^2$ and $g=-\frac{1}{27}b_2^3+b_6$. In the second case we have $g=0$ identically, which implies $b_6=-\frac{2}{27}(b_2^3-9b_2b_4)$ and $f=-\frac{1}{3}b_2^2+2b_4$. The constraints on $b_2,b_4,b_6$ turn into constraints on $b,c_0,\ldots,c_3$ given the restrictions in \eqref{eq:bMP}. In either case, we find that any simple ans\"atze respecting these constraints would lead to branch cuts.

For the D-type, the progenitor of \eqref{heto}, i.e. $(E_8,SO(2N+8))$, reads \cite{Aspinwall:1997ye}:
\begin{align}\label{eq:fgDeltaE8SO2N}
f&=-3u^4v^2(1-v^N) \ , \\
g&=2u^5v^3(u+v^m)\ , \\
\Delta &= 108 u^{10} v^{6+N} (3u^2-3u^2 v^N+u^2 v^{2N}+2uv^{m-N}+v^{2m-N})\ .
\end{align}
Again, we find that any simple ans\"atze respecting these constraints would lead to branch cuts.


\section{Conclusions} \label{sec:CONC}

One of the important features of a quantum field theory is its flavor symmetries.
In this paper we have developed a general set of methods for extracting the
global $U(1)$ symmetries of a 6D SCFT. At a broad level, all of the symmetries we have identified descend from a small set of progenitor theories. Since all known 6D SCFTs originate from ``fission moves'' on the progenitors followed by fusion operations on the resulting fission products, by and large these $U(1)$ symmetries have a common origin (excluding emergent $U(1)$'s). We have also presented a general prescription for
how to read off the $U(1)$ symmetries for any 6D SCFT using just the data available from its tensor branch. Lastly, we have shown
that in the F-theory description of progenitor theories for 6D SCFTs, there are two geometric origins for such symmetries. One is the non-abelian flavor symmetry of a seven-brane. The other is the appearance of an additional section in a non-compact model, as occurs in some progenitor theories with an A-type flavor symmetry. In the remainder of this section we discuss some further avenues of investigation.

In this paper we have presented a combination of bottom-up and top-down techniques for extracting the global symmetries of a 6D SCFT, demonstrating agreement between various different approaches. One interesting outcome of this analysis is that of the progenitor theories, only some of the theories with an A-type global symmetry have an additional
generator of the Mordell-Weil group. What we have not undertaken here is a direct analysis of the resulting elliptic threefolds obtained from fission and fusion of the progenitors. In principle, the resulting $U(1)$ symmetries we have identified via field theory may have a geometric origin as additional generators of the Mordell-Weil group in the corresponding F-theory models. It would be quite interesting to see whether the
identification between Higgs branch deformations of a 6D SCFT and complex structure deformations of a Calabi-Yau could be used to establish a geometric origin of these symmetries as well.

It would also be interesting to analyze the fate of these $U(1)$ symmetries in compactifications to 5D SCFTs. If the non-abelian part of the 6D SCFT flavor symmetry is known, by looking at triple intersection numbers among surfaces in the associated resolved Calabi-Yau geometry, one can compute the total rank of the flavor symmetry, thus determining residual abelian factors \cite{Apruzzi:2018nre, Apruzzi:2019opn}. This would provide a complementary approach compared with directly analyzing the Mordell-Weil group of the associated F-theory model.

Indeed, with our improved understanding of how to determine $U(1)$ symmetries in 6D SCFTs, it is natural to ask about the fate of these symmetries upon compactification to lower-dimensional quantum field theories. This is especially important in the context of 4D $\mathcal{N} = 1$ SCFTs where the infrared R-symmetry is typically a linear combination of the UV R-symmetry and global $U(1)$ symmetries \cite{Intriligator:2003jj}. It would be interesting to analyze this case, as well as the sense in which compactification defines an RG flow from six to four dimensions (see e.g. \cite{Razamat:2016dpl, Apruzzi:2018oge, Kim:2018lfo, Apruzzi:2018xkw}).

The primary aim of this paper has been the analysis of continuous abelian global symmetries. Another important question in the study of 6D SCFTs is to determine possible discrete global symmetries. This data can in principle be recovered by tracking the group theoretic data associated with vevs triggering Higgs branch deformations. For example, a large class of RG flows are triggered by operators vevs which are nilpotent in the flavor symmetry algebra. In the flavor symmetry group, these are associated with unipotent elements. It would be interesting to see whether a more refined analysis is capable of tracking this discrete data.

Along the same lines, it is natural to ask whether this information is encoded in the structure
of the tensor branch of a 6D SCFT. Doing so would likely also require a better understanding
of the gauge groups present there, rather than just the associated gauge algebras. It also be
instructive to track the geometric origin of such symmetries in the associated F-theory model.

Finally, we have also discussed at a broad level the sense in which the symmetries of a 6D SCFT are inherited from data of a progenitor theory. Given this, it is natural to ask about the fate of other symmetries such as the R-symmetry of a 6D SCFT. This is broken along the trajectory of an RG flow, but is recovered at an IR fixed point. It would be interesting to see whether the techniques developed here could be adapted to this class of questions.

\newpage

\section*{Acknowledgements}

We thank M. Cveti\v{c}, M. Dierigl, and C. Lawrie for
helpful discussions. We thank the 2019
Simons Summer workshop for hospitality during part of this work.
The work of FA is supported by the ERC Consolidator Grant number
682608 ``Higgs bundles: Supersymmetric
Gauge Theories and Geometry (HIGGSBNDL)''.
The work of MF is supported in part by the European Union’s Horizon 2020 research and innovation programme under the Marie Skłodowska-Curie grant agreement No.~754496~-~FELLINI. The work of JJH is supported by NSF CAREER grant PHY-1756996.  The work of TR is
supported by NSF grant PHY-1911298 and the Roger Dashen Membership. FA, MF and TR thank the UPenn theory group for hospitality during part of this work. MF also thanks the Technion and the Weizmann Institute of Science for hospitality during part of this work.



\appendix

\section{Anomaly Polynomials and $U(1)$ Flavor Symmetries} 
\label{app:A}

In this Appendix we discuss some additional aspects of how abelian flavor symmetries can contribute to the anomaly polynomial of a 6D SCFT.
Recall that the anomalies of a 6D SCFT are encoded in an anomaly 8-form $I_8$. As discussed in \cite{Ohmori:2014kda}, this anomaly polynomial can be divided into two parts: a one-loop contribution (which can be computed on the tensor branch), and a Green-Schwarz contribution:
\begin{equation}
I_{\text{tot}} = I_{\text{1-loop}}+ I_{\text{GS}} .
\label{eq:Itot}
\end{equation}
The one-loop contribution is simply a sum of contributions from
free tensor multiplets, vector multiplets, hypermultiplets, and E-strings. These
contributions are, respectively \cite{Ohmori:2014kda}:\footnote{Here, we present only the formula for a rank-1 E-string. The formula for a more general rank-$k$ E-string can be found in \cite{Ohmori:2014kda}.}
\begin{align}
I_{\text{tensor}} &= \frac{c_{2}(R)^{2}}{24} + \frac{c_{2}(R)p_{1}(T)}{48}
+ \frac{23 p_{1}(T)^{2} -116 p_{2}(T)}{5760},\label{eq:tensor}\\
I_{\text{vector}} &= -\frac{\Tr_{\text{adj}} F^{4} + 6 c_{2}(R) \Tr_{\text{adj}%
} F^{2} + d_{G} c_{2}(R)^{2}}{24}   - \frac{\Tr_{\text{adj}}F^{2}+d_{G}
c_{2}(R)p_{1}(T)}{48}\ + \nonumber\\
&\ \ \ \,  - d_{G} \frac{7 p_{1}(T)^{2} - 4 p_{2}(T)}{5760},\label{eq:vector}\\
I_{\text{hyper}} &= \frac{\Tr_{\rho} F^{4} }{24} + \frac{\Tr_{\text{adj}}F^{2}
p_{1}(T)}{48}   + d_{\rho}\frac{7 p_{1}(T)^{2} - 4 p_{2}(T)}{5760}
\label{eq:hyper}\\
I_{\text{E-string}}^{(1)} &=
\frac{13}{24} c_2(R)^2 - \frac{11}{48} c_2(R) p_1(T) +\frac{203}{5760} p_1(T)^2 - \frac{29}{1440} p_2(T) - \frac{1}{4} c_2(R) \Tr F_{E_8}^2\ + \nonumber \\
&\ \ \ \, + \frac{1}{16} p_1(T) \Tr F_{E_8}^2 + \frac{1}{32} (\Tr F_{E_8}^2)^2.
\label{eq:Estring}
\end{align}
Here, $\Tr_{\rho}$ is the trace in the representation $\rho$, $d_{\rho}$ is
the dimension of the representation $\rho$, and $d_{G}$ is the dimension of
the group $G$. In computing the anomaly polynomial, one should express the
traces over arbitrary representations in terms of the trace in a defining
representation. For a given simple Lie algebra $\mf{g}$, we can write
\begin{align}
\Tr_{\rho}F^{4} &= x_{\rho}\, \mathrm{Tr} F^{4} + y_{\rho}(\, \mathrm{Tr}
F^{2})^{2}\\
\Tr_{\rho}F^{3} &= c_{\rho} \Tr F^3\\
\Tr_{\rho}F^{2} &= \text{Ind}_{\rho}\, \mathrm{Tr} F^{2} ,
\end{align}
with $x_{\rho}$, $y_{\rho}$, $c_{\rho}$, and $\text{Ind}_{\rho}$ known group theory constants. $c_{\rho}$ here is defined as in Table \ref{tab:GTCC}, and throughout this paper the values of $x_{\rho}$, $y_{\rho}$, and $\text{Ind}_{\rho}$ agree with the conventions of \cite{Ohmori:2014kda, Apruzzi:2017nck, Heckman:2018jxk}.\footnote{In particular, note that $\text{Ind}_{\text{fund}} = 1/2$ for the fundamental of $SU(N)$.}

The one-loop contributions to the anomaly polynomial generalize straightforwardly to the case of a $U(1)$ symmetry, whether gauge or global. For a gauge symmetry, we have the contribution from a single vector multiplet with $d_G = 1$. The vector multiplet does not contribute to the gauge anomaly for the $U(1)$ because, as opposed to the case of a non-abelian gauge group, there are no W-bosons charged under the Cartan of a $U(1)$. Thus we have
\begin{equation}
I_{\text{vector}}^{U(1)} = -\frac{ c_{2}(R)^{2}}{24}    - \frac{
c_{2}(R)p_{1}(T)}{48}- \frac{7 p_{1}(T)^{2} - 4 p_{2}(T)}{5760}
\end{equation}
The case of a hypermultiplet is likewise analogous to the previous case. Consider a hypermultiplet in a representation $R$ of some semi-simple, non-abelian lie algebra $\mf{g}$ (which could be trivial, or a direct sum of simple lie algebras) with field strength $P$, which is also charged under some combination of $U(1)$'s with field strengths $F_i$ and associated charges $q_i$. Then, in (\ref{eq:hyper}), we have
\begin{align}
\Tr_{\rho} F^4 &=  \Tr_R P^4 + 4 \Tr_R P^3 q_i F_i + 6 \Tr_R P^2 q_i q_j F_i F_j + q_i q_j q_k q_l F_i F_j F_k F_l \\
\Tr_{\rho} F^2 &=  \Tr_R F^2 + q_i q_j F_i F_j .
\end{align}
Here, the sum over repeated indices is implied, $q_i F_i := \sum_i q_i F_i$. Note that we have not included a $\Tr_R P$ term, as this necessarily vanishes for a semi-simple Lie algebra.

To compute the abelian contribution to the anomaly from an E-string, we suppose that the $E_8$ global symmetry of the E-string has been decomposed into a product of subgroups, which may be either gauge or global, abelian or non-abelian:
\begin{equation}
    E_8 \supset G_1 \times ... \times G_k.
\end{equation}
Then, we simply decompose terms in (\ref{eq:Estring}) involving the $E_8$ field strength into a sum over the field strengths of the subgroups:
\begin{equation}
    \Tr F_{E_8}^2 \rightarrow \sum_{i=1}^k \Tr F_{G_k}^2.
\end{equation}

Anomalies present at one-loop can be canceled via the Green-Schwarz mechanism. Note that not every type of anomaly presents a problem for a 6D SCFT, however. We can sort the terms in the anomaly polynomial according to the number of gauge currents~/~global currents appearing in them. Terms of the form $F_{\rm gauge}^4$, with purely gauge anomalies, lead to a sickness and must be canceled. Terms of the form $F_{\rm gauge}^3 F_{\rm global}$ represent an ABJ anomaly, which leads to a divergence in the global anomaly current and thus violates the putative global symmetry. Terms of the form $F_{\rm gauge}^2 F_{\rm global}^2$ and $F_{\rm gauge} F_{\rm global}^3$ are also forbidden in 6D SCFTs but for a different reason: there is no multiplet in the 6D superconformal algebra that would give rise to such terms \cite{Cordova:2018cvg,Cordova:2016emh}. Finally, $F_{\rm global}^4$ terms represent 't Hooft anomalies, which are benign and do not need to be canceled.

There is a unique Green-Schwarz mechanism available for canceling non-abelian gauge anomalies in six dimensions, which was introduced in \cite{Green:1984bx}. The Green-Schwarz contribution to the anomaly polynomial takes the form
\begin{equation}
I_{\text{GS}} = \frac{1}{2} \Omega_{\alpha \beta} I^\alpha I^\beta =  \frac{1}{2} \Omega^{\alpha \beta} I_\alpha I_\beta.
\end{equation}
Here, $\Omega_{\alpha \beta}$ is the Dirac pairing on the string charge lattice, and indices are raised via the inverse $\Omega^{\alpha \beta} = (\Omega^{-1})^{\alpha\beta}$. This comes from a term in the action of the form
\begin{equation}
\Omega^{\alpha \beta} B_\alpha \wedge I_\beta,
\end{equation}
with $B_\alpha$ a 2-form and $I_\beta$ a 4-form, whose non-abelian part was given in \cite{Apruzzi:2017iqe}:
\begin{equation}
I_\alpha = h_{G_\alpha}^\vee c_2(R) + \frac{n_\alpha-2}{4}  p_1(T) + \frac{n_\alpha}{4}\Tr F_\alpha^2 - \frac{1}{4} \sum_{\beta \in \textrm{nn}} \Tr F_\beta^2.
\end{equation}
Here, $G_\alpha$ is the gauge group associated with the $\alpha$th tensor multiplet, $F_\alpha$ is its field strength, $h_{G_\alpha}^\vee$ is its dual Coxeter number, and $n_\alpha := -\Omega_{\alpha \alpha}$ (indices not summed) is the $\alpha$th string charge. The expression ``$\textrm{nn}$'' refers to nearest neighbors in the corresponding intersection pairing of curves.

For $U(1)$ anomalies, the above anomaly cancelation mechanism is still valid, but there is an additional Green-Schwarz mechanism available \cite{Park:2011wv}: we may include in the action a term of the form
\begin{equation}
C \wedge X_6,
\end{equation}
with $X_6$ a 6-form and $C$ a St\"uckelburg 0-form that couples to a $U(1)$ gauge boson $A^i_\mu$ with a coupling of the form
\begin{equation}
(\partial_\mu C - A_\mu)^2.
\end{equation}
This leads to an anomaly polynomial contribution of the form
\begin{equation}
F \wedge X_6,
\end{equation}
which can be used to cancel any remaining abelian gauge anomalies. However, this comes at a price: the St\"uckelburg mechanism gives a mass to the $U(1)$ vector boson, thereby removing the associated gauge symmetry \cite{Park:2011wv}. Note that the anomaly involving the cube of a non-abelian gauge field strength and the first power of an abelian field strength can only be canceled by this latter Green-Schwarz mechanism. Such anomalies plague all would-be $U(1)$ gauge groups in 6D SCFTs, and as a result all abelian gauge groups are removed from the low-energy theory on the tensor branch.

\section{$U(1)$'s from Group Theory}\label{sec:grouptheory}

In this Appendix we review the analysis of $U(1)$ symmetries expected from symmetry breaking patterns, and their associated group theoretic structures.

Two very large classes of 6D SCFTs are in one-to-one correspondence with two particular classes of homomorphisms: namely, $\text{Hom}(\Gamma,E_8)$ and $\text{Hom}(\ksu(2),\mf{g})$, where $\Gamma$ is a discrete subgroup of $SU(2)$ and $\mf{g}$ is a simple Lie algebra.

The relationship between 6D SCFTs and homomorphisms follows from their M/F-theory constructions. Consider the progenitor theory of $k$ M5-branes probing a $\mathbb{C}^2/\Gamma$ orbifold singularity as well as an $E_8$ wall. This theory is labeled by $k$ and $\Gamma$ as well as a flat connection of $E_8$ at the infinity of $\mathbb{C}^2/\Gamma \simeq S^3/\Gamma$. These flat connections, and hence this class of 6D SCFTs, are in one-to-one correspondence with $\text{Hom}(\Gamma,E_8)$.

Given a particular homomorphism $\rho \in$ $\text{Hom}(\Gamma,E_8)$, we may consider the image $\rho(\Gamma)$ in $E_8$. The elements of $E_8$ that commute with $\rho(\Gamma)$ form a group $H$ known as the commutant of the homomorphism $\rho$. This commutant subgroup $H \subset E_8$ translates in field theory terms to the global symmetry of the associated 6D SCFT.

More precisely, for $\Gamma$ of D/E-type, the full global symmetry of the theory is given by
\begin{equation}
    G_{\text{global}} = H \times G_{\Gamma},
\end{equation}
with $G_\Gamma$ the D/E-type Lie group associated with the discrete group $\Gamma$ via the McKay correspondence.

For $\Gamma = \mathbb{Z}_N, N \neq 2$, we instead have
\begin{equation}
    G_{\text{global}} = H \times SU(N) \times U(1).
\end{equation}
The extra $U(1)$ factor is inherited from the progenitor theory, which is associated with the isometry of the $\mathbb{C}^2/\Gamma$ orbifold singularity, as discussed in section \ref{sec:prog}.

For $\Gamma = \mathbb{Z}_2$, the isometry enhances, and there is a further enhancement $U(1) \rightarrow SU(2)$:
\begin{equation}
    G_{\text{global}} = H \times SU(2) \times SU(2).
    \label{eqn:E8homglobalsu2}
\end{equation}

Unfortunately, within the mathematics literature, the full set of homomorphisms and their allowed commutants is only known for $\Gamma = \mathbb{Z}_N$ \cite{MR739850} and  $\Gamma = SL(2,5)$ \cite{FREY, Frey:2018vpw}. While the classification of the latter is very involved, the former can be labeled in a simple combinatorial fashion in terms of the extended Dynkin diagram of $E_8$. Namely, if we label the nodes of this diagram as follows:
\begin{equation}
\begin{tikzpicture}
\begin{scope}[start chain]
{
\dnode{1}
\dnode{2}
\dnode{3}
\dnode{4}
\dnode{5}
\dnode{6}
\dnode{4'}
\dnode{2'}
}
\end{scope}
\begin{scope}[start chain=br going above]
\chainin (chain-6);
\dnodebr{3'}
\end{scope}
\end{tikzpicture}
\end{equation}
then homomorphisms $\mathbb{Z}_N \rightarrow E_8$ are in one-to-one correspondence with lists of nodes such that the sum of the numbers of these nodes equals $N$, where any given node may be used multiple times.  For instance, for $N=4$, we have the following choices of nodes:
\begin{equation}
1+1+1+1,~~1+1+2,~~1+1+2',~~1+3,~~1+3',~~2+2,~~2+2',~~2'+2',~~4,~~4'.
\end{equation}
The commutant $H$ of the homomorphism is then given simply by the diagram remaining after deleting the corresponding nodes from the affine $E_8$ Dynkin diagram, with additional $U(1)$'s added as necessary to ensure that the rank of the commutant is always 8. For instance, we have
$$
1+1+1+1 ~~\leftrightarrow~~ \begin{tikzpicture}
\begin{scope}[start chain]
{
\dnode{2}
\dnode{3}
\dnode{4}
\dnode{5}
\dnode{6}
\dnode{4'}
\dnode{2'}
}
\end{scope}
\begin{scope}[start chain=br going above]
\chainin (chain-5);
\dnodebr{3'}
\end{scope}
\end{tikzpicture}~~
\leftrightarrow~~H=E_8
$$
$$
1+1+2' ~~\leftrightarrow ~~\begin{tikzpicture}
\begin{scope}[start chain]
{
\dnode{2}
\dnode{3}
\dnode{4}
\dnode{5}
\dnode{6}
\dnode{4'}
}
\end{scope}
\begin{scope}[start chain=br going above]
\chainin (chain-5);
\dnodebr{3'}
\end{scope}
\end{tikzpicture}
~~\leftrightarrow~~
H=[SO(14) \times U(1)]
$$
$$
1+3 ~~\leftrightarrow ~~\begin{tikzpicture}
\begin{scope}[start chain]
{
\dnode{2}
}
\end{scope}
\end{tikzpicture}~~~~~~~
\begin{tikzpicture}
\begin{scope}[start chain]
{
\dnode{4}
\dnode{5}
\dnode{6}
\dnode{4'}
\dnode{2'}
}
\end{scope}
\begin{scope}[start chain=br going above]
\chainin (chain-3);
\dnodebr{3'}
\end{scope}
\end{tikzpicture}
~~\leftrightarrow~~
H=E_6 \times SU(2) \times U(1)
$$

In the remainder of the cases, $G_\Gamma \neq SU(N), E_8$, the match with 6D SCFTs has been used previously to give the first conjectured classification of $\text{Hom}(\Gamma, E_8)$ \cite{Frey:2018vpw}. While this is an exciting application of M-theory to pure mathematics, the lack of mathematical literature on the topic somewhat limits our ability to learn new properties of 6D SCFTs from this correspondence.

Fortunately, the other class of homomorphisms, $\text{Hom}(\ksu(2),\mf{g})$, is much more familiar to mathematicians, as these homomorphisms are in one-to-one correspondence with nilpotent orbits of $\mf{g}$. This permits us to use the vast body of mathematical literature on nilpotent orbits to learn about global symmetries and RG flows of 6D SCFTs \cite{DelZotto:2014hpa, Heckman:2016ssk, Mekareeya:2016yal,Cabrera:2019izd}.

While nilpotent orbits of all simple Lie algebras have been classified, they are simplest to describe in the case of the classical algebras, $\mf{su}(N)$, $\mf{so}(N)$, and $\mf{sp}(N)$. For $\mf{su}(N)$, nilpotent orbits are labeled simply by partitions of $N$. For $\mf{so}(N)$, they are labeled by partitions of $N$ subject to the constraint that each even number must appear an even number of times. For $\mf{sp}(N)$, they are labeled by partitions of $2N$ subject to the constraint that any odd number must appear an even number of times. The commutant subalgebra $\mf{h} \subset \mf{g}$ left unbroken by the nilpotent orbit is then given in terms of the partition. Given a partition $\mu = [\mu_1^{d_1}, \mu_2^{d_2}, \mu_3^{d_3},...]$ in which the entry $\mu_{i}$ has multiplicity $d_{i}$, the commutant is given by
\begin{align}
\mathfrak{su}  &  :\mf{h}=\mathfrak{s}\left(
\underset{i}{\oplus}\mathfrak{u}(d_{i})\right) \label{suflavor}\\
\mathfrak{so}  &  :\mf{h}=\underset{i\text{
odd}}{\oplus}\mathfrak{so}\left(  d_{i}\right)  \oplus\underset{i\text{
even}}{\oplus}\mathfrak{sp}\left(  d_{i}/2\right) \label{eq:soflavor} \\
\mathfrak{sp}  &  :\mf{h}=\underset{i\text{
even}}{\oplus}\mathfrak{so}\left(  d_{i}\right)  \oplus\underset{i\text{
odd}}{\oplus}\mathfrak{sp}\left(  d_{i}/2\right) , \label{eq:spflavor}
\end{align}
where in the above ``$i$ odd'' or ``$i$ even'' is shorthand
for indicating that $\mu_i$ is odd or even, respectively.

For instance, for $\mf{g}=\mf{su}(4)$, there are five nilpotent orbits, labeled by the partitions $\mu=[1^4]$, $[2, 1^2]$, $[2^2]$, $[3,1]$, and $[4]$. The associated commutator subalgebras are given by $\mf{su}(4)$, $\mf{su}(2) \times \mf{u}(1)$, $\mf{su}(2)$, $\mf{u}(1)$, and the trivial subalgebra, respectively.

In F-theory, these homomorphisms arise as ``T-brane'' data for two intersecting stacks of seven-branes probing a $\mathbb{C}^2/\mathbb{Z}_N$ singularity (for a partial list of references to the T-brane literature, see references
\cite{Aspinwall:1998he, Donagi:2003hh,Cecotti:2009zf,Cecotti:2010bp,Donagi:2011jy,Anderson:2013rka,
Collinucci:2014qfa,Cicoli:2015ylx,Heckman:2016ssk,Collinucci:2016hpz,Bena:2016oqr,
Marchesano:2016cqg,Anderson:2017rpr,Collinucci:2017bwv,Cicoli:2017shd,Marchesano:2017kke,
Heckman:2018pqx, Apruzzi:2018xkw, Cvetic:2018xaq, Collinucci:2018aho, Carta:2018qke, Marchesano:2019azf, Bena:2019rth, Barbosa:2019bgh, Hassler:2019eso}). In particular, we can turn on T-brane data for either stack of seven-branes, resulting in
a theory labeled by a pair of homomorphisms $\rho_1, \rho_2 \in \text{Hom}(\ksu(2),\mf{g})$.

Once again, the global symmetry of a given theory is related to the commutant $H$ of a given homomorphism inside $G$. In this case, since a theory is labeled by two homomorphisms $\rho_1, \rho_2$, its global symmetry will be determined by the respective commutants $H_1$, $H_2$. In particular, for $\mathfrak{g}$ of D/E-type, the global symmetry is given by
\begin{equation}
    G_{\text{global}} = H_1 \times H_2.
\end{equation}
For $\mf{g} = \mf{su}(N)$, there is an additional $U(1)$ factor, inherited from the isometry of the $\mathbb{C}^2/\mathbb{Z}_N$ orbifold of the progenitor theory (see section \ref{sec:prog}):
\begin{equation}
    G_{\text{global}} = H_1 \times H_2 \times U(1).
    \label{eqn:nilpotentglobal}
\end{equation}
For $\mathfrak{g} = \mf{su}(2)$, the isometry of the progenitor enhances, so the global $U(1)$ enhances to $SU(2)$
\begin{equation}
    G_{\text{global}} = H_1 \times H_2 \times SU(2).
\end{equation}

The tensor branch descriptions of 6D SCFTs corresponding to $\text{Hom}(\Gamma, E_8)$ have been studied in
\cite{DelZotto:2014hpa, Heckman:2015bfa, Mekareeya:2017jgc, Heckman:2018pqx, Frey:2018vpw, Cabrera:2019izd}, while the tensor branch descriptions of 6D SCFTs corresponding to $\text{Hom}(\ksu(2), \mf{g})$ have been studied in \cite{DelZotto:2014hpa, Heckman:2016ssk, Mekareeya:2016yal, Heckman:2018pqx, Bourget:2019aer}. From these descriptions, we can use the match between 6D SCFTs and homomorphisms  as a sort of ``Rosetta Stone": $U(1)$ symmetries appearing in the commutant $H$ of a given homomorphism translate into $U(1)$ symmetries of the SCFT quiver, which allows us to verify the field theory rules for $U(1)$ symmetries in an SCFT quiver.

\section{Resolution of A-type Progenitor Geometries}
 \label{app:TBres}

In this Appendix we determine the tensor branch associated with heterotic $E_8$ small instanton probes of A-type singularities. In particular, we focus on the role of the $I_1$ component of the discriminant locus in the associated F-theory model.

In order to understand the meaning of the additional $I_1$ component of the discriminant \eqref{eq:fgDeltaE8SUn}, we need to resolve the base. This introduces additional $\mathbb{P}^1$'s, in particular we review the results of \cite{Aspinwall:1997ye} with particular focus on the $I_1$, see also \cite{Cvetic:2018xaq, Heckman:2018jxk} for similar computations. For convenience, let us repeat the coefficients of the relevant Weierstrass model:
\begin{align}
f&=-3u^4(1-v^N), \\
g&=2u^5(u+v^m), \\
\Delta &= 108 u^{10} v^N (3u^2-3u^2 v^N+u^2 v^{2N}+2uv^{m-N}+v^{2m-N}). \label{eq:DeltaI1}
\end{align}
The resolution procedure is locally implemented by the following shift
\begin{align} \label{eq:BU}
&{\rm Patch\; 1:}\quad \{u \mapsto u, \; v\mapsto \zeta u,\; y\mapsto u^3 y,\; x \mapsto u^2 x,\; f \mapsto u^4 f,\; g \mapsto u^6 g \}\\
&{\rm Patch\; 2:}\quad \{u \mapsto \eta v, \; v\mapsto v,\; y\mapsto v^3 y,\; x \mapsto v^2 x,\;  f \mapsto v^4 f,\; g \mapsto v^6 g \} \nonumber
\end{align}
where $[\zeta : \eta]$ are the homogeneous coordinates of the resolution $\mathbb P^1$, such that
\begin{equation}
\zeta= \frac{1}{\eta}.
\end{equation}
As usual, $x,y$ are fiber coordinates in the (local) Weierstrass model.

In order to understand the fate of the $I_1$ locus in the discriminant of equation \eqref{eq:DeltaI1},
\begin{equation}
I_1:\quad(3u^2-3u^2 v^N+u^2 v^{2N}+2uv^{m-N}+v^{2m-N}),
\end{equation}
we look at some explicit examples. Let us start with $k=m-N=0$. We then apply \eqref{eq:BU}. The first blowup of the model is
\begin{align}
&{\rm P1:}&\quad & f_1=-3(1-u^N\zeta_1^N), \quad
g_1=2(1+u^{N-1}\zeta_1^N),\;\\
& & &\Delta_1 = 108 u^{N-1} \zeta_1^N (2+3u+u^{N-1}\zeta_1^N(1-3u^2+u^{N+2}\zeta_1^N) )\nonumber \\
&{\rm P2:}& \quad &f_2=-3\eta_1^4(1-v^N), \quad
g_2=2\eta_1^5(\eta_1+v^{N-1}),\nonumber\\
& &  & \Delta_2 = 108 v^{N-1} \eta_1^{10} (v^{N-1}+2\eta_1+\eta_1^2 v(3-3v^N+v^{2N})), \nonumber
\end{align}
where P1 and P2 denote the two patches. We note that $\Delta_1=0$ when $\zeta_1=0$ and $u\neq0$, in agreement with the interpretation that the extra $I_1$ locus leads to a decoupled $SU(N)\times SU(1)$ hypermultiplet in the $k=0$ case, which was proposed in section \ref{sec:E8sum}. Moreover, the $I_1$ component also intersects the resolution at $\eta_1=v=0$, however this locus needs to be blown up again since the $(f_2,g_2, \Delta_2)$ vanish with degree higher than $(4,6,12)$ (non-minimal singularity). If we perform a full blowup of the base,
\begin{align}
&{\rm P1:}&\quad & f_1=-3(1-u^N\zeta_1^N), \quad
g_1=2(1+u^{N-1}\zeta_1^N),\;\\
& & &\Delta_1 = 108 u^{N-1} \zeta_1^N (2+3u+u^{N-1}\zeta_1^N(1-3u^2+u^{N+2}\zeta_1^N) )\nonumber \\
&{\rm P(i+1):}& \quad &f_i=-3(1-\zeta_{i+1}^N\zeta_i^N), \quad
g_i=2(1+\zeta_{i+1}^{N-i}\eta_i^{N-1-i}), \nonumber \\
& & &\Delta_i = 108 \zeta_{i+1}^{N-i} \eta_i^{N-i-1} (2+3\zeta_{i+1}^i \eta_i^{i+1}-3 \zeta_{i+1}^{N+i} \eta_i^{N+i+1} +  \zeta_{i+1}^{N-i} \eta_i^{N-i-1}+  \zeta_{i+1}^{2N+i} \eta_i^{2N+i+1} )\nonumber\\
& & & 1\leq i \leq N-2\nonumber\\
&{\rm PN:}& \quad &f_N=-3\eta_{N}^4(1-v^N), \quad
g_N=2\eta_N^5(1+\eta_N),\nonumber\\
& &  & \Delta_2 = 108\eta_N^{10} (1+2\eta_N+\eta_N^2 v(3-3v^N+v^{2N})), \nonumber
\end{align}
we notice that the only other intersection point of the $I_1$ with the resolved base is on the ``last'' $\mathbb P^1$, which has self-intersection $(-1)$ and no gauge fiber enhancement; this however does not lead to any additional charged matter.

Let us now analyze the $k=1$, $m=n+1$, case. The first blowup follows from \eqref{eq:BU}, and it reads
\begin{align}
&{\rm P1:}&\quad & f_1=-3(1-u^N\zeta_1^N), \quad
g_1=2(1+u^{N}\zeta_1^{N+1}),\;\\
& & &\Delta_1 = 108 u^{N} \zeta_1^N (2+3\zeta_1+u^{N}\zeta_1^N(-3+\zeta_1^2+u^{N}\zeta_1^{N}) );\nonumber \\
&{\rm P2:}& \quad &f_2=-3\eta_1^4(1-v^N), \quad
g_2=2\eta_1^5(\eta_1+v^{N}),\nonumber\\
& &  & \Delta_2 = 108 v^{N} \eta_1^{10} (v^{N}+2\eta_1+\eta_1^2 (3-3v^N+v^{2N})). \nonumber
\end{align}
$\Delta_1=0$ when $u=0$ and $\zeta_1\neq 0$. This implies that $\Delta_2$ in the second patch is meaningful. We then blow up the non-minimal singularity in P2, and we obtain
\begin{align}
&{\rm P2:}&\quad & f_2=-3(1-\zeta_2^N\eta_1^N), \quad
g_2=2(1+\zeta_2^{N}\eta_1^{N-1}),\;\\
& & &\Delta_2= 108 \zeta_2^{N} \eta_1^{N-1} (2+3\eta_1+\zeta_2^{N}\eta_1^{N-1}(1-3\eta_1^2+\zeta_2^{N}\eta_1^{N+2}) )\nonumber \\
&{\rm P3:}& \quad &f_3=-3\eta_2^4(1-v^N), \quad
g_3=2\eta_2^5(\eta_2+v^{N-1}),\nonumber\\
& &  & \Delta_3 = 108 v^{N-1} \eta_2^{10} (v^{N-1}+2\eta_2+\eta_2^2 v (3-3v^N+v^{2N})). \nonumber
\end{align}
We observe that $\Delta_2=0$ if $\zeta_2=0$ and $\eta_1=\frac{1}{\zeta_1}\neq 0$. Therefore the $I_1$ locus intersects the first resolution $\mathbb P^1$, as stated in section \ref{sec:E8sum}. Again, we have seen that in order to understand how the $I_1$ locus interacts with the SCFT, we needed to blow up twice. The $I_1$ locus still contains $v=\eta_2=0$, but this point is still a non-minimal singularity. For this reason we need to apply sequentially the resolution procedure of line \eqref{eq:BU}. We will show in \eqref{eq:fullBUk} that the $I_1$ will still meet the ``last'' self-intersection $(-1)$ $\mathbb P^1$, but this intersection does not carry any nontrivial extra matter.

At last we analyze one more example, i.e $k>1$. For simplicity we consider $k=2$, but the argument will straightforwardly generalize to higher $k$. We repeat the previous strategy where we sequentially blow up the base. In order to see how the $I_1$ locus behaves we have to blow up $k+1$ times. In this case we perform three blowups, and we get:
\begin{align}
&{\rm P1:}&\quad & f_1=-3(1-u^N\zeta_1^N), \quad
g_1=2(1+u^{N}\zeta_1^{N+1}), \\
& & &\Delta_1 = 108 u^{N} \zeta_1^N (2+3 u\zeta^2_1+u^{N}\zeta_1^N(-3+\zeta_1^4 u^2+u^{N}\zeta_1^{N}) ); \nonumber \\
&{\rm P2:}& \quad &f_2=-3(1-\zeta_2^N\eta_1^N), \quad
g_2=2(1+\zeta_2^{N+1}\eta_1^{N}),\nonumber \\
& & &\Delta_2 = 108 \zeta_2^{N} \eta_1^N (3+2 \zeta_2 +\zeta_2^N \eta_1^N(-3+\zeta_2^2 \eta_1^N+\zeta_2^{N}\eta_1^{N}) ) ;\nonumber \\
&{\rm P3:}&\quad & f_3=-3(1-\zeta_3^N\eta_2^N), \quad
g_3=2(1+\zeta_3^{N}\eta_1^{N-1}), \nonumber\\
& & &\Delta_3 = 108 \zeta_3^{N} \eta_2^{N-1} (2+3 \eta_2+\zeta_3^N\eta_2^{N-1}(1-3\eta_2^2+\zeta_3^{N}\eta_2^{N+2}) ); \nonumber \\
&{\rm P4:}& \quad &f_4=-3\eta_3^4(1-v^N), \quad
g_4=2\eta_3^5(\eta_3+v^{N-1}),\nonumber\\
& &  & \Delta_4 = 108 v^{N-1} \eta_3^{10} (v^{N-1}+2\eta_3+\eta_3^2 v (3-3v^N+v^{2N})). \nonumber
\end{align}
The $I_1$ locus intersects at $\eta_1=0, \zeta_3=0, \eta_2 = \frac{1}{\zeta_2}\neq 0$, which means that it meets the second resolution $\mathbb P^1$.

For generic $k$ the $I_1$ locus intersects at the $k$-th resolution $\mathbb P^1$, see also \cite[Fig. 7]{Aspinwall:1997ye}. This becomes manifest when we fully resolve the base for $k>0$,
\begin{align} \label{eq:fullBUk}
&{\rm P1:}&\quad & f_1=-3(1-u^N\zeta_1^N), \quad
g_1=2(1+u^{k+N-1}\zeta_1^N),\;\\
& & &\Delta_1 = 108 u^{N} \zeta_1^N (3+2\zeta_1^k u^{k-1}+u^{2k+N-2}\zeta_1^{2k+N}-3u^N \zeta_1^N+u^{2N}\zeta_1^{2N} )\nonumber \\
&{\rm P(i+1):}& \quad &f_i=-3(1-\zeta_{i+1}^N\zeta_i^N), \quad
g_i=2(1+\zeta_{i+1}^{N+k-i}\eta_i^{N+k-1-i}), \nonumber \\
& & &\Delta_i = 108 \zeta_{i+1}^{N} \eta_i^{N} (3+ 2 \zeta_{i+1}^{k-i}\eta_{i}^{k-1-i} +\zeta_{i+1}^{2k+N-2i}\eta_{i}^{2k+N-2-2i} +\zeta_{i+1}^{N} \eta_i^{N} (-3 + \zeta_{i+1}^{N} \eta_i^{N}))\nonumber\\
& & & 1\leq i \leq k-2\nonumber\\
&{\rm P(k+i+1):}& \quad &f_{k+i}=-3(1-\zeta_{k+i+1}^N\zeta_{k+i}^N), \quad
g_i=2(1+\zeta_{k+i+1}^{N-i}\eta_{k+i}^{N-1-i}), \nonumber \\
& & &\Delta_i = 108 \zeta_{k+i+1}^{N-i} \eta_{k+i}^{N-i-1} (2+3\zeta_{k+i+1}^i \eta_{k+i}^{i+1}-3 \zeta_{k+i+1}^{N+i} \eta_{k+i}^{N+i+1} + \zeta_{k+i+1}^{N-i} \eta_{k+i}^{N-i-1}+\nonumber \\
& & &  +\zeta_{k+i+1}^{2N+i} \eta_{k+i}^{2N+i+1} )\quad
 1\leq i \leq N-2 \nonumber\\
&{\rm P(N+k):}& \quad &f_{k+N}=-3\eta_{k+N}^4(1-v^N), \quad
g_{k+N}=2\eta_{k+N}^5(1+\eta_{k+N}),\nonumber\\
& &  & \Delta_{k+N} = 108\eta_{k+N}^{10} (1+2\eta_{k+N}+\eta_{k+N}^2 v(3-3v^N+v^{2N})), \nonumber
\end{align}
Similarly to $k=0$, the $I_1$ meets the (self-intersection $-1$) curve, but this does not add any matter to the SCFT spectrum.

What we have seen admits a dual interpretation in the Type IIA brane configurations of \cite{Hanany:1997gh, Brunner:1997gk, Gaiotto:2014lca, Bah:2017wxp, Apruzzi:2017nck}, namely the extra $I_1$ corresponds to a single D8-brane intersecting $N$ D6-branes.

\bibliographystyle{utphys}
\bibliography{fabelian}

\end{document}